\documentclass[12pt]{iopart}
\usepackage{iopams} 
\usepackage{mathbbol} 
\usepackage[colorlinks, linkcolor =blue]{hyperref} 
\usepackage{amsthm} 
\usepackage{mathrsfs} 
\usepackage{graphicx} 
\usepackage{esint}
\usepackage{color}
\usepackage{epsf}
\usepackage{subfigure}
\usepackage{epstopdf}
\usepackage{caption}
\newcommand{\real}{\mathrm{Re}}
\newcommand{\imag}{\mathrm{Im}}
\newcommand{\bigO}{\mathcal{O}}

\newcommand{\lambdabar}{\mathchar'26\mkern-9mu \lambda}

\newtheorem*{conjecture*}{Conjecture} 

\begin{document}


\title{Phase-space representation of diffraction in time: Analytic results}

\author{Maximilien Barbier$^{1,2}$ and Arseni Goussev$^3$}

\address{$^1$Max-Planck-Institut f\"ur Physik komplexer Systeme, N\"othnitzer Str. 38, D-01187 Dresden, Germany \\
$^2$Center for Nonlinear Phenomena and Complex Systems,Universit\'e Libre de Bruxelles (ULB), Code Postal 231, Campus Plaine, 
B-1050 Brussels, Belgium \\
$^3$School of Mathematics and Physics, University of Portsmouth, Portsmouth PO1 3HF, United Kingdom}

\ead{mbarbier@pks.mpg.de}


\begin{abstract}
Diffraction in time manifests itself as the appearance of probability-density fringes when a matter wave passes through an opaque screen with abrupt temporal variations of transmission properties. Here we analytically describe the phase-space structure of diffraction-in-time fringes for a class of smooth time gratings. More precisely, we obtain an analytic expression for the Husimi distribution representing the state of the system in the case of time gratings comprising a succession of Lorentzian-like slits. In particular, for a double-slit scenario, we derive a simple and intuitive expression that accurately captures the position of interference fringes in phase space.
\end{abstract}

\vspace{2pc}


\section{Introduction}\label{Intro_sec}

When a matter wave passes through a barrier with abrupt spatial variations of transmission properties (e.g., an opaque screen with one or multiple apertures) it exhibits interference/diffraction \footnote{Quoting Richard Feynman~\cite{Feynman}, ``No one has ever been able to define the difference between interference and diffraction satisfactorily. It is just a question of usage, and there is no specific, important physical difference between them.''}, which manifests itself as the appearance of characteristic fringes in the probability density. The most iconic example that illustrates such a behavior is the celebrated Youg's double-slit scenario, whose experimental implementation for single electrons~\cite{Jon61,TEM89} has even been informally referred to as ``the most beautiful experiment in physics''~\cite{Cre02}. Similar probability-density fringes appear when a matter wave interacts with a screen whose transmission properties are spatially uniform but change abruptly in time (e.g., a neutron shatter). This phenomenon, commonly referred to as diffraction in time, was discovered by Moshinsky~\cite{Mos52} and received much attention in the literature (see Ref.~\cite{CGM09} for a review). In particular, there have been several experimental realizations of this effect~\cite{SGA96, ASD96, HFG98, LSW05, CMP05, Paulus}.

There are different approaches to describing diffraction in time analytically, each involving its own assumptions and approximations. We outline them here in the context of the following physical problem, which is at the heart of the present work. We consider a (nonrelativistic and stuctureless) quantum particle moving in one dimension along the $x$ axis, in the presence of a shutter at $x=0$. The particle starts from the $x < 0$ region and moves towards the shutter. The latter is an infinitesimally thin barrier that can be opened and closed (suddenly or gradually) in the course of time. During the time intervals when the barrier is fully open, the system behaves simply as a free particle on a line. On the other hand, a completely closed barrier blocks the matter-wave exchange between the incident ($x<0$) and transmission ($x>0$) regions.

Our main goal is to analytically address the appearance and structure of probability-density fringes caused by the time variations of the shutter transparency. Needless to say, physical properties of the shutter, e.g. its reflectivity and absorptivity, strongly depend on the particular realization of the particle-shutter scenario, and, essentially, all existing analytic approaches to this problem differ from one another in how they mathematically describe the shutter.

One apparent way to model the shutter is to introduce a time-dependent Dirac delta-potential $V(x,t) = \omega(t) \delta(x)$ in the Hamiltonian that governs the dynamics of the particle. The positive function $\omega(t)$ ranges between $0$ and $\infty$, and controls the shutter transparency: the shutter is open when $\omega=0$ and closed when $\omega = \infty$. The main difficulty with this approach though is that analytic solutions to the corresponding time-dependent Schr\"odinger equation are only known in very few special cases: $\omega(t) = \mbox{constant}$~\cite{GS86, Bli88}, $\omega(t) \propto t$~\cite{Dem64}, $\omega(t) \propto 1/t$~\cite{SK88}, and $\omega(t) \propto (a t^2 + b t + c)^{-1/2}$ with $a$, $b$, $c$ being some constants~\cite{DMN92}. This severely limits the range of diffraction-in-time systems amenable to analytic investigation. For instance, the time-domain version of the double-slit scenario does not appear to be analytically accessible within the delta-potential approach. 

An alternative analytic approach to investigating the particle-shutter system is to focus entirely on the transmission, i.e. $x > 0$, region. The particle's wave function $\Psi(x,t)$ satisfies the free-particle Schr\"odinger equation for all $x>0$ and $t>0$. Initially, at $t = 0$, the particle is assumed to have a zero probability to be found in the transmission region: $\Psi(x,0) = 0$ for all $x > 0$. The action of the shutter is then introduced by means of a time-dependent boundary condition imposed on the wave function at $x=0$, namely $\Psi(0,t) = a(t)$ for all $t > 0$. Here $a(t)$ is a given (complex-valued) function that essentially contains all information about the initial wave packet in the incident ($x < 0$) region, as well as the time protocol of the shutter opening. This approach is very popular in diffraction-in-time studies and has been extensively used in the literature (see, e.g., Refs.~\cite{HFG98, BZ97, CMM07, GOC07, TMB11, DM15}). What it does not take into account however is the back-action of the transmitted matter wave on the incoming wave in the incident region. Such a back-action may arise, for instance, as a consequence of the so-called quantum backflow effect (see Refs.~\cite{All69, BM94} for pioneering works on quantum backflow and the introduction section in Ref.~\cite{Bra21} for the most up-to-date review of the literature).  

Another approach, which was originally developed in Refs.~\cite{Gou12, Gou13} and that we hereinafter refer to as the aperture function model, is of particular relevance to the present work. It models a shutter that can absorb but not scatter the incident matter wave. The shutter's transparency can depend arbitrarily on time, and this dependence is represented by an aperture function $\chi(t)$. The latter ranges between 0 (perfectly absorbing, closed shutter) and 1 (perfectly transmitting, open shutter). In the aperture function model, the shutter is modelled by means of discontinuous time-dependent matching conditions, involving $\chi(t)$, that connect the value of the wave function $\Psi$ and its spatial derivative $\partial \Psi / \partial x$ across the shutter. These matching conditions are closely related to Kottler's matching conditions used to justify Kirchhoff's diffraction theory in stationary-wave optics~\cite{Kot23, Kot65, NHL95}, the latter being known to yield experimentally relevant predictions in the transmission region~\cite{NHL95}. The aperture function model then allows one to express the particle's wave function in the transmission region as an integral involving the initial state $\Psi(x,0)$ and the aperture function $\chi(t)$. This transmitted wave function constructed from the aperture function model has been explicitly shown to be consistent with the wave function obtained from a first-principle analysis, based on a delta-potential $\omega(t) \delta(x)$, of physically relevant atom-optics systems~\cite{BBG15}.

In this paper, we analytically describe the interference fringes obtained in a class of diffraction-in-time scenarios by extending the aperture function model into phase space. More precisely, assuming that the incident particle is characterized by a fast localized (Gaussian) wave packet, we derive an expression for the Husimi quasiprobability distribution representing the part of the matter wave transmitted through the shutter. We then evaluate this expression analytically for the case when the shutter aperture function $\chi(t)$ consists of a finite number of Lorentzian-like slits. Our final analytic result, obtained for narrow slits, offers a simple and intuitive description of (multiple) slit interference/diffraction in the time domain. Special attention is devoted to the single- and double-slit cases.

The paper is organized as follows. We first recall in section~\ref{AFM_sec} the definition of the aperture function model. We see in particular that the Husimi distribution associated to the transmitted state can be expressed as an integral over a finite time interval. We then discuss in section~\ref{complex_integral_sec} how the latter can be written as a contour integral in the complex plane that allows to apply Cauchy's residue theorem. We see in particular that this approach requires one to compute a residue at an essential singularity. This challenging technical difficulty is addressed in section~\ref{apodization_sec}. Here we explicitly compute this residue, and thus the resulting Husimi distribution, for a particular class of aperture functions $\chi_n$ of the form $\chi_n (t) = 1/[1+\nu^n (t-T)^n]$, with $n$ an even integer. Such Lorentzian-like functions describe apodization barriers, which smoothly open around the time $T$ and have a width $1/\nu$. The underlying integral yielding the Husimi distribution being by construction linear in $\chi$, our results can thus be straightforwardly extended to aperture functions that consist in an arbitrary sequence of such $\chi_n$: taking different opening times $T$ then effectively describes a time grating. We treat the particular (Lorentzian) case $n=2$ in section~\ref{n_2_sec}. Considering the slit regime $\nu \gg 1$ of narrow Lorentzian barriers $\chi_2$ allows us to considerably simplify our analytic expression of the Husimi distribution. In particular, we are able to derive a simple analytic expression of the position of the interference fringes in the phase space. Our work is summarized and concluding remarks are drawn in section~\ref{concl_sec}. Additional technical details are deferred to the appendices.


\section{The aperture function model of matter wave absorption}\label{AFM_sec}

In this section we discuss how the dynamics of quantum wave packets in the presence of a barrier can be adequately described by a particular model of matter-wave absorption. The latter was originally devised in~\cite{Gou12, Gou13} and is hereinafter referred to as the aperture function model. We consider a nonrelativistic structureless quantum particle of mass $m$ that moves in one dimension along the $x$ axis, and whose state at time $\tau$ is described by a wave function $\Psi (x,\tau)$. The absorbing barrier is taken to be time dependent and pointlike, located at position $x=0$.

The particle is assumed to be prepared at the initial time $\tau=0$ in the minimum-uncertainty Gaussian wave packet $\psi_{\sigma, \, x_0, \, v_0} (x,0)$, i.e.
\begin{eqnarray}
\Psi(x,0) = \psi_{\sigma, \, x_0, \, v_0} (x,0) \, ,
\label{Psi_0_def}
\end{eqnarray}
given by
\begin{eqnarray}
\psi_{\sigma, \, x_0, \, v_0} (x,0) \equiv \left( \frac{1}{\pi \sigma^2} \right)^{1/4} \exp \left[ - \frac{(x-x_0)^2}{2 \sigma^2} + i \frac{m v_0}{\hbar} (x-x_0) \right] \, ,
\label{init_Gaussian_def}
\end{eqnarray}
where $x_0$ and $v_0$ correspond to the initial mean position and mean velocity, respectively, of the particle, while $\sigma > 0$ characterizes the width of the wave packet. Throughout this work, we use velocities $v$ rather than momenta $p=mv$. The freely evolved state $\psi_{\sigma, \, x_0, \, v_0} (x,\tau) = \int dx' \, K_{0} (x-x',\tau) \, \psi_{\sigma, \, x_0, \, v_0} (x',0)$ at some time $\tau$, with
\begin{eqnarray}
K_{0} (\xi,\tau) = \sqrt{\frac{m}{2 i \pi \hbar \tau}} \, \exp \left(i \frac{m\xi^2}{2 \hbar \tau} \right)
\label{free_propa_def}
\end{eqnarray}
the well-known free-particle propagator (see e.g.~\cite{Schul}), is then given by
\begin{eqnarray}
\psi_{\sigma, \, x_0, \, v_0} (x,\tau) = \left( \frac{1}{\pi \sigma_{\tau}^2} \right)^{1/4} \exp \left[ - \frac{(x-x_{\tau})^2}{2 \sigma_{\tau}^2} + i \mathcal{S}(x,\tau) \right] \, ,
\label{free_Gaussian_expr}
\end{eqnarray}
where
\begin{eqnarray}
x_{\tau} \equiv x_0 + v_0 \tau \qquad \mbox{and} \qquad \sigma_{\tau} \equiv \sigma \sqrt{1 + \left( \frac{\hbar \tau}{m \sigma^2} \right)^2}
\label{sigma_tau_def}
\end{eqnarray}
denote the mean position and the width, respectively, while the phase $\mathcal{S}$ is given by
\begin{eqnarray}
\fl \mathcal{S}(x,\tau) \equiv \frac{(x-x_{\tau})^2}{2 \sigma_{\tau}^2} \, \frac{\hbar \tau}{m \sigma^2} + \frac{m v_0}{\hbar} (x-x_{\tau}) + \frac{m v_0^2 \tau}{2 \hbar} - \frac{1}{2} \mathrm{Arctan} \left( \frac{\hbar \tau}{m \sigma^2} \right) \, .
\label{S_x_tau_def}
\end{eqnarray}

We now discuss the dynamics of the particle in the presence of the time-dependent absorbing barrier. Before we introduce the aperture function model in section~\ref{model_sec}, we first briefly discuss in section~\ref{semi_clas_reg_sec} the so-called frozen Gaussian regime in which this model has been shown~\cite{BBG15} to yield accurate physical predictions. Finally, section~\ref{Husimi_sec} is devoted to defining and writing the Husimi distribution, the quantity that is at the heart of the present work, within this model.


\subsection{Frozen Gaussian regime}\label{semi_clas_reg_sec}

Here we discuss the particular dynamical regime that we consider throughout this work, and that can be summarized by the set of assumptions
\begin{eqnarray}
\sigma \ll -x_0 = |x_0| \lesssim x_t = v_0 (t-t_{\mathrm{c}}) \ll \frac{m \sigma^2 v_0}{\hbar} \, ,
\label{semiclassical_reg_def}
\end{eqnarray}
where
\begin{eqnarray}
t_{\mathrm{c}} \equiv \frac{\left| x_0 \right|}{v_0} = - \frac{x_0}{v_0}
\label{semiClassicalTime_def}
\end{eqnarray}
corresponds to the time needed for a classical free particle initially located at position $x_0$ and moving with the velocity $v_0$ to reach the barrier at $x=0$. For this reason, we hereinafter refer to $t_{\mathrm{c}}$ as the classical hitting time. Here and in the sequel we take the final time of propagation $t > t_{\mathrm{c}}$ of the particle to be a fixed parameter.


\begin{figure}[ht]
\centering
\includegraphics[width=0.45\textwidth]{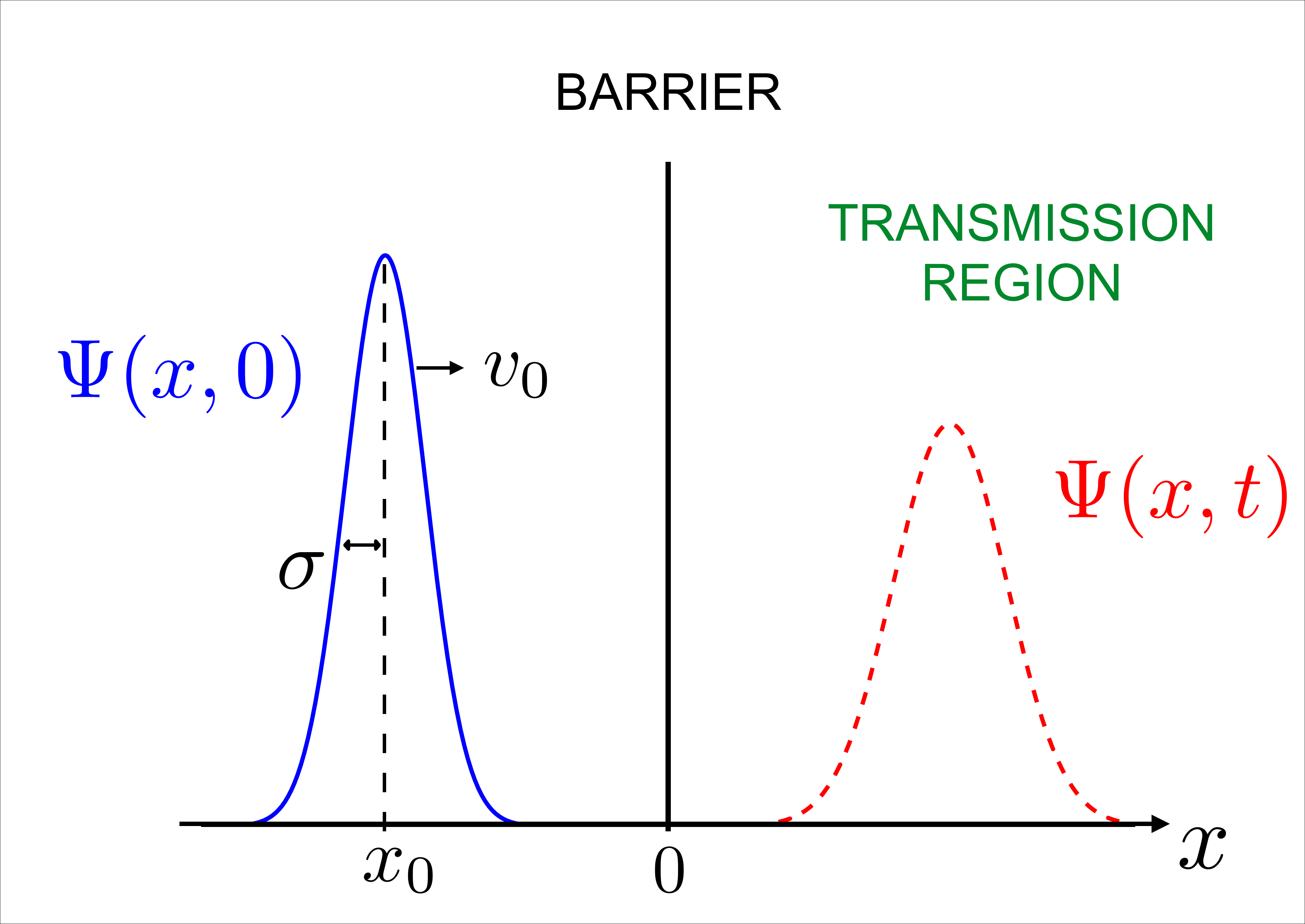}
\caption{Physical picture of the dynamical regime described by~\eref{semiclassical_reg_def}, with the initial Gaussian state $\Psi(x,0)$ (blue) and a sketch of the final transmitted state $\Psi(x,t)$ (red).}
\label{phys_pict_fig}
\end{figure}


The regime~\eref{semiclassical_reg_def} allows for an intuitive picture, schematized on figure~\ref{phys_pict_fig}, of the dynamics of the particle. The leftmost condition in~\eref{semiclassical_reg_def} states that the particle is initially localized on the left of the barrier. It then moves towards the latter with a positive velocity $v_0>0$, crosses the barrier around the classical hitting time $t_{\mathrm{c}}$ before it reaches the (mean) position $x_t$ that, according to the second condition in~\eref{semiclassical_reg_def}, is in the transmission region $x>0$. Finally, the rightmost condition in~\eref{semiclassical_reg_def} ensures that the particle is localized on the right of the barrier at the final time $t$.

Indeed, the latter condition yields in particular
\begin{eqnarray}
\frac{\hbar t}{m \sigma^2} \ll 1 \, .
\label{frozen_Gaussian_cond}
\end{eqnarray}
This ensures, in view of~\eref{sigma_tau_def}, that the freely evolved wave packet $\psi_{\sigma, \, x_0, \, v_0} (x,t)$ [which, as sketched on figure~\ref{phys_pict_fig}, is known~\cite{BBG15,Bar17} to be representative of the actual transmitted state $\Psi(x,t)$] does not spread in the time $t$. This characterizes the so-called frozen Gaussian approximation~\cite{Hel81}, hence the name frozen Gaussian regime that we give to~\eref{semiclassical_reg_def}. The latter is well within the reach of experiments using ultracold atoms (such as e.g.~\cite{FCG11,JMR12,CFV13}).

With the frozen Gaussian regime~\eref{semiclassical_reg_def} in hand we can now introduce the particular model that we consider in order to describe the time-dependent absorbing barrier.


\subsection{Model}\label{model_sec}

Here we introduce the aperture function model, which is a model of time-dependent absorption that was originally devised in~\cite{Gou12,Gou13}. The presence of the barrier is taken into account by imposing discontinuous time-dependent matching conditions on both the wave function $\Psi$ and its spatial derivative $\partial \Psi / \partial x$ at $x=0$.

The problem can be equivalently formulated in terms of the propagator $K$, which fully specifies the dynamics of the quantum particle between the initial time $\tau = 0$ and the final time $\tau = t$. In view of the frozen Gaussian regime~\eref{semiclassical_reg_def}, the propagator here corresponds to a function $K(x,x',t)$ that relates the transmitted state $\Psi(x,t)$ of the particle at time $t$ to its initial state $\Psi(x,0)$ through
\begin{eqnarray}
\Psi(x,t) = \int_{- \infty}^{0} dx' \, K(x,x',t) \Psi(x',0) \, .
\label{transm_state_def}
\end{eqnarray}
We emphasize that the integration range in~\eref{transm_state_def} is restricted to $x'<0$ since we assume the particle to be initially localized on the left of the barrier. That is, the variable $x'$ of $K$ can be restricted to take negative-only values.

The propagator $K$ is then constructed as follows~\cite{Gou12,Gou13}. It is first required to obey the free-particle time-dependent Schr\"odinger equation on both sides of the barrier, i.e.
\begin{eqnarray}
\left[ i \hbar \frac{\partial}{\partial \tau} + \frac{\hbar ^2}{2m} \frac{\partial ^2}{\partial x ^2} \right] K (x,x',\tau) = 0
\label{TDSE_K}
\end{eqnarray}
for any $0 < \tau < t$ and $x,x' \neq 0$. Then, it is set to satisfy the usual initial condition
\begin{eqnarray}
K (x,x',0^+) = \delta (x-x') \, ,
\label{IC_K}
\end{eqnarray}
where $0^+$ merely means the limit $\epsilon \to 0$ with $\epsilon > 0$, as well as Dirichlet boundary conditions at $x \to \pm \infty$ for negative imaginary times, i.e.
\begin{eqnarray}
K (x \to \pm \infty ,x',\tau) = 0 \qquad \mbox{for} \qquad \tau = -i | \tau | \, .
\label{Dirichlet_K}
\end{eqnarray}
Finally, discontinuous time-dependent matching conditions are imposed on the propagator and its spatial derivative at $x=0$, reading (since $x'<0$ here)
\begin{eqnarray}
\left. K (x,x',\tau) \right|_{x=0^-}^{x=0^+} = - \left[ 1-\chi (\tau) \right] \left. K_0 (x-x',\tau) \right|_{x=0}
\label{matchingCond_K}
\end{eqnarray}
and
\begin{eqnarray}
\left. \frac{\partial}{\partial x} K (x,x',\tau) \right|_{x=0^-}^{x=0^+} = - \left[ 1-\chi (\tau) \right] \left. \frac{\partial}{\partial x} K_0 (x-x',\tau) \right|_{x=0} \, ,
\label{matchingCond_derivK}
\end{eqnarray}
for any $0 < \tau < t$. The real-valued time-dependent function $\chi(\tau)$ in~\eref{matchingCond_K}-\eref{matchingCond_derivK} embeds the absorbing properties of the barrier. It is required to satisfy $0 \leqslant \chi(\tau) \leqslant 1$ at any time $\tau$, with $\chi=0$ ($\chi=1$) corresponding to a fully absorbing (fully transparent) barrier. For this reason, here and in the sequel $\chi$ is referred to as the aperture function of the barrier, and thus the model itself as the aperture function model. The function $\chi$ has been explicitly connected to the (time-dependent) intensity of a laser beam in~\cite{BBG15}.

It can then be shown~\cite{Gou13} that the propagator $K (x,x',t)$, unique solution to the well-posed problem formed by~\eref{TDSE_K}-\eref{matchingCond_derivK}, is given by
\begin{eqnarray}
K (x,x',t) = \int_{0}^{t} d\tau \, \frac{\chi (\tau)}{2} \left( \frac{x}{t-\tau} - \frac{x'}{\tau} \right) K_0 (x,t-\tau) K_0 (x',\tau)
\label{K_afm_expr}
\end{eqnarray}
in the transmission region, i.e. for $x>0$, and for an \textit{arbitrary} $\chi(\tau)$. Substituting~\eref{Psi_0_def} and~\eref{K_afm_expr} into~\eref{transm_state_def} then yields the following expression of $\Psi(x,t)$~\cite{Gou15}:
\begin{eqnarray}
\Psi (x,t) = \int_{0}^{t} d\tau \, \frac{\chi (\tau)}{2} \left[ \frac{x}{t-\tau} + \frac{v_0 \, \alpha_{\tau}}{\alpha_{t_{\mathrm{c}}}} \right] K_0 (x,t-\tau) \psi_{\sigma, \, x_0, \, v_0} (0,\tau) \, ,
\label{transm_state_expr}
\end{eqnarray}
with
\begin{eqnarray}
\alpha_{\tau} \equiv \frac{1}{2 \sigma^2} \frac{1}{1 + i \hbar \tau / m \sigma^2} = \frac{1}{2 \sigma_{\tau}^2} \left( 1 - i \frac{\hbar \tau}{m \sigma^2} \right) \, .
\label{alpha_T_def}
\end{eqnarray}
The transmitted state $\Psi(x,t)$ is in general not normalized as a consequence of absorption: that is, as soon as $\chi(\tau) \neq 1$, we have $\int dx \left| \Psi \right|^2 \leqslant 1$.

Now, the structure of~\eref{transm_state_expr} suggests to consider not $\Psi$ itself, but rather a particular phase-space representation of $\Psi$ given by the Husimi distribution, as we now discuss.


\subsection{Husimi distribution}\label{Husimi_sec}

In this section we construct the Husimi distribution associated with the transmitted state $\Psi(x,t)$ given by~\eref{transm_state_expr}. The former is a well-known phase-space representation of a quantum state (see e.g.~\cite{Lee95,LCohen}). We hereinafter denote by $\widetilde{x}$ and $\widetilde{v}$ the phase-space variables corresponding to position and velocity, respectively. We also introduce for convenience the (reduced) de Broglie wavelengths $\lambdabar$, $\widetilde{\lambdabar}$ and time scale $\widetilde{t}$ given by
\begin{eqnarray}
\lambdabar \equiv \frac{\hbar}{m v_0} \qquad , \qquad \widetilde{\lambdabar} \equiv \frac{\hbar}{m \widetilde{v}} \qquad \mbox{and} \qquad \widetilde{t} \equiv \frac{\widetilde{x}}{\widetilde{v}} \, .
\label{lambdaBar_t_tilde_def}
\end{eqnarray}

The Husimi distribution, which we denote by $F$, is a nonnegative function of $\widetilde{x}$ and $\widetilde{v}$ that can be written as
\begin{eqnarray}
F (\widetilde{x},\widetilde{v},t) \equiv \left| f (\widetilde{x},\widetilde{v},t) \right| ^2
\label{Husimi_distr_from_Husimi_ampl}
\end{eqnarray}
in terms of the complex-valued function $f$ that we define by
\begin{eqnarray}
f (\widetilde{x},\widetilde{v},t) \equiv \int_{- \infty}^{\infty} dx \left[ \psi_{\sigma, \, \widetilde{x}, \, \widetilde{v}} (x,0) \right]^{*} \Psi (x,t) \, ,
\label{Husimi_ampl_def}
\end{eqnarray}
where the asterisk denotes complex conjugation. In view of~\eref{Husimi_distr_from_Husimi_ampl} we hence refer to $f$ as the Husimi amplitude. As is clear from~\eref{Husimi_ampl_def}, $f$ is by construction the overlap between $\Psi$ and a minimum-uncertainty Gaussian wave packet $\psi_{\sigma, \, \widetilde{x}, \, \widetilde{v}} (x,0)$ with mean position $\widetilde{x}$, mean velocity $\widetilde{v}$ and width $\sigma$ [indeed identical to the width of the initial state~\eref{Psi_0_def}].

The frozen Gaussian regime~\eref{semiclassical_reg_def} ensures that $F (\widetilde{x},\widetilde{v},t)$ only takes non negligible values in the phase-space quadrant where $\widetilde{x}>0$ and $\widetilde{v}>0$. This is discussed in details in~(\cite{Bar17}, Section 5.1). Substituting the expression~\eref{transm_state_expr} of $\Psi (x,t)$ into~\eref{Husimi_ampl_def}, it can be shown~\cite{Gou15} that $f(\widetilde{x},\widetilde{v},t)$ is then given by
\begin{eqnarray}
\fl f(\widetilde{x},\widetilde{v},t) = \int_{0}^{t} d\tau \, \frac{\chi (\tau)}{2} \left[ \frac{\widetilde{v} \, \alpha_{t-\tau}}{\alpha_{\widetilde{t}}} + \frac{v_0 \, \alpha_{\tau}}{\alpha_{t_{\mathrm{c}}}} \right] \psi_{\sigma, \, \widetilde{x}, \, - \widetilde{v}} (0,t-\tau) \psi_{\sigma, \, x_0, \, v_0} (0,\tau) \, .
\label{Husimi_ampl_final_expr}
\end{eqnarray}
This expression yields a formal solution for the Husimi amplitude corresponding to the transmitted state $\Psi(x,t)$. The physics that is embedded in $f$ can be for instance accessed by numerically computing the integral in~\eref{Husimi_ampl_final_expr}. Here the challenge, and the main aim of our study, is to \textit{analytically} compute this integral. As we discuss in the remaining part of this work, we perform this challenging task for a particular class of physically relevant aperture functions $\chi(\tau)$ by means of complex analysis.

To this end, we rewrite the Husimi amplitude~\eref{Husimi_ampl_final_expr} in an alternative form. Combining~\eref{Husimi_ampl_final_expr} with~\eref{free_Gaussian_expr}-\eref{S_x_tau_def} and~\eref{alpha_T_def}, we show (details may be found in~\ref{f_froz_app}) that, in the frozen Gaussian regime~\eref{semiclassical_reg_def}, $f$ can be written in the form
\begin{eqnarray}
f(\widetilde{x},\widetilde{v},t) = \left[ 1 + \bigO \left( \frac{\hbar t}{m \sigma^2} \right) \right] f_{\mathrm{froz}} (\widetilde{x},\widetilde{v},t) \, ,
\label{Husimi_ampl_h_sc_def}
\end{eqnarray}
where we introduced the frozen Gaussian Husimi amplitude
\begin{eqnarray}
f_{\mathrm{froz}} (\widetilde{x},\widetilde{v},t) \equiv \int_{0}^{t} d\tau \, \widetilde{f}_{\mathrm{froz}}(\tau) \, ,
\label{Husimi_ampl_sc_def}
\end{eqnarray}
with the function $\widetilde{f}_{\mathrm{froz}}$ (whose dependence on $\widetilde{x}, \widetilde{v}$ is dropped for compactness) given by
\begin{eqnarray}
\widetilde{f}_{\mathrm{froz}}(\tau) \equiv \frac{1}{4 \sqrt{\pi} \sigma^3} \left( \frac{\widetilde{v}}{\alpha_{\widetilde{t}}} + \frac{v_0}{\alpha_{t_{\mathrm{c}}}} \right) \chi (\tau) \, e^{\varphi (\tau)} \, .
\label{h_sc_def}
\end{eqnarray}
Here the dimensionless complex-valued function $\varphi(\tau)$ is defined by
\begin{eqnarray}
\varphi (\tau) \equiv \frac{T_0}{\tau - \tau_0} + \frac{T_1}{\tau - \tau_1} - \frac{1}{2} \left( \frac{\sigma}{\widetilde{\lambdabar}} \right)^2 - \frac{1}{2} \left( \frac{\sigma}{\lambdabar} \right)^2
\label{varPhi_def}
\end{eqnarray}
in terms of the complex quantities
\begin{eqnarray}
\tau_0 \equiv i \frac{m \sigma^2}{\hbar} \qquad \mbox{and} \qquad \tau_1 \equiv t - i \frac{m \sigma^2}{\hbar} \, ,
\label{tau_0_tau_1_def}
\end{eqnarray}
\begin{eqnarray}
\fl T_0 \equiv - \frac{i}{2} \left( \frac{\sigma}{\lambdabar} \right)^2 \left( 1 + i \frac{\lambdabar \left| x_0 \right|}{\sigma^2} \right)^2 \frac{m \sigma^2}{\hbar} \qquad \mbox{and} \qquad T_1 \equiv \frac{i}{2} \left( \frac{\sigma}{\widetilde{\lambdabar}} \right)^2 \left( 1 + i \frac{\widetilde{\lambdabar} \widetilde{x}}{\sigma^2} \right)^2 \frac{m \sigma^2}{\hbar} \, .
\label{T_0_T_1_def}
\end{eqnarray}

The form~\eref{Husimi_ampl_h_sc_def}-\eref{Husimi_ampl_sc_def} of the Husimi amplitude is well suited to an analytic evaluation by means of complex analysis. Our strategy is thus to resort to Cauchy's residue theorem in order to explicitly compute the integral in~\eref{Husimi_ampl_sc_def}, as we now discuss.


\section{The Husimi amplitude as a residue}\label{complex_integral_sec}

In this section we express the frozen Gaussian Husimi amplitude~\eref{Husimi_ampl_sc_def} as a contour integral in the complex plane that allows to apply the residue theorem.


\begin{figure}[ht]
\centering
\includegraphics[width=0.40\textwidth]{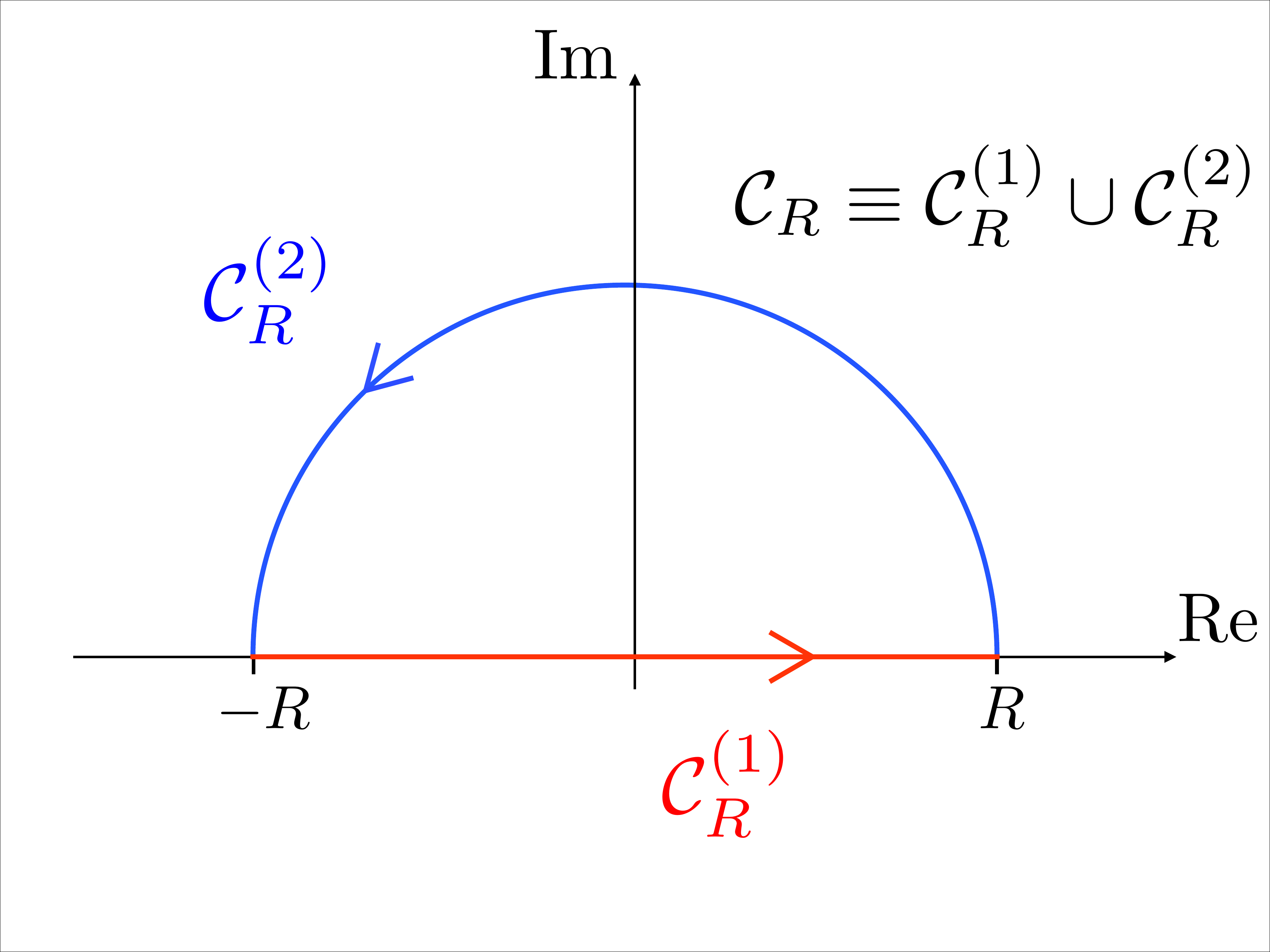}
\caption{Contours $\mathcal{C}_R^{(1)}$, corresponding to the segment line $\left[ -R , R \right]$ of the real axis, and $\mathcal{C}_R^{(2)}$, corresponding to the upper half-circle with center the origin and radius $R$, defining the simple closed contour $\mathcal{C}_R \equiv \mathcal{C}_R^{(1)} \cup \mathcal{C}_R^{(2)}$ run in the positive direction.}
\label{contour_fig}
\end{figure}


In view of~\eref{Husimi_ampl_sc_def} we introduce the contour integral $f_{\mathrm{froz}}^{\ointctrclockwise}$ defined by the limit
\begin{eqnarray}
\fl f_{\mathrm{froz}}^{\ointctrclockwise} \equiv \lim\limits_{R \to \infty} \ointctrclockwise_{\mathcal{C}_R} d z \, \widetilde{f}_{\mathrm{froz}} (z) = \int_{- \infty}^{\infty} d \tau \, \widetilde{f}_{\mathrm{froz}} (\tau) + \lim\limits_{R \to \infty} \left[ i R \int_{0}^{\pi} d \theta \, e^{i \theta} \widetilde{f}_{\mathrm{froz}} \left( R \, e^{i \theta} \right) \right] \, ,
\label{complexInt_f_froz_def}
\end{eqnarray}
with $\mathcal{C}_R \equiv \mathcal{C}_R^{(1)} \cup \mathcal{C}_R^{(2)}$ the simple closed contour, run in the positive direction, that is depicted on figure~\ref{contour_fig}. Comparing~\eref{complexInt_f_froz_def} with~\eref{Husimi_ampl_sc_def} hence shows that
\begin{eqnarray}
f_{\mathrm{froz}} (\widetilde{x},\widetilde{v},t) = f_{\mathrm{froz}}^{\ointctrclockwise} - I_{\mathcal{C}_{\infty}^{(2)}} - I^{(-)} - I^{(+)} \, ,
\label{I_fr_decomposition_complexInt}
\end{eqnarray}
where we introduced the integrals $I_{\mathcal{C}_{\infty}^{(2)}}$, $I^{(-)}$ and $I^{(+)}$ defined by
\begin{eqnarray}
I_{\mathcal{C}_{\infty}^{(2)}} \equiv \lim\limits_{R \to \infty} \left[ i R \int_{0}^{\pi} d \theta \, e^{i \theta} \widetilde{f}_{\mathrm{froz}} \left( R \, e^{i \theta} \right) \right] \, ,
\label{I_C_R_def}
\end{eqnarray}
\begin{eqnarray}
I^{(-)} \equiv \int_{- \infty}^{0} d \tau \, \widetilde{f}_{\mathrm{froz}} (\tau) \qquad \mbox{and} \qquad I^{(+)} \equiv \int_{t}^{\infty} d \tau \, \widetilde{f}_{\mathrm{froz}} (\tau) \, .
\label{I_minus_I_plus_def}
\end{eqnarray}

We now use the decomposition~\eref{I_fr_decomposition_complexInt} to explicitly compute $f_{\mathrm{froz}}$. Our strategy is as follows. We first show in section~\ref{half_circle_sec} that the integral $I_{\mathcal{C}_{\infty}^{(2)}}$ vanishes under suitable conditions imposed on the aperture function $\chi$. We then analyze the line integrals $I^{(-)}$ and $I^{(+)}$ in section~\ref{upper_bound_sec} and determine an upper bound for $|I^{(-)} + I^{(+)}|$. Finally, we express in section~\ref{res_sec} the contour integral $f_{\mathrm{froz}}^{\ointctrclockwise}$ by means of the residue theorem, for which the main difficulty is seen to arise from the presence of an essential singularity.


\subsection{The half-circle integral \texorpdfstring{$I_{\mathcal{C}_{\infty}^{(2)}}$}{along the upper half circle}}\label{half_circle_sec}

Here we determine a sufficient condition that the aperture function $\chi$ must satisfy for the integral $I_{\mathcal{C}_{\infty}^{(2)}}$ along the upper half circle to vanish. 

Combining~\eref{I_C_R_def} with~\eref{h_sc_def}-\eref{varPhi_def} yields for $I_{\mathcal{C}_{\infty}^{(2)}}$
\begin{eqnarray}
\fl I_{\mathcal{C}_{\infty}^{(2)}} = i \frac{1}{4 \sqrt{\pi} \sigma^3} \left( \frac{\widetilde{v}}{\alpha_{\widetilde{t}}} + \frac{v_0}{\alpha_{t_{\mathrm{c}}}} \right) \exp \left[ - \frac{1}{2} \left( \frac{\sigma}{\widetilde{\lambdabar}} \right)^2 - \frac{1}{2} \left( \frac{\sigma}{\lambdabar} \right)^2 \right] \nonumber\\
\times \int_{0}^{\pi} d \theta \lim\limits_{R \to \infty} \left[ R \, e^{i \theta} \chi \left( R \, e^{i \theta} \right) \mathrm{exp} \left( \frac{T_0}{R \, e^{i \theta} - \tau_0} + \frac{T_1}{R \, e^{i \theta} - \tau_1} \right) \right] \, .
\label{I_C_R_explicit_def}
\end{eqnarray}
We then Taylor expand, in powers of $1/R$, the exponential term in the integrand in the right-hand side of~\eref{I_C_R_explicit_def}, and we have
\begin{eqnarray}
\mathrm{exp} \left( \frac{T_0}{R \, e^{i \theta} - \tau_0} + \frac{T_1}{R \, e^{i \theta} - \tau_1} \right) = 1 + \frac{T_0 + T_1}{R \, e^{i \theta}} + \bigO \left[ \left( \frac{T_0 + T_1}{R \, e^{i \theta}} \right) ^2 \right] \, .
\label{exponent_Taylor_final}
\end{eqnarray}
Substituting~\eref{exponent_Taylor_final} into~\eref{I_C_R_explicit_def} then readily shows that a sufficient condition for having $I_{\mathcal{C}_{\infty}^{(2)}} = 0$ is that the aperture function $\chi$ satisfies
\begin{eqnarray}
\lim\limits_{R \to \infty} \left[ R \, e^{i \theta} \chi \left( R \, e^{i \theta} \right) \right] = 0 \qquad \mbox{for any} \qquad 0 \leqslant \theta \leqslant \pi \, .
\label{chi_sufficient_cond}
\end{eqnarray}

Therefore, the integral $I_{\mathcal{C}_{\infty}^{(2)}}$ plays no role in the expression~\eref{I_fr_decomposition_complexInt} of $f_{\mathrm{froz}}$ under the condition that the aperture function $\chi$ satisfies~\eref{chi_sufficient_cond}. We now derive an adequate upper bound regarding the line integrals $I^{(-)}$ and $I^{(+)}$.


\subsection{Upper bound for \texorpdfstring{$\left| I^{(-)} + I^{(+)} \right|$}{the complementary real axis integrals}}\label{upper_bound_sec}

From the triangle inequality and the definition~\eref{I_minus_I_plus_def} of $I^{(-)}$ and $I^{(+)}$ we have
\begin{eqnarray}
\fl \left| I^{(-)} + I^{(+)} \right| \leqslant \left| I^{(-)} \right| + \left| I^{(+)} \right| \leqslant \int_{- \infty}^{0} d \tau \, \left| \widetilde{f}_{\mathrm{froz}} (\tau) \right| + \int_{t}^{\infty} d \tau \, \left| \widetilde{f}_{\mathrm{froz}} (\tau) \right| \, ,
\label{mod_I_minus_I_plus_triangIneq}
\end{eqnarray}
Combining~\eref{mod_I_minus_I_plus_triangIneq} with~\eref{h_sc_def}, we show (details may be found in~\ref{upper_bound_app}) that $|I^{(-)} + I^{(+)}|$ admits the following global upper bound:
\begin{eqnarray}
\fl \left| I^{(-)} + I^{(+)} \right| \leqslant \Gamma_{\mathrm{up}}(\widetilde{x},\widetilde{v}) \, \mathcal{N}_{\widetilde{x}, \, - \widetilde{v}} \left[ \mathrm{max} \left( t \, , \, \widetilde{t} \right) \right] \mathrm{max} \left[ e^{ - \frac{1}{2} \left( \frac{\sigma}{\lambdabar} \right) ^2} \, , e^{ - \frac{1}{2} \left( \frac{x_0}{\sigma} \right) ^2} \right] \int_{- \infty}^{0} d \tau \, \left| \chi (\tau) \right| \nonumber\\[2mm]
+ \Gamma_{\mathrm{up}}(\widetilde{x},\widetilde{v}) \, \mathcal{N}_{x_0, \, v_0} (t) \, \mathrm{max} \left[ e^{ - \frac{1}{2} \left( \frac{\widetilde{x}}{\sigma} \right) ^2} \, , e^{ - \frac{1}{2} \left( \frac{\sigma}{\widetilde{\lambdabar}} \right) ^2} \right] \int_{t}^{\infty} d \tau \, \left| \chi (\tau) \right| \, ,
\label{sum_I_minus_I_plus_UPPER_BOUND}
\end{eqnarray}
where $\Gamma_{\mathrm{up}}$ is the algebraic function
\begin{eqnarray}
\Gamma_{\mathrm{up}}(\widetilde{x},\widetilde{v}) \equiv \frac{1}{2 \sqrt{\pi} \sigma} \left[ \widetilde{v} \sqrt{1 + \left( \frac{\widetilde{\lambdabar} \widetilde{x}}{\sigma^2} \right) ^2} + v_0 \sqrt{1 + \left( \frac{\lambdabar \left| x_0 \right|}{\sigma^2} \right) ^2} \right] \, ,
\label{Gamma_up_def}
\end{eqnarray}
while $\mathcal{N}$ is the Gaussian function
\begin{eqnarray}
\mathcal{N}_{\xi_1,\xi_2}(\tau) \equiv \mathrm{exp} \left[ - \frac{1}{2 \sigma_{\tau}^2} \left( \xi_1 + \xi_2 \tau \right)^2 \right]
\label{N_Gaussian_def}
\end{eqnarray}
and $\mathrm{max} (\xi_1 \, , \xi_2)$ denotes the maximum between $\xi_1$ and $\xi_2$.

The actual values of the integrals in~\eref{sum_I_minus_I_plus_UPPER_BOUND} must be estimated, either numerically or analytically, for any particular $\chi$ that we may consider. These integrals can reasonably be expected to exist in view of the condition~\eref{chi_sufficient_cond} that any valid $\chi$ must satisfy. In particular, the apodization barriers $\chi_n$ that we consider in sections~\ref{apodization_sec} and~\ref{n_2_sec} below allow us to analytically derive upper bounds of these integrals.

Combining now the expression~\eref{I_fr_decomposition_complexInt} of $f_{\mathrm{froz}}$ with $I_{\mathcal{C}_{\infty}^{(2)}} = 0$ yields in particular
\begin{eqnarray}
\left| f_{\mathrm{froz}} (\widetilde{x},\widetilde{v},t) - f_{\mathrm{froz}}^{\ointctrclockwise} \right| = \left| I^{(-)} + I^{(+)} \right| \, .
\label{mod_h_sc_min_I}
\end{eqnarray}
Therefore, we can take the upper bound~\eref{sum_I_minus_I_plus_UPPER_BOUND} as quantifying the maximum error that we make when identifying $f_{\mathrm{froz}}$ to the contour integral $f_{\mathrm{froz}}^{\ointctrclockwise}$.

We hence showed at this point that the terms $I_{\mathcal{C}_{\infty}^{(2)}}$, $I^{(-)}$ and $I^{(+)}$ can be safely neglected in the expression~\eref{I_fr_decomposition_complexInt} of $f_{\mathrm{froz}}$. Therefore, we now discuss how we can compute the remaining (and most important) term in~\eref{I_fr_decomposition_complexInt}, namely the contour integral $f_{\mathrm{froz}}^{\ointctrclockwise}$.


\subsection{Residues}\label{res_sec}

In this section we describe our strategy for analytically computing the contour integral $f_{\mathrm{froz}}^{\ointctrclockwise}$ that, in view of the definition~\eref{complexInt_f_froz_def}, is defined by
\begin{eqnarray}
f_{\mathrm{froz}}^{\ointctrclockwise} = \ointctrclockwise_{\mathcal{\gamma}} d z \, \widetilde{f}_{\mathrm{froz}} (z) \qquad \mbox{with} \qquad \gamma \equiv \lim\limits_{R \to \infty} \mathcal{C}_R \, .
\label{I_def}
\end{eqnarray}

First, substituting~\eref{alpha_T_def} and~\eref{varPhi_def} into~\eref{h_sc_def}, we write the function $\widetilde{f}_{\mathrm{froz}}$ in the form
\begin{eqnarray}
\widetilde{f}_{\mathrm{froz}} (z) = \Omega \, \chi (z) \, \exp \left( \frac{T_0}{z - \tau_0} + \frac{T_1}{z - \tau_1} \right) \, ,
\label{h_frozen_Omega_notation_chap}
\end{eqnarray}
where we introduced the quantity
\begin{eqnarray}
\Omega \equiv \frac{1}{2 \sqrt{\pi}} \left[ \frac{\widetilde{v}}{\sigma} \left( 1 +i \frac{\widetilde{\lambdabar} \widetilde{x}}{\sigma^2} \right) + \frac{v_0}{\sigma} \left( 1 +i \frac{\lambdabar \left| x_0 \right|}{\sigma^2} \right) \right] e^{- \frac{1}{2} \left( \frac{\sigma}{\widetilde{\lambdabar}} \right)^2 - \frac{1}{2} \left( \frac{\sigma}{\lambdabar} \right)^2} \, .
\label{Omega_def}
\end{eqnarray}
For convenience, we also introduce the notations $\mathcal{R}_{\gamma}$ and $\mathcal{C} ( \widetilde{z} \, , \rho )$, with
\begin{eqnarray}
\mathcal{R}_{\gamma} \equiv \left\{ z \in \mathbb{C} \mid z \; \; \mbox{enclosed by} \; \; \gamma \right\}
\label{R_gamma_def}
\end{eqnarray}
denoting the interior of the oriented closed contour $\gamma$ and
\begin{eqnarray}
\mathcal{C} ( \widetilde{z} \, , \rho ) \equiv \left\{ z \in \mathbb{C} \mid \left| z - \widetilde{z} \right| = \rho \right\}
\label{circle_Z_rho_def}
\end{eqnarray}
denoting the circle of center $\widetilde{z}$ and radius $\rho$.

We now use Cauchy's residue theorem (see e.g.~\cite{Ablowitz,Appel}) to compute the integral~\eref{I_def}. As is clear from~\eref{h_frozen_Omega_notation_chap} $\tau_0$ and $\tau_1$ are \textit{essential singularities} of $\widetilde{f}_{\mathrm{froz}}$. However,~\eref{tau_0_tau_1_def} ensures that their imaginary parts satisfy $\imag \, (\tau_0) > 0$ and $\imag \, (\tau_1) < 0$. Therefore, the integration contour $\gamma$ in~\eref{I_def} only encloses $\tau_0$. If we further assume that the function $\chi (z)$ only possesses isolated singularities $Z_j$ (as is indeed the case for the apodization barriers $\chi_n$ considered in sections~\ref{apodization_sec} and~\ref{n_2_sec} below), then the Cauchy residue theorem states that the contour integral~\eref{I_def} can be written as
\begin{eqnarray}
f_{\mathrm{froz}}^{\ointctrclockwise} = 2 \pi i \left\{ \mathrm{Res} \left[ \widetilde{f}_{\mathrm{froz}} (z) \, , \tau_0 \right] + \sum_{j \atop Z_j \in \mathcal{R}_{\gamma}} \mathrm{Res} \left[ \widetilde{f}_{\mathrm{froz}} (z) \, , Z_j \right] \right\} \, ,
\label{I_residue_general_expr}
\end{eqnarray}
where $\mathrm{Res} \, [\widetilde{f}_{\mathrm{froz}} (z) \, , \widetilde{z} \,]$ denotes the residue of the function $\widetilde{f}_{\mathrm{froz}}$ at the point $\widetilde{z}$.

The term $\mathrm{Res} \, [\widetilde{f}_{\mathrm{froz}} (z) \, , \tau_0]$ in~\eref{I_residue_general_expr} corresponds to a residue at an essential singularity and is thus a priori the most challenging to treat. Indeed, no general method exists to compute it other than to resort to its very definition: it is the coefficient of the term $1/(z-\tau_0)$ in the Laurent series of $\widetilde{f}_{\mathrm{froz}}$ about $\tau_0$. To this end, it turns out that the exponential term in~\eref{h_frozen_Omega_notation_chap} can be transformed into the generating function of the Bessel functions of the first kind: this is what we discuss in the remaining part of this section.

By definition of a residue, we have~\cite{Ablowitz,Appel}
\begin{eqnarray}
\mathrm{Res} \left[ \widetilde{f}_{\mathrm{froz}} (z) \, , \tau_0 \right] = \frac{1}{2 \pi i} \ointctrclockwise_{\gamma_{\tau_0}} d z \, \widetilde{f}_{\mathrm{froz}} (z) \, ,
\label{residue_h_fr_tau_0_def}
\end{eqnarray}
where $\gamma_{\tau_0}$ is any simple closed contour enclosing $\tau_0$ such that $\widetilde{f}_{\mathrm{froz}} (z)$ is analytic on and inside $\gamma_{\tau_0}$ except at $z=\tau_0$. For convenience, we choose $\gamma_{\tau_0} = \mathcal{C} ( \tau_0 \, , r )$, and we introduce for concreteness the positive number $r_{\tau_0}$ defined by
\begin{eqnarray}
r_{\tau_0} \equiv \mathrm{min} \left[ \left| \tau_0 - \tau_1 \right| \, , \left( \left| \tau_0 - Z_j \right| \right)_{j} \right] \, ,
\label{r_tau_0_def}
\end{eqnarray}
where $\mathrm{min}$ denotes the minimum function. It is thus clear that, in order for $\tau_0$ to be the only singularity of $\widetilde{f}_{\mathrm{froz}}$ enclosed by $\mathcal{C} ( \tau_0 \, , r )$, the radius $r$ must satisfy
\begin{eqnarray}
0 < r < r_{\tau_0} \, .
\label{r_less_r_tau_0}
\end{eqnarray}
Substituting~\eref{h_frozen_Omega_notation_chap} into~\eref{residue_h_fr_tau_0_def} hence yields
\begin{eqnarray}
\mathrm{Res} \left[ \widetilde{f}_{\mathrm{froz}} (z) \, , \tau_0 \right] = \frac{\Omega}{2 \pi i} \ointctrclockwise_{\mathcal{C} ( \tau_0 \, , r )} d z' \, \chi (z') \, \exp \left( \frac{T_0}{z' - \tau_0} + \frac{T_1}{z' - \tau_1} \right) \, .
\label{residue_h_fr_tau_0_explicit_def}
\end{eqnarray}

As we discuss in details in~\ref{Mobius_app}, applying first the M\"obius transformation
\begin{eqnarray}
w = \frac{- z' + \tau_1}{z' - \tau_0} \, ,
\label{Mobius_transform_def}
\end{eqnarray}
followed by the change of variable
\begin{eqnarray}
z = \sqrt{\frac{T_0}{T_1}} w \, ,
\label{change_of_var_def}
\end{eqnarray}
we show that the residue~\eref{residue_h_fr_tau_0_explicit_def} can be written as
\begin{eqnarray}
\mathrm{Res} \left[ \widetilde{f}_{\mathrm{froz}} (z) \, , \tau_0 \right] = \frac{\Omega \left( \tau_1 - \tau_0 \right)}{2 \pi i Z} \, e^{\frac{T_0 - T_1}{\tau_1 - \tau_0}} \ointctrclockwise_{\mathcal{C} \left( - Z \, , \left| Z \right| \frac{\left| \tau_0-\tau_1 \right|}{r} \right)} d z \, g_{\chi} (z) g (z) \, ,
\label{residue_h_fr_tau_0_after_change_of_var}
\end{eqnarray}
where we introduced the complex number $Z$ given by
\begin{eqnarray}
Z \equiv \sqrt{\frac{T_0}{T_1}} \, ,
\label{Z_def}
\end{eqnarray}
as well as the functions $g_{\chi} (z)$ and $g (z)$ defined by
\begin{eqnarray}
g_{\chi} (z) \equiv \frac{1}{\left( z/Z + 1 \right) ^2} \, \chi \left( \frac{\tau_0 z/Z + \tau_1}{z/Z + 1} \right)
\label{f_Mobius_def}
\end{eqnarray}
and
\begin{eqnarray}
g (z) \equiv \exp \left[ \frac{\sqrt{T_0 T_1}}{\tau_1 - \tau_0} \left( z - \frac{1}{z} \right) \right] = \sum_{k=-\infty}^{\infty} J_{k} \left( \frac{2\sqrt{T_0 T_1}}{\tau_1 - \tau_0} \right) z^k \, ,
\label{g_Mobius_def}
\end{eqnarray}
the latter hence corresponding to the generating function of the Bessel functions of the first kind $J_{k} (z)$ (see e.g.~\cite{Arfken}).

In view of~\eref{r_less_r_tau_0}, it is clear that the integration contour in~\eref{residue_h_fr_tau_0_after_change_of_var} encloses the essential singularity $z=0$ of $g(z)$, but not its other essential singularity $z=\infty$. In addition, the number of singularities of $g_{\chi}(z)$ enclosed by the integration contour in~\eref{residue_h_fr_tau_0_after_change_of_var} must, for consistency, be independent of the particular value of $r$.

While the expression~\eref{residue_h_fr_tau_0_after_change_of_var} is valid for an arbitrary aperture function $\chi$, to evaluate the remaining integral can a priori only be done in some specific cases. We now consider a class of functions $\chi$ for which the integral in~\eref{residue_h_fr_tau_0_after_change_of_var} can be evaluated explicitly.


\section{Apodization barriers}\label{apodization_sec}

The analysis performed in section~\ref{complex_integral_sec} above is valid for general aperture functions $\chi$. Here and in the sequel, we consider a particular class of such functions, which we denote by $\chi_n$ with $n$ a strictly positive even integer. We define the function $\chi_n (\tau)$ by
\begin{eqnarray}
\chi_n (\tau) \equiv \frac{1}{1+\nu^n (\tau-T_{\mathrm{op}})^n} \qquad , \qquad n \enspace \mbox{even} \enspace , \quad \nu > 0 \enspace , \quad T_{\mathrm{op}} \in \mathbb{R} \, .
\label{chi_n_tau_def}
\end{eqnarray}

\begin{figure}[ht]
\centering
\includegraphics[width=1.0\textwidth]{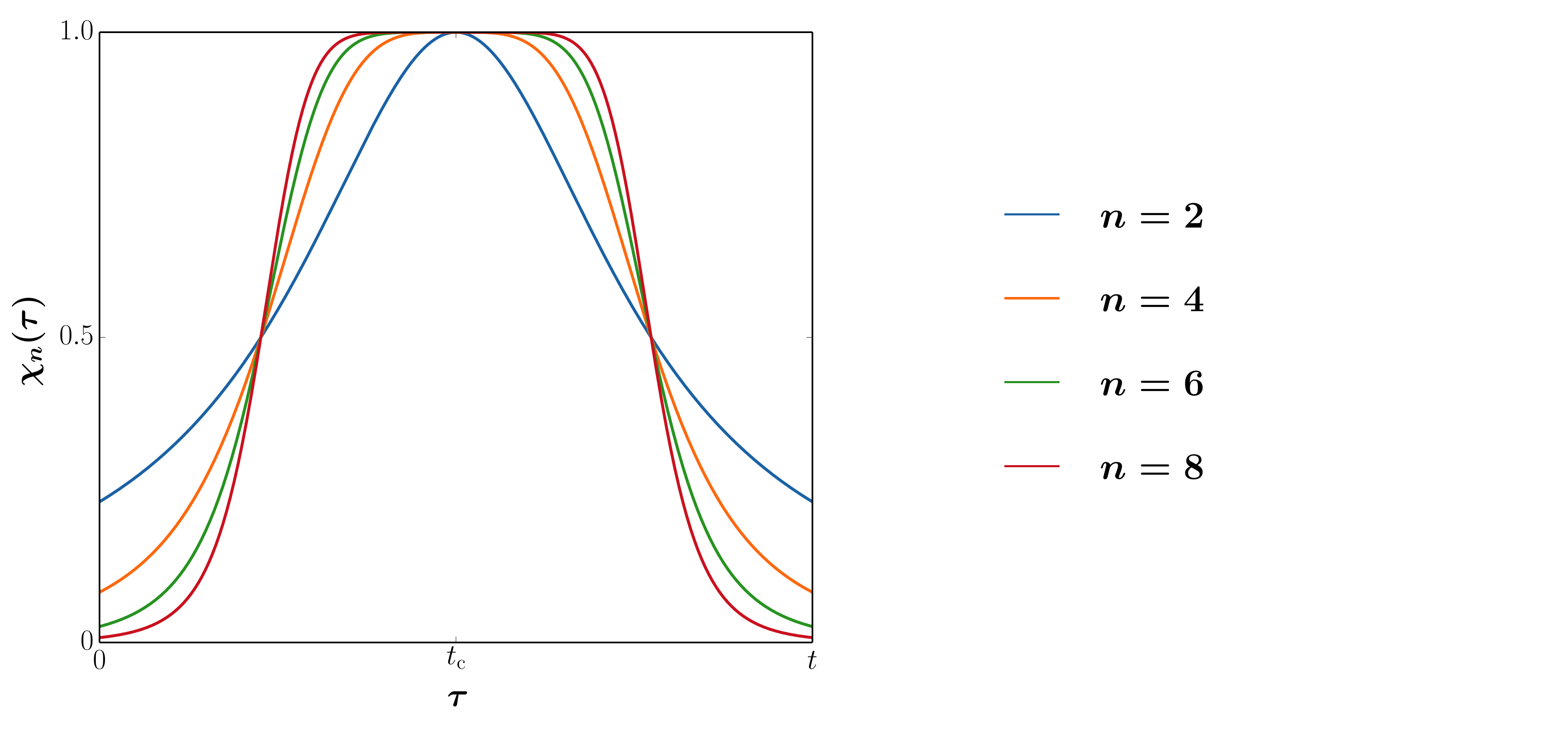}
\caption{Apodization barriers~\eref{chi_n_tau_def} as functions of $\tau$ for $n=2$ (blue), $n=4$ (orange), $n=6$ (green) and $n=8$ (red). Here we chose the numerical parameters to be $\nu=5$ (in dimensionless units), $T_{\mathrm{op}}=t_{\mathrm{c}}$ and $t=2t_{\mathrm{c}}$.}
\label{apod_barriers_fig}
\end{figure}

The requirement of an even $n$ ensures that $0 \leqslant \chi_n (\tau) \leqslant 1$ for any $\tau$, so that~\eref{chi_n_tau_def} indeed represents a valid class of aperture functions. As is illustrated on figure~\ref{apod_barriers_fig}, $\chi_{n}$ models a smooth algebraic window, or apodization barrier. Because $\chi_{n} (T_{\mathrm{op}}) = 1$, we call $T_{\mathrm{op}}$ the opening time of the barrier. We assume hereinafter that it satisfies
\begin{eqnarray}
0< T_{\mathrm{op}} < t \, .
\label{assumption_T_op}
\end{eqnarray}
This indeed ensures that the barriers~\eref{chi_n_tau_def} open when a substantial part of the incident wave packet reaches the barrier. The quantity $\nu$ characterizes the inverse of the width of the barrier, with $\nu \ll 1$ ($\nu \gg 1$) corresponding to a broad (narrow) barrier.


Before analyzing the residue~\eref{residue_h_fr_tau_0_after_change_of_var}, we must check that the apodization barriers of the form~\eref{chi_n_tau_def} satisfy our complex formulation of the Husimi amplitude discussed in section~\ref{complex_integral_sec} above. First, since $\chi_n (R \, e^{i \theta}) \sim 1 / \nu^n R^n \, e^{i n \theta}$ as $R \to \infty$, we have
\begin{eqnarray}
\lim\limits_{R \to \infty} \left[ R \, e^{i \theta} \chi_n \left( R \, e^{i \theta} \right) \right] = \lim\limits_{R \to \infty} \left[ \frac{1}{\nu^n \left( R \, e^{i \theta} \right)^{n-1}} \right] = 0
\label{chi_n_criterion}
\end{eqnarray}
for any $n \geqslant 2$. Therefore, $\chi_n (z)$ indeed satisfies condition~\eref{chi_sufficient_cond} for any $n \geqslant 2$.

We then evaluate the two integrals that appear in the right-hand side of~\eref{sum_I_minus_I_plus_UPPER_BOUND} for $\chi = \chi_n$. As we show in details in~\ref{apod_up_bound_app}, we have the upper bounds
\begin{eqnarray}
\int_{- \infty}^{0} d \tau \, \chi_{n} (\tau) \leqslant \frac{1}{\nu} \left[ \frac{\pi}{2} - \mathrm{Arctan} \left( \nu T_{\mathrm{op}} \right) \right] \quad , \quad \mbox{if} \; \; \nu \geqslant \frac{1}{T_{\mathrm{op}}} \, ,
\label{neg_integral_chi_2k_up_bound}
\end{eqnarray}
and
\begin{eqnarray}
\int_{t}^{\infty} d \tau \, \chi_{n} (\tau) \leqslant \frac{1}{\nu} \left\{ \frac{\pi}{2} - \mathrm{Arctan} \left[ \nu \left( t-T_{\mathrm{op}} \right) \right] \right\} \quad , \quad \mbox{if} \; \; \nu \geqslant \frac{1}{t-T_{\mathrm{op}}} \, ,
\label{pos_integral_chi_2k_up_bound}
\end{eqnarray}
for any even integer $n \geqslant 2$. It is thus clear from~\eref{neg_integral_chi_2k_up_bound}-\eref{pos_integral_chi_2k_up_bound} that we can make the upper bound~\eref{sum_I_minus_I_plus_UPPER_BOUND} arbitrarily small by making $\nu$ large enough. Therefore, the apodization barriers~\eref{chi_n_tau_def} are indeed perfectly suited to our complex formulation.

We now explicitly compute the residues in~\eref{I_residue_general_expr} for the apodization barriers~\eref{chi_n_tau_def}. We first investigate in section~\ref{struct_chi_n_f_n_sec} the analytic structure of the two functions $\chi_n$ and $g_{\chi_n}$. We then derive in section~\ref{Taylor_f_n_sec} the Taylor expansion of $g_{\chi_n}$ about $z=0$. We use the latter in section~\ref{residue_ess_sing_sec} to construct, by means of Cauchy products, the relevant part of the Laurent series of $g_{\chi_n} g$ about $z=0$: the term $1/z$ in the latter series then yields the residue $\mathrm{Res} \left[ g_{\chi_n} (z) g (z) \, , 0 \right]$. After obtaining this residue at the essential singularity, we conclude in section~\ref{residues_poles_sec} by computing the remaining, simple residues at poles.


\subsection{Analytic structure of \texorpdfstring{$\chi_n (z)$}{chi} and \texorpdfstring{$g_{\chi_n} (z)$}{f}}\label{struct_chi_n_f_n_sec}

In this section we study the analytic structure of the two functions $\chi_n (z)$, given by~\eref{chi_n_tau_def}, and $g_{\chi_n} (z)$, obtained from~\eref{f_Mobius_def} for $\chi = \chi_n$, for an arbitrary even integer $n \geqslant 2$. These two functions being rational functions, their only singularities are thus poles~\cite{Ablowitz}. Technical details are deferred to~\ref{struct_chi_n_f_n_app}.

As is clear from~\eref{chi_n_tau_def}, $\chi_n$ admits the $n$ simple poles $Z_j^{(n)}$ given by
\begin{eqnarray}
Z_j^{(n)} = T_{\mathrm{op}} + \frac{1}{\nu} e^{(2j+1) \frac{i \pi}{n}} \qquad \mbox{with} \qquad 0 \leqslant j \leqslant n-1 \, .
\label{poles_chi_n_expr}
\end{eqnarray}
We now use~\eref{poles_chi_n_expr} to identify which of these poles must be taken into account in~\eref{I_residue_general_expr}, namely those that have a positive imaginary part. Since $n$ is even by assumption, it is clear from~\eref{poles_chi_n_expr} that $\mathrm{Im} \, [Z_j^{(n)}] > 0$ if and only if $0 \leqslant j \leqslant n/2 - 1$. Therefore, the contour integral~\eref{I_residue_general_expr}, which we relabel $f_{\mathrm{froz}}^{\ointctrclockwise(n)}$ here to emphasize that it corresponds to the aperture function $\chi_n$, is given by
\begin{eqnarray}
\fl f_{\mathrm{froz}}^{\ointctrclockwise(n)} \equiv \ointctrclockwise_{\mathcal{\gamma}} d z \, \widetilde{f}_{\mathrm{froz}}^{\,(n)} (z) = 2 \pi i \left\{ \mathrm{Res} \left[ \widetilde{f}_{\mathrm{froz}}^{\,(n)} (z) \, , \tau_0 \right] + \sum_{j=0}^{n/2-1} \mathrm{Res} \left[ \widetilde{f}_{\mathrm{froz}}^{\,(n)} (z) \, , Z_j^{(n)} \right] \right\} \, ,
\label{I_n_expr_with_poles_specified}
\end{eqnarray}
where $\widetilde{f}_{\mathrm{froz}}^{\,(n)}$ is obtained upon substituting $\chi = \chi_n$ into~\eref{h_frozen_Omega_notation_chap}, that is
\begin{eqnarray}
\widetilde{f}_{\mathrm{froz}}^{\,(n)} (z) \equiv \Omega \, \chi_n (z) \, \exp \left( \frac{T_0}{z - \tau_0} + \frac{T_1}{z - \tau_1} \right) = \frac{\Omega \, \exp \left( \frac{T_0}{z - \tau_0} + \frac{T_1}{z - \tau_1} \right)}{\nu^n \prod\limits_{j=0}^{n-1} \left[ z - Z_j^{(n)} \right]} \, .
\label{h_frozen_Omega_apodization_def}
\end{eqnarray}

We then denote by $z_j^{(n)}$, with $j$ an integer, the poles of the function $g_{\chi_n}$ that is obtained upon substituting~\eref{chi_n_tau_def} into~\eref{f_Mobius_def}, that is
\begin{eqnarray}
g_{\chi_n} (z) = \frac{\left( z/Z + 1 \right)^{n-2}}{\left( z/Z + 1 \right)^{n} + \nu^n \left[ \left( \tau_0 - T_{\mathrm{op}} \right) z/Z + \tau_1 - T_{\mathrm{op}} \right]^n} \, .
\label{f_n_z_expr}
\end{eqnarray}
As is clear from~\eref{f_n_z_expr}, $g_{\chi_n}$ admits the $n$ simple poles
\begin{eqnarray}
\fl z_j^{(n)} = - Z \, \left[ \tau_1 - T_{\mathrm{op}} - \frac{1}{\nu} e^{(2j+1) \frac{i \pi}{n}} \right] \left[ \tau_0 - T_{\mathrm{op}} - \frac{1}{\nu} e^{(2j+1) \frac{i \pi}{n}} \right]^{-1} \enspace \mbox{with} \enspace 0 \leqslant j \leqslant n-1 \, .
\label{poles_f_n_temp_expr}
\end{eqnarray}
Combining~\eref{poles_f_n_temp_expr} with~\eref{poles_chi_n_expr} shows that the poles $z_j^{(n)}$ are related to the poles $Z_j^{(n)}$ through
\begin{eqnarray}
z_j^{(n)} = - Z \, \left[ \tau_1 - Z_j^{(n)} \right] \left[ \tau_0 - Z_j^{(n)} \right]^{-1} \qquad \mbox{with} \qquad 0 \leqslant j \leqslant n-1 \, .
\label{poles_f_n_final_expr}
\end{eqnarray}

We can now express the residue $\mathrm{Res} \, [ \widetilde{f}_{\mathrm{froz}}^{\,(n)} (z) \, , \tau_0 ]$ obtained from~\eref{residue_h_fr_tau_0_after_change_of_var} for $\chi=\chi_n$, i.e.
\begin{eqnarray}
\mathrm{Res} \left[ \widetilde{f}_{\mathrm{froz}}^{\,(n)} (z) \, , \tau_0 \right] = \frac{\Omega \left( \tau_1 - \tau_0 \right)}{2 \pi i Z} \, e^{\frac{T_0 - T_1}{\tau_1 - \tau_0}} \ointctrclockwise_{\mathcal{C} \left( - Z \, , \left| Z \right| \frac{\left| \tau_0-\tau_1 \right|}{r} \right)} d z \, g_{\chi_n} (z) g (z) \, .
\label{residue_f_tilde_froz_n_gen_expr}
\end{eqnarray}
Indeed, combining~\eref{poles_f_n_final_expr} with~\eref{r_less_r_tau_0} shows that the distance $| - Z - z_j^{(n)} |$ between $z=-Z$ and any pole $z_j^{(n)}$ satisfies the strict inequality
\begin{eqnarray}
\left| - Z - z_j^{(n)} \right| < \left| Z \right| \frac{\left| \tau_0 - \tau_1 \right|}{r} \, .
\label{distance_smaller_radius}
\end{eqnarray}
This shows that the circle $\mathcal{C} \left( - Z \, , \, \left| Z \right| \left| \tau_0-\tau_1 \right| / r \right)$ encloses \textit{all} the poles $z_j^{(n)}$ of $g_{\chi_n}$. Furthermore, we already saw in section~\ref{res_sec} that it also encloses the essential singularity $z=0$ of $g$. The integral in~\eref{residue_f_tilde_froz_n_gen_expr} can thus itself be computed by means of the Cauchy Residue Theorem, and we get
\begin{eqnarray}
\fl \mathrm{Res} \left[ \widetilde{f}_{\mathrm{froz}}^{\,(n)} (z) \, , \tau_0 \right] = \frac{\Omega \left( \tau_1 - \tau_0 \right)}{Z} \, \exp \left( \frac{T_0 - T_1}{\tau_1 - \tau_0} \right) \nonumber\\
\times \left\{ \mathrm{Res} \left[ g_{\chi_n} (z) g (z) \, , 0 \right] + \sum_{j = 0}^{n-1} \mathrm{Res} \left[ g_{\chi_n} (z) g (z) \, , z_j^{(n)} \right] \right\} \, .
\label{residue_h_fr_tau_0_as_sum_residues}
\end{eqnarray}

The terms $\mathrm{Res} \, [g_{\chi_n} (z) g (z) \, , z_j^{(n)}]$ in the right-hand side of~\eref{residue_h_fr_tau_0_as_sum_residues} correspond to residues at simple poles, and are thus straightforward to compute. The main challenge hence arises from the residue $\mathrm{Res} \, [g_{\chi_n} (z) g (z) \, , 0]$ at the essential singularity. The latter requires to construct the Laurent series of $g_{\chi_n} g$ about $z=0$. Since we already know the Laurent series of $g$ [see~\eref{g_Mobius_def}], we now compute the Taylor series of $g_{\chi_n} (z)$ about $0$.


\subsection{Taylor expansion of \texorpdfstring{$g_{\chi_n} (z)$}{f}}\label{Taylor_f_n_sec}

In view of the results obtained in section~\ref{struct_chi_n_f_n_sec}, we can write~\eref{f_n_z_expr} as
\begin{eqnarray}
g_{\chi_n} (z) = \frac{Z^2}{1+\nu^n (\tau_0-T_{\mathrm{op}})^n} \, \left( z+Z \right)^{n-2} \left\{ \prod\limits_{j=0}^{n-1} \left[ z - z_j^{(n)} \right] \right\}^{-1} \, .
\label{f_n_z_factor_expr}
\end{eqnarray}
We can then write a partial fraction decomposition of~\eref{f_n_z_factor_expr} of the form
\begin{eqnarray}
g_{\chi_n} (z) = \frac{Z^2}{1+\nu^n (\tau_0-T_{\mathrm{op}})^n} \sum_{j=0}^{n-1} \frac{A_j^{(n)}}{z - z_j^{(n)}} \, ,
\label{f_n_z_partial_frac_def}
\end{eqnarray}
where the $n$ complex numbers $A_j^{(n)}$ must be explicitly determined. We propose the following hypothesis regarding the expression of these coefficients $A_j^{(n)}$:
\begin{eqnarray}
\fl A_j^{(n)} = \left[ z_j^{(n)} + Z \right]^{n-2} \left\{ \prod\limits_{j'=0 \atop j' \neq j}^{n-1} \left[ z_j^{(n)} - z_{j'}^{(n)} \right] \right\}^{-1} \qquad , \qquad 0 \leqslant j \leqslant n-1 \, .
\label{Taylor_coef_f_n_expr_guess}
\end{eqnarray}
While we have explicitly checked that the expression~\eref{Taylor_coef_f_n_expr_guess} of $A_j^{(n)}$ is valid for even integers $n$ up to 10, we have been unable to show that~\eref{Taylor_coef_f_n_expr_guess} holds for an arbitrary $n$. Indeed, the difficulty arises from explicitly expanding factorized polynomials of arbitrary degrees. We emphasize however that~\eref{Taylor_coef_f_n_expr_guess} is exact for the $n=2$ case that we analyze in section~\ref{n_2_sec}.

We now use~\eref{f_n_z_partial_frac_def} to write the Taylor series of $g_{\chi_n}$ about $z=0$. We easily get
\begin{eqnarray}
\frac{1}{z-\zeta} = - \frac{1}{\zeta} \sum_{k=0}^{\infty} \left( \frac{z}{\zeta} \right)^k \, ,
\label{Taylor_1_ov_zMinZeta}
\end{eqnarray}
which converges absolutely for any $\zeta \neq 0$ and any $z \in \mathbb{C}$ such that $\left| z/\zeta \right| < 1$. Substituting~\eref{Taylor_1_ov_zMinZeta} into~\eref{f_n_z_partial_frac_def} hence readily yields
\begin{eqnarray}
g_{\chi_n} (z) = - \frac{Z^2}{1+\nu^n (\tau_0-T_{\mathrm{op}})^n} \sum_{j=0}^{n-1} \frac{A_j^{(n)}}{z_j^{(n)}} \sum_{k=0}^{\infty} \left[ \frac{z}{z_j^{(n)}} \right]^k \, .
\label{f_n_z_Taylor_series}
\end{eqnarray}

We can now use the Taylor series~\eref{f_n_z_Taylor_series} to construct the Laurent series of $g_{\chi_n} g$ about $z=0$, which is necessary in order to obtain the residue $\mathrm{Res} \, [ g_{\chi_n} (z) g (z) \, , 0 ]$, and thus $\mathrm{Res} \, [ \widetilde{f}_{\mathrm{froz}}^{\,(n)} (z) \, , \tau_0 ]$ in view of~\eref{residue_h_fr_tau_0_as_sum_residues}, at the essential singularity.


\subsection{The residue at the essential singularity}\label{residue_ess_sing_sec}

Here we evaluate the residue $\mathrm{Res} \, [ g_{\chi_n} (z) g (z) \, , 0 ]$ from its very definition~\cite{Ablowitz,Appel}: it corresponds to the coefficient of the $1/z$ term in the Laurent series of $g_{\chi_n} g$ about $z=0$.

Combining the Taylor series~\eref{f_n_z_Taylor_series} of $g_{\chi_n}$ with the Laurent series~\eref{g_Mobius_def} of $g$ yields the Laurent series of $g_{\chi_n} g$ about 0, and we have
\begin{eqnarray}
\fl g_{\chi_n} (z) g(z) = - \frac{Z^2}{1+\nu^n (\tau_0-T_{\mathrm{op}})^n} \sum_{j=0}^{n-1} \frac{A_j^{(n)}}{z_j^{(n)}} \sum_{k=0}^{\infty} \left[ \frac{z}{z_j^{(n)}} \right]^k \sum_{k'=-\infty}^{\infty} J_{k'} \left( \frac{2\sqrt{T_0 T_1}}{\tau_1 - \tau_0} \right) z^{k'} \, .
\label{f_n_g_z_Laurent_def}
\end{eqnarray}
As we show in details in~\ref{residue_zero_app}, using Cauchy products to express the product of the two series in~\eref{f_n_g_z_Laurent_def} allows us to identify the term proportional to $1/z$. We hence obtain the following expression of $\mathrm{Res} \, [ g_{\chi_n} (z) g (z) \, , 0 ]$:
\begin{eqnarray}
\fl \mathrm{Res} \left[ g_{\chi_n} (z) g (z) \, , 0 \right] = - \frac{Z^2}{1+\nu^n (\tau_0-T_{\mathrm{op}})^n} \sum_{j=0}^{n-1} A_j^{(n)} \sum_{k=1}^{\infty} \frac{(-1)^k}{\left[ z_j^{(n)} \right]^k} J_{k} \left( \frac{2\sqrt{T_0 T_1}}{\tau_1 - \tau_0} \right) \, .
\label{residue_f_n_g_z_final_expr}
\end{eqnarray}

We now substitute~\eref{residue_f_n_g_z_final_expr} into the expression~\eref{residue_h_fr_tau_0_as_sum_residues} of $\mathrm{Res} \, [ \widetilde{f}_{\mathrm{froz}}^{\,(n)} (z) \, , \tau_0 ]$ to get
\begin{eqnarray}
\fl \mathrm{Res} \left[ \widetilde{f}_{\mathrm{froz}}^{\,(n)} (z) \, , \tau_0 \right] = \frac{\Omega \left( \tau_1 - \tau_0 \right)}{Z} \, \exp \left( \frac{T_0 - T_1}{\tau_1 - \tau_0} \right) \left\{ \vphantom{\sum_{k=1}^{\infty} \frac{(-1)^k}{\left[ z_j^{(n)} \right]^k} J_{k} \left( \frac{2\sqrt{T_0 T_1}}{\tau_1 - \tau_0} \right)} \sum_{j = 0}^{n-1} \mathrm{Res} \left[ g_{\chi_n} (z) g (z) \, , z_j^{(n)} \right]  \right. \nonumber\\[0.2cm]
\left. - \frac{Z^2}{1+\nu^n (\tau_0-T_{\mathrm{op}})^n} \sum_{j=0}^{n-1} A_j^{(n)} \sum_{k=1}^{\infty} \frac{(-1)^k}{\left[ z_j^{(n)} \right]^k} J_{k} \left( \frac{2\sqrt{T_0 T_1}}{\tau_1 - \tau_0} \right) \right\} \, .
\label{residue_h_fr_tau_0_poles_remaining}
\end{eqnarray}
Finally, we substitute~\eref{residue_h_fr_tau_0_poles_remaining} into the expression~\eref{I_n_expr_with_poles_specified} of the contour integral $f_{\mathrm{froz}}^{\ointctrclockwise(n)}$ to get
\begin{eqnarray}
\fl f_{\mathrm{froz}}^{\ointctrclockwise(n)} = \left\{ \frac{\Omega \left( \tau_1 - \tau_0 \right)}{Z} \, e^{\frac{T_0 - T_1}{\tau_1 - \tau_0}} \left( - \frac{Z^2}{1+\nu^n (\tau_0-T_{\mathrm{op}})^n} \sum_{j=0}^{n-1} A_j^{(n)} \sum_{k=1}^{\infty} \frac{(-1)^k}{\left[ z_j^{(n)} \right]^k} J_{k} \left( \frac{2\sqrt{T_0 T_1}}{\tau_1 - \tau_0} \right) \right. \right. \nonumber\\[0.3cm]
+ \left. \left. \sum_{j = 0}^{n-1} \mathrm{Res} \left[ g_{\chi_n} (z) g (z) \, , z_j^{(n)} \right] \vphantom{\sum_{k=1}^{\infty} \frac{(-1)^k}{\left[ z_j^{(n)} \right]^k} J_{k} \left( \frac{2\sqrt{T_0 T_1}}{\tau_1 - \tau_0} \right)} \right) + \sum_{j=0}^{n/2-1} \mathrm{Res} \left[ \widetilde{f}_{\mathrm{froz}}^{\,(n)} (z) \, , Z_j^{(n)} \right] \right\} 2 \pi i  \, .
\label{I_n_expr_no_more_essential_sing}
\end{eqnarray}
The remaining residues in~\eref{I_n_expr_no_more_essential_sing} are at poles and are now easily computed.


\subsection{The residues at poles}\label{residues_poles_sec}

Here we explicitly compute the remaining residues in~\eref{I_n_expr_no_more_essential_sing}.

First, it is clear from~\eref{h_frozen_Omega_apodization_def} that any $Z_j^{(n)}$ is a simple pole of $\widetilde{f}_{\mathrm{froz}}^{\,(n)}$. Therefore, the residue $\mathrm{Res} \, [ \widetilde{f}_{\mathrm{froz}}^{\,(n)} (z) \, , Z_j^{(n)} ]$ can be simply computed from the limit~\cite{Ablowitz}
\begin{eqnarray}
\mathrm{Res} \left[ \widetilde{f}_{\mathrm{froz}}^{\,(n)} (z) \, , Z_j^{(n)} \right] = \lim_{z \to Z_j^{(n)}} \left\{ \left[ z - Z_j^{(n)} \right] \widetilde{f}_{\mathrm{froz}}^{\,(n)} (z) \right\} \, .
\label{res_h_sc_Z_j_n_def}
\end{eqnarray}
We hence get, in view of~\eref{h_frozen_Omega_apodization_def},
\begin{eqnarray}
\fl \mathrm{Res} \left[ \widetilde{f}_{\mathrm{froz}}^{\,(n)} (z) \, , Z_j^{(n)} \right] = \frac{\Omega}{\nu^n} \, \left\{ \prod\limits_{j'=0 \atop j' \neq j}^{n-1} \left[ Z_j^{(n)} - Z_{j'}^{(n)} \right] \right\}^{-1} \mathrm{exp} \left[ \frac{T_0}{Z_j^{(n)} - \tau_0} + \frac{T_1}{Z_j^{(n)} - \tau_1} \right] \, ,
\label{residue_h_fr_explicit}
\end{eqnarray}
for any $0 \leqslant j \leqslant n-1$.

We then have from~\eref{g_Mobius_def} and~\eref{f_n_z_factor_expr}
\begin{eqnarray}
\fl g_{\chi_n}(z) g(z) = \frac{Z^2 \left( z+Z \right)^{n-2}}{1+\nu^n (\tau_0-T_{\mathrm{op}})^n} \, \left\{ \prod\limits_{j=0}^{n-1} \left[ z - z_{j}^{(n)} \right] \right\}^{-1} \, \exp \left[ \frac{\sqrt{T_0 T_1}}{\tau_1 - \tau_0} \left( z - \frac{1}{z} \right) \right] \, .
\label{f_n_g_expr}
\end{eqnarray}
This shows that any $z_j^{(n)}$ is a simple pole of $g_{\chi_n} g$, so that
\begin{eqnarray}
\mathrm{Res} \left[ g_{\chi_n} (z) g(z) \, , z_j^{(n)} \right] = \lim_{z \to z_j^{(n)}} \left\{ \left[ z - z_j^{(n)} \right] g_{\chi_n} (z) g(z) \right\} \, ,
\label{res_f_n_g_z_j_n_def}
\end{eqnarray}
and thus we get from~\eref{f_n_g_expr}
\begin{eqnarray}
\fl \mathrm{Res} \left[ g_{\chi_n} (z) g(z) \, , z_j^{(n)} \right] = \frac{Z^2}{1+\nu^n (\tau_0-T_{\mathrm{op}})^n} \, \frac{\left[ z_j^{(n)} + Z \right]^{n-2}}{\prod\limits_{j'=0 \atop j' \neq j}^{n-1} \left[ z_j^{(n)} - z_{j'}^{(n)} \right]} \, e^{\frac{\sqrt{T_0 T_1}}{\tau_1 - \tau_0} \left[ z_j^{(n)} - \frac{1}{z_j^{(n)}} \right]} \, ,
\label{residue_f_n_g_explicit}
\end{eqnarray}
for any $0 \leqslant j \leqslant n-1$.

We then set $f_{\mathrm{froz}}^{\ointctrclockwise}=f_{\mathrm{froz}}^{\ointctrclockwise(n)}$ into~\eref{I_fr_decomposition_complexInt} to get the corresponding frozen Gaussian Husimi amplitude
\begin{eqnarray}
f_{\mathrm{froz}}^{(n)} (\widetilde{x},\widetilde{v},t) \equiv f_{\mathrm{froz}}^{\ointctrclockwise(n)} - I^{(-)} - I^{(+)} \, ,
\label{h_sc_n_def}
\end{eqnarray}
where we used $I_{\mathcal{C}_{\infty}^{(2)}} = 0$ [see section~\ref{half_circle_sec} as well as condition~\eref{chi_n_criterion}]. Combining~\eref{h_sc_n_def} with~\eref{I_n_expr_no_more_essential_sing},~\eref{residue_h_fr_explicit} and~\eref{residue_f_n_g_explicit} hence yields
\begin{eqnarray}
\fl f_{\mathrm{froz}}^{(n)} (\widetilde{x},\widetilde{v},t) = 2 \pi i \sum_{j=0}^{n/2-1} \frac{\Omega}{\nu^n} \, \frac{1}{\prod\limits_{j'=0 \atop j' \neq j}^{n-1} \left[ Z_j^{(n)} - Z_{j'}^{(n)} \right]} e^{\frac{T_0}{Z_j^{(n)} - \tau_0} + \frac{T_1}{Z_j^{(n)} - \tau_1}} \nonumber\\
\fl + 2 \pi i \frac{\Omega \left( \tau_1 - \tau_0 \right)}{Z} \, e^{\frac{T_0 - T_1}{\tau_1 - \tau_0}} \left\{ \vphantom{\frac{\left( z_j^{(n)} + Z \right)^{n-2}}{\prod\limits_{j'=0 \atop j' \neq j}^{n-1} \left( z_j^{(n)} - z_{j'}^{(n)} \right)} \, e^{\frac{\sqrt{T_0 T_1}}{\tau_1 - \tau_0} \left( z_j^{(n)} - \frac{1}{z_j^{(n)}} \right)}} - \frac{Z^2}{1+\nu^n (\tau_0-T_{\mathrm{op}})^n} \sum_{j=0}^{n-1} A_j^{(n)} \sum_{k=1}^{\infty} \frac{(-1)^k}{\left[ z_j^{(n)} \right]^k} J_{k} \left( \frac{2\sqrt{T_0 T_1}}{\tau_1 - \tau_0} \right) \right. \nonumber\\[0.3cm]
\fl + \left. \sum_{j = 0}^{n-1} \frac{Z^2}{1+\nu^n (\tau_0-T_{\mathrm{op}})^n} \, \frac{\left[ z_j^{(n)} + Z \right]^{n-2}}{\prod\limits_{j'=0 \atop j' \neq j}^{n-1} \left[ z_j^{(n)} - z_{j'}^{(n)} \right]} \, e^{\frac{\sqrt{T_0 T_1}}{\tau_1 - \tau_0} \left[ z_j^{(n)} - \frac{1}{z_j^{(n)}} \right]} \right\} - I^{(-)} - I^{(+)} \, .
\label{h_sc_n_explicit_expr_general}
\end{eqnarray}
Finally, setting $f_{\mathrm{froz}} = f_{\mathrm{froz}}^{(n)}$ into~\eref{Husimi_ampl_h_sc_def} defines the Husimi amplitude $f^{(n)}$, namely
\begin{eqnarray}
f^{(n)} (\widetilde{x},\widetilde{v},t) \equiv \left[ 1 + \bigO \left( \frac{\hbar t}{m \sigma^2} \right) \right] f_{\mathrm{froz}}^{(n)} (\widetilde{x},\widetilde{v},t) \, .
\label{Husimi_ampl_n_h_sc_n_def}
\end{eqnarray}

The result~\eref{h_sc_n_explicit_expr_general}, combined with~\eref{Husimi_ampl_n_h_sc_n_def}, is the main result of our work. Indeed, it provides an analytic expression of the Husimi amplitude $f^{(n)} (\widetilde{x},\widetilde{v},t)$ for an apodization barrier $\chi_{n}(\tau)$ of the form~\eref{chi_n_tau_def}. This expression is valid for an arbitrary opening time $T_{\mathrm{op}}$, (inverse) width $\nu$ and even integer $n \geqslant 2$.

Since the expression~\eref{h_sc_n_explicit_expr_general} is rather intricate in general, we now illustrate this important analytic result on the simplest, Lorentzian case $n=2$.


\section{The Lorentzian case \texorpdfstring{$n=2$}{n=2}}\label{n_2_sec}

In this final section we consider the apodization barriers~\eref{chi_n_tau_def} in the simplest case $n=2$, i.e. we consider Lorentzian aperture functions $\chi_2$. In addition to simplifying the expressions~\eref{h_sc_n_explicit_expr_general}-\eref{Husimi_ampl_n_h_sc_n_def}, we also see that it embeds all the ingredients that are necessary to exhibit interesting physical behaviors such as diffraction.

Setting $n=2$ into~\eref{h_sc_n_explicit_expr_general} hence yields the frozen Gaussian Husimi amplitude
\begin{eqnarray}
\fl f_{\mathrm{froz}}^{(2)} (\widetilde{x},\widetilde{v},t;T_{\mathrm{op}}) = 2 \pi i \frac{\Omega}{\nu^2} \, \frac{1}{Z_0^{(2)} - Z_{1}^{(2)}} e^{\frac{T_0}{Z_0^{(2)} - \tau_0} + \frac{T_1}{Z_0^{(2)} - \tau_1}} \nonumber\\
\fl + 2 \pi i \frac{\Omega \left( \tau_1 - \tau_0 \right)}{Z} \, e^{\frac{T_0 - T_1}{\tau_1 - \tau_0}} \left\{ \vphantom{\sum_{j = 0}^{1} \frac{Z^2}{1+\nu^2 (\tau_0-T_{\mathrm{op}})^2} \, A_j^{(2)} \, e^{\frac{\sqrt{T_0 T_1}}{\tau_1 - \tau_0} \left[ z_j^{(2)} - \frac{1}{z_j^{(2)}} \right]}} - \frac{Z^2}{1+\nu^2 (\tau_0-T_{\mathrm{op}})^2} \sum_{j=0}^{1} A_j^{(2)} \sum_{k=1}^{\infty} \frac{(-1)^k}{\left[ z_j^{(2)} \right]^k} J_{k} \left( \frac{2\sqrt{T_0 T_1}}{\tau_1 - \tau_0} \right) \right. \nonumber\\[0.3cm]
\fl + \left. \vphantom{\sum_{k=1}^{\infty} \frac{(-1)^k}{\left[ z_j^{(2)} \right]^k} J_{k} \left( \frac{2\sqrt{T_0 T_1}}{\tau_1 - \tau_0} \right)} \sum_{j = 0}^{1} \frac{Z^2}{1+\nu^2 (\tau_0-T_{\mathrm{op}})^2} \, A_j^{(2)} \, e^{\frac{\sqrt{T_0 T_1}}{\tau_1 - \tau_0} \left[ z_j^{(2)} - \frac{1}{z_j^{(2)}} \right]} \right\} - I^{(-)} - I^{(+)} \, ,
\label{h_sc_2_explicit_expr_general}
\end{eqnarray}
where the dependence on $T_{\mathrm{op}}$ is explicitly written for later convenience.

The expression~\eref{h_sc_2_explicit_expr_general} remains rather intricate, notably because of the series of Bessel functions. Therefore, we first discuss in section~\ref{nu_large_subsec} how this expression can be significantly simplified in the limit of a large $\nu$. This so-called slit regime is then applied in section~\ref{dble_slit_subsec} to the case of a double barrier. This allows us in particular to exhibit diffraction, and to analytically describe the phase-space structure of the resulting diffraction pattern.


\subsection{The slit regime \texorpdfstring{$\nu \gg 1$}{nu large}}\label{nu_large_subsec}

Hereinafter we consider the particular regime of large $\nu$, which we write for convenience $\nu \gg 1$ (irrespective of units). This makes of $\chi_2$ a time slit that is open at time $T_{\mathrm{op}}$.

As we discuss in details in~\ref{expand_nu_app}, we show that i) the first term in the right-hand side of~\eref{h_sc_2_explicit_expr_general} is of order $1/\nu$, whereas ii) all the other terms in the right-hand side of~\eref{h_sc_2_explicit_expr_general} are of order $1/\nu^2$. Therefore, we can write $f_{\mathrm{froz}}^{(2)}$ in the simple form
\begin{eqnarray}
f_{\mathrm{froz}}^{(2)} (\widetilde{x},\widetilde{v},t;T_{\mathrm{op}}) = f_{\mathrm{1slit}} (\widetilde{x},\widetilde{v},t;T_{\mathrm{op}}) + \bigO \left( \frac{1}{\nu^2} \right) \, ,
\label{h_sc_2_leading_order_nu_final}
\end{eqnarray}
where we introduced the slit Husimi amplitude $f_{\mathrm{1slit}}$ defined by
\begin{eqnarray}
f_{\mathrm{1slit}} (\widetilde{x},\widetilde{v},t;T_{\mathrm{op}}) \equiv \frac{\pi}{\nu} \Omega \exp \left[ \gamma_0 (T_{\mathrm{op}}) + \gamma_1 (\widetilde{x},\widetilde{v},t;T_{\mathrm{op}}) \right] \, ,
\label{h_slit_gamma_expr}
\end{eqnarray}
with the quantities $\gamma_{0,1} \equiv T_{0,1}/(T_{\mathrm{op}} - \tau_{0,1})$, i.e. in view of~\eref{lambdaBar_t_tilde_def} and~\eref{tau_0_tau_1_def}-\eref{T_0_T_1_def}
\begin{eqnarray}
\mathrm{Re} \left[ \gamma_0 (T_{\mathrm{op}}) \right] = \frac{1}{2 \sigma^2} \, \frac{2 T_{\mathrm{op}} |x_0| v_0 - x_0^2 + m^2 \sigma^4 v_0^2/\hbar^2}{1 + \hbar^2 T_{\mathrm{op}}^2 / m^2 \sigma^4} \, ,
\label{real_gamma_0_def}
\end{eqnarray}
\begin{eqnarray}
\mathrm{Im} \left[ \gamma_0 (T_{\mathrm{op}}) \right] = \frac{m}{2 \hbar} \, \frac{2 |x_0| v_0 - T_{\mathrm{op}} \left( v_0^2 - \hbar^2 x_0^2/m^2 \sigma^4 \right)}{1 + \hbar^2 T_{\mathrm{op}}^2 / m^2 \sigma^4}
\label{imag_gamma_0_def}
\end{eqnarray}
and
\begin{eqnarray}
\mathrm{Re} \left[ \gamma_1 (\widetilde{x},\widetilde{v},t;T_{\mathrm{op}}) \right] = - \frac{1}{2 \sigma^2} \, \frac{2 \left( T_{\mathrm{op}} - t \right) \widetilde{x} \widetilde{v} + \widetilde{x}^2 - m^2 \sigma^4 \widetilde{v}^2/\hbar^2}{1 + \hbar^2 \left( T_{\mathrm{op}} - t \right)^2 / m^2 \sigma^4} \, ,
\label{real_gamma_1_def}
\end{eqnarray}
\begin{eqnarray}
\mathrm{Im} \left[ \gamma_1 (\widetilde{x},\widetilde{v},t;T_{\mathrm{op}}) \right] = \frac{m}{2 \hbar} \, \frac{2 \widetilde{x} \widetilde{v} + \left( T_{\mathrm{op}} - t \right) \left( \widetilde{v}^2 - \hbar^2 \widetilde{x}^2/m^2 \sigma^4 \right)}{1 + \hbar^2 \left( T_{\mathrm{op}} - t \right)^2 / m^2 \sigma^4} \, .
\label{imag_gamma_1_def}
\end{eqnarray}

In view of~\eref{Husimi_distr_from_Husimi_ampl}, we hence define the slit Husimi distribution $F_{\mathrm{1slit}}$ by
\begin{eqnarray}
F_{\mathrm{1slit}} (\widetilde{x},\widetilde{v},t;T_{\mathrm{op}}) \equiv \left| f_{\mathrm{1slit}} (\widetilde{x},\widetilde{v},t;T_{\mathrm{op}}) \right| ^2 \, ,
\label{Husimi_dist_slit_def}
\end{eqnarray}
that is using~\eref{h_slit_gamma_expr}
\begin{eqnarray}
\fl F_{\mathrm{1slit}} (\widetilde{x},\widetilde{v},t;T_{\mathrm{op}}) = \frac{\pi^2}{\nu^2} \left| \Omega \right|^2 \exp \left\{ 2 \mathrm{Re} \left[ \gamma_0 (T_{\mathrm{op}}) \right] + 2 \mathrm{Re} \left[ \gamma_1 (\widetilde{x},\widetilde{v},t;T_{\mathrm{op}}) \right] \right\} \, .
\label{Husimi_dist_slit_gen_expr}
\end{eqnarray}

While~\eref{real_gamma_0_def}-\eref{imag_gamma_1_def} are valid for an arbitrary $T_{\mathrm{op}}$, the effect of the barrier is magnified when $T_{\mathrm{op}}$ is close to $t_{\mathrm{c}}$. In addition, to further assume that $t - T_{\mathrm{op}} = T_{\mathrm{op}}$ allows to considerably simplify~\eref{real_gamma_0_def}-\eref{imag_gamma_1_def}. Therefore, here we assume that
\begin{eqnarray}
T_{\mathrm{op}} = t_{\mathrm{c}} \qquad \mbox{and} \qquad t = 2 t_{\mathrm{c}} = 2 T_{\mathrm{op}} \, .
\label{op_time_fin_time_assumption}
\end{eqnarray}
Combining~\eref{Husimi_dist_slit_gen_expr} with~\eref{lambdaBar_t_tilde_def},~\eref{Omega_def},~\eref{real_gamma_0_def}-\eref{imag_gamma_1_def} and~\eref{op_time_fin_time_assumption} hence yields
\begin{eqnarray}
\fl F_{\mathrm{1slit}} (\widetilde{x},\widetilde{v},2t_{\mathrm{c}};t_{\mathrm{c}}) = \frac{\pi}{4 \sigma^2 \nu^2} \left[ \left( \widetilde{v} + v_0 \right)^2 + \frac{\hbar ^2}{m^2 \sigma^4} \left( \widetilde{x} + |x_0| \right)^2 \right] \exp \left[ - \frac{(\widetilde{x}-\widetilde{v} t_{\mathrm{c}})^2}{\sigma_{t_{\mathrm{c}}}^2} \right] \, .
\label{Husimi_dist_slit_fin_expr}
\end{eqnarray}
Therefore, it is clear from~\eref{Husimi_dist_slit_fin_expr} that $F_{\mathrm{1slit}}$ does not exhibit any diffraction pattern. This is in agreement with~\cite{BZ97}, see more precisely Eq.~(36) in~\cite{BZ97} for a single rectangular slit in space and in time: here we consider a time slit with a very small width, so that the nearest diffraction peak (whose position depends on the inverse of the width of the time slit) is essentially sent to infinity.

The expression~\eref{h_slit_gamma_expr} of $f_{\mathrm{1slit}}$ is valid for an arbitrary narrow Lorentzian aperture function $\chi_2$. It can be adequately used to construct the Husimi amplitude obtained in the more interesting case of a double slit in time, as we now discuss.


\subsection{Double slit}\label{dble_slit_subsec}

In this final section, we consider a double-slit scenario characterized by the aperture function $\chi_{\mathrm{2slit}}$ given by a superposition of two narrow Lorentzian functions $\chi_2$ that open at two different times $T_{\mathrm{op}}^{(0)}$ and $T_{\mathrm{op}}^{(1)}$, with $T_{\mathrm{op}}^{(0)} < T_{\mathrm{op}}^{(1)}$, that is
\begin{eqnarray}
\chi_{\mathrm{2slit}}(\tau) \equiv \frac{1}{2} \left\{ \frac{1}{1+\nu^2 \left[ \tau-T_{\mathrm{op}}^{(0)} \right]^2} + \frac{1}{1+\nu^2 \left[ \tau-T_{\mathrm{op}}^{(1)} \right]^2} \right\} \, .
\label{chi_double_slit_def}
\end{eqnarray}
The factor $1/2$ in~\eref{chi_double_slit_def} is added in order to ensure that $0 \leqslant \chi_{\mathrm{2slit}}(\tau) \leqslant 1$ at any time $\tau$. Similarly to section~\ref{nu_large_subsec} above, we still consider the slit regime $\nu \gg 1$ here.

As is clear from~\eref{Husimi_ampl_final_expr}, the Husimi amplitude is by construction linear in $\chi$. Therefore, the double-slit Husimi amplitude $f_{\mathrm{2slit}}$ that corresponds to~\eref{chi_double_slit_def} can be constructed from the single-slit Husimi amplitude $f_{\mathrm{1slit}}$, and we merely have
\begin{eqnarray}
\fl f_{\mathrm{2slit}} \left[ \widetilde{x},\widetilde{v},t;T_{\mathrm{op}}^{(0)},T_{\mathrm{op}}^{(1)} \right] = \frac{1}{2} \left\{ f_{\mathrm{1slit}} \left[ \widetilde{x},\widetilde{v},t;T_{\mathrm{op}}^{(0)} \right] + f_{\mathrm{1slit}} \left[ \widetilde{x},\widetilde{v},t;T_{\mathrm{op}}^{(1)} \right] \right\} \, .
\label{h_diff_def}
\end{eqnarray}
The corresponding double-slit Husimi distribution $F_{\mathrm{2slit}}$ is then defined by~\eref{Husimi_distr_from_Husimi_ampl}, i.e.
\begin{eqnarray}
F_{\mathrm{2slit}} \left[ \widetilde{x},\widetilde{v},t;T_{\mathrm{op}}^{(0)},T_{\mathrm{op}}^{(1)} \right] \equiv \left| f_{\mathrm{2slit}} \left[ \widetilde{x},\widetilde{v},t;T_{\mathrm{op}}^{(0)},T_{\mathrm{op}}^{(1)} \right] \right|^2 \, .
\label{Husimi_dist_double_slit_def}
\end{eqnarray}
Therefore, combining~\eref{Husimi_dist_double_slit_def} with~\eref{lambdaBar_t_tilde_def},~\eref{Omega_def},~\eref{h_slit_gamma_expr} and~\eref{h_diff_def} shows that $F_{\mathrm{2slit}}$ reads
\begin{eqnarray}
\fl F_{\mathrm{2slit}} \left[ \widetilde{x},\widetilde{v},t;T_{\mathrm{op}}^{(0)},T_{\mathrm{op}}^{(1)} \right] = \frac{\pi}{8 \sigma^2 \nu^2} g_1 (\widetilde{x},\widetilde{v}) \, \exp \left\{ g_2 \left[ \widetilde{x},\widetilde{v},t;T_{\mathrm{op}}^{(0)},T_{\mathrm{op}}^{(1)} \right] \right\} \nonumber\\[0.2cm]
\times \left( \cosh \left\{ f_1 \left[ \widetilde{x},\widetilde{v},t;T_{\mathrm{op}}^{(0)},T_{\mathrm{op}}^{(1)} \right] \right\} + \cos \left\{ f_2 \left[ \widetilde{x},\widetilde{v},t;T_{\mathrm{op}}^{(0)},T_{\mathrm{op}}^{(1)} \right] \right\} \right) \, ,
\label{Husimi_dist_double_slit_expr}
\end{eqnarray}
where we introduced the notations
\begin{eqnarray}
g_1 \left( \widetilde{x},\widetilde{v} \right) \equiv \left( \widetilde{v} + v_0 \right)^2 + \frac{\hbar ^2}{m^2 \sigma^4} \left( \widetilde{x} + |x_0| \right)^2 \, ,
\label{g_1_def}
\end{eqnarray}
\begin{eqnarray}
\fl g_2 \left[ \widetilde{x},\widetilde{v},t;T_{\mathrm{op}}^{(0)},T_{\mathrm{op}}^{(1)} \right] \equiv \mathrm{Re} \left\{ \gamma_0 \left[ T_{\mathrm{op}}^{(0)} \right] \right\} + \mathrm{Re} \left\{ \gamma_1 \left[ \widetilde{x},\widetilde{v},t;T_{\mathrm{op}}^{(0)} \right] \right\} \nonumber\\
+ \mathrm{Re} \left\{ \gamma_0 \left[ T_{\mathrm{op}}^{(1)} \right] \right\} + \mathrm{Re} \left\{ \gamma_1 \left[ \widetilde{x},\widetilde{v},t;T_{\mathrm{op}}^{(1)} \right] \right\} - \frac{m^2 \sigma^2}{\hbar^2} \left( \widetilde{v}^2 + v_0^2 \right) \, ,
\label{g_2_def}
\end{eqnarray}
\begin{eqnarray}
\fl f_1 \left[ \widetilde{x},\widetilde{v},t;T_{\mathrm{op}}^{(0)},T_{\mathrm{op}}^{(1)} \right] \equiv \mathrm{Re} \left\{ \gamma_0 \left[ T_{\mathrm{op}}^{(0)} \right] \right\} + \mathrm{Re} \left\{ \gamma_1 \left[ \widetilde{x},\widetilde{v},t;T_{\mathrm{op}}^{(0)} \right] \right\} \nonumber\\
- \mathrm{Re} \left\{ \gamma_0 \left[ T_{\mathrm{op}}^{(1)} \right] \right\} - \mathrm{Re} \left\{ \gamma_1 \left[ \widetilde{x},\widetilde{v},t;T_{\mathrm{op}}^{(1)} \right] \right\} \, ,
\label{f_1_def}
\end{eqnarray}
\begin{eqnarray}
\fl f_2 \left[ \widetilde{x},\widetilde{v},t;T_{\mathrm{op}}^{(0)},T_{\mathrm{op}}^{(1)} \right] \equiv \mathrm{Im} \left\{ \gamma_0 \left[ T_{\mathrm{op}}^{(0)} \right] \right\} + \mathrm{Im} \left\{ \gamma_1 \left[ \widetilde{x},\widetilde{v},t;T_{\mathrm{op}}^{(0)} \right] \right\} \nonumber\\
- \mathrm{Im} \left\{ \gamma_0 \left[ T_{\mathrm{op}}^{(1)} \right] \right\} - \mathrm{Im} \left\{ \gamma_1 \left[ \widetilde{x},\widetilde{v},t;T_{\mathrm{op}}^{(1)} \right] \right\} \, .
\label{f_2_def}
\end{eqnarray}

It seems reasonable to expect the occurrence of diffraction in the case where the time difference $T_{\mathrm{op}}^{(1)} - T_{\mathrm{op}}^{(0)}$ between the two slits is small enough, as it is the analog of the distance that separates the two slits in Young's classic double-slit scenario. In view of this, i) we assume for convenience that the two opening times $T_{\mathrm{op}}^{(0)}$ and $T_{\mathrm{op}}^{(1)}$ are taken symmetrically with respect to the classical hitting time $t_{\mathrm{c}}$, that is
\begin{eqnarray}
T_{\mathrm{op}}^{(0)} = t_{\mathrm{c}} (1-\epsilon) \qquad \mbox{and} \qquad T_{\mathrm{op}}^{(1)} = t_{\mathrm{c}} (1+\epsilon) \, ,
\label{sym_times_dble_slit}
\end{eqnarray}
and ii) we further assume that the dimensionless parameter $\epsilon$ is small, i.e.
\begin{eqnarray}
\epsilon \ll 1 \, .
\label{epsilon_small}
\end{eqnarray}
This ensures that both slits open when a significant part of the incident wave packet reaches the barrier. Furthermore, in order to make the algebra as simple as possible while at the same time preserving the essence of diffraction, similarly to~\eref{op_time_fin_time_assumption} we here again assume that the final time $t$ is simply twice the classical hitting time $t_{\mathrm{c}}$,
\begin{eqnarray}
t = 2t_{\mathrm{c}} \, .
\label{t_2_t_c}
\end{eqnarray}
Combining~\eref{sym_times_dble_slit} with~\eref{t_2_t_c} yields the symmetries
\begin{eqnarray}
t - T_{\mathrm{op}}^{(0)} = T_{\mathrm{op}}^{(1)} \qquad \mbox{and} \qquad t - T_{\mathrm{op}}^{(1)} = T_{\mathrm{op}}^{(0)} \, .
\label{sym_time_differences}
\end{eqnarray}
The expressions of the functions $g_2$, $f_1$ and $f_2$ defined by~\eref{g_2_def}-\eref{f_2_def} that result from~\eref{sym_times_dble_slit} and~\eref{t_2_t_c} are written explicitly in~\ref{expr_funct_app}.


\begin{figure}[ht]
\centering
\includegraphics[width=1.0\textwidth]{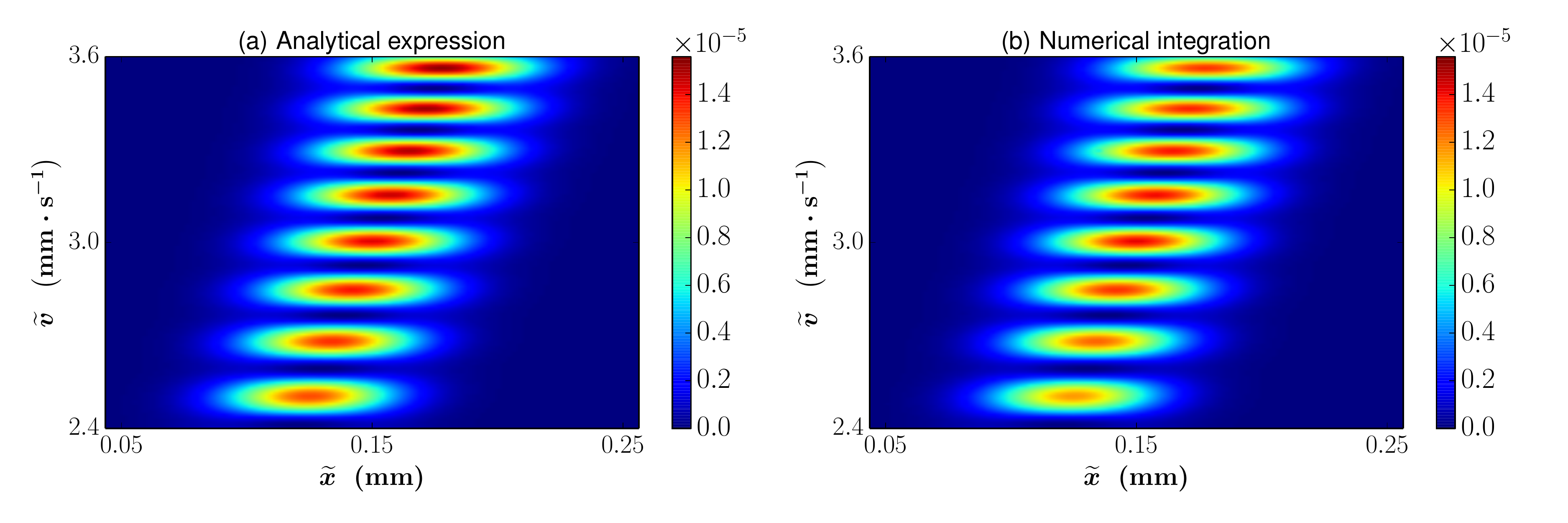}
\caption{(a) Analytic expression~\eref{Husimi_dist_double_slit_expr} of the double-slit Husimi distribution $F_{\mathrm{2slit}}$ (see text for numerical parameters); (b) Corresponding Husimi distribution obtained by means of a fully numerical evaluation of the integral~\eref{Husimi_ampl_final_expr} for the aperture function~\eref{chi_double_slit_def} and for the exact same parameters as in (a).}
\label{an_vs_num_H_diff_fig}
\end{figure}


We now compute the Husimi distribution~\eref{Husimi_dist_double_slit_expr} in view of~\eref{sym_times_dble_slit}-\eref{t_2_t_c}. The results are shown on figure~\ref{an_vs_num_H_diff_fig}(a) for a $^{87} \mathrm{Rb}$ atom of mass $m_{\mathrm{Rb}} = 86.9091805 \, \mathrm{u}$. We choose as numerical parameters $x_0 = - 0.15 \, \mathrm{mm}$, $\sigma = 30 \, \mu \mathrm{m}$, $v_0 = 3 \, \mathrm{mm} / \mathrm{s}$, $t_{\mathrm{c}} \equiv | x_0 | / v_0 = 50 \, \mathrm{ms}$ and $t = 2 t_{\mathrm{c}} = 100 \, \mathrm{ms}$ (similar parameters have been e.g. used by Jendrzejewski \textit{et al.} in their study of the coherent backscattering of ultracold atoms of $^{87} \mathrm{Rb}$ \cite{JMR12}). This set of parameters is designed so as to satisfy the frozen Gaussian regime~\eref{semiclassical_reg_def} since we have $v_0 \left( t-t_{\mathrm{c}} \right) = |x_0| = 0.15 \, \mathrm{mm}$, while $m \sigma^2 v_0 / \hbar \simeq 3.7 \, \mathrm{mm} > 20 \, v_0 \left( t-t_{\mathrm{c}} \right)$. The dimensionless parameter $\epsilon$ in~\eref{sym_times_dble_slit} is set to $\epsilon = 0.1$, hence yielding the opening times $T_{\mathrm{op}}^{(0)}=45 \, \mathrm{ms}$ and $T_{\mathrm{op}}^{(1)}=55 \, \mathrm{ms}$. Finally, we have set $\nu \approx 36.537 \, \mathrm{kHz}$ (namely $\nu=5000$ in the corresponding numerical set of dimensionless parameters). It is clear from figure~\ref{an_vs_num_H_diff_fig}(a) that the double-slit Husimi distribution~\eref{Husimi_dist_double_slit_expr} exhibits a succession of peaks in the phase space: this is indeed a clear signature of diffraction.

In order to check that the function $F_{\mathrm{2slit}}$ given by~\eref{Husimi_dist_double_slit_expr} indeed provides an accurate analytic description of the actual state of the system, we confront it to a fully numerical evaluation of the original integral~\eref{Husimi_ampl_final_expr}, which hence gives the actual Husimi amplitude, for the double-slit aperture function~\eref{chi_double_slit_def}. The results are shown on figure~\ref{an_vs_num_H_diff_fig}(b) for the exact same numerical parameters as the ones used on figure~\ref{an_vs_num_H_diff_fig}(a). We can readily see that the agreement between the numerical and analytic results is excellent: while our analytic expression~\eref{Husimi_dist_double_slit_expr} apparently slightly overestimates the amplitude of the peaks, the phase-space structure of the Husimi distribution is indeed remarkably predicted by~\eref{Husimi_dist_double_slit_expr}. Therefore, we can now adequately use the latter to analytically investigate the phase-space structure of the diffraction pattern.

We are for instance able to infer an analytic expression of the position of the interference fringes in phase space. To this end, we first note on figure~\ref{an_vs_num_H_diff_fig} that the peaks are seemingly arranged along a line $\widetilde{v} = \alpha \widetilde{x} + \beta$, for some $\alpha > 0$ and $\beta \in \mathbb{R}$. In view of the mathematical structure~\eref{Husimi_dist_double_slit_expr} of the double-slit Husimi distribution, our strategy is thus as follows (additional technical details may be found in~\ref{peaks_app}): i) First, we substitute the ansatz $\widetilde{v} = \alpha \widetilde{x} + \beta$ into the expression~\eref{f_1_def} of $f_1$. ii) We then require the resulting expression of $f_1$ to vanish, i.e. $ f_1 \left[ \widetilde{x},\alpha \widetilde{x} + \beta \right] = 0$ (what is also suggested by numerical investigations). This allows us to obtain $\alpha$ and $\beta$, and we find
\begin{eqnarray}
\alpha = \frac{1}{t_{\mathrm{c}}} \left[ 1 + \bigO \left( \frac{\hbar^2 t_{\mathrm{c}}^2}{m^2 \sigma^4} \epsilon^2 \right) \right] \qquad \mbox{and} \qquad \beta = v_0 \, \bigO \left( \frac{\hbar^2 t_{\mathrm{c}}^2}{m^2 \sigma^4} \epsilon^2 \right) \, .
\label{alpha_beta_expr}
\end{eqnarray}
This ensures that in the diffraction regime~\eref{epsilon_small}, we can safely take $\alpha = 1/t_{\mathrm{c}}=v_0/|x_0|$ and $\beta = 0$. iii) Finally, we substitute the resulting ansatz $\widetilde{v} = \widetilde{x} / t_{\mathrm{c}} = v_0 \widetilde{x} / |x_0|$ into the expression~\eref{f_2_def} of $f_2$, and require $\sin f_2$ to vanish, i.e. $ \sin \left[ f_2 \left( \widetilde{x},\widetilde{x}/t_{\mathrm{c}} \right) \right] = 0$. The latter condition hence precisely yields a countable family of solutions $\{\widetilde{x}_k^{(2)},\widetilde{v}_k^{(2)}\}$ with $k \in \mathbb{Z}$, and we find
\begin{eqnarray}
\fl \left\{ \widetilde{x}_k^{(2)},\widetilde{v}_k^{(2)} \right\} = \left\{ \left| x_0 \right| \sqrt{1 + \frac{2 \hbar}{m v_0^2 \left[ T_{\mathrm{op}}^{(1)} - T_{\mathrm{op}}^{(0)} \right]} k \pi} \, , \, v_0 \sqrt{1 + \frac{2 \hbar}{m v_0^2 \left[ T_{\mathrm{op}}^{(1)} - T_{\mathrm{op}}^{(0)} \right]} k \pi} \right\} \, .
\label{x_k_v_k_def}
\end{eqnarray}
This result precisely yields the position of the interference fringes in phase space. This is in agreement with~\cite{BZ97}, see more precisely Eq.~(44) in~\cite{BZ97} for a single rectangular slit in space and a double rectangular slit in time.


\begin{figure}[ht]
\centering
\includegraphics[width=1.0\textwidth]{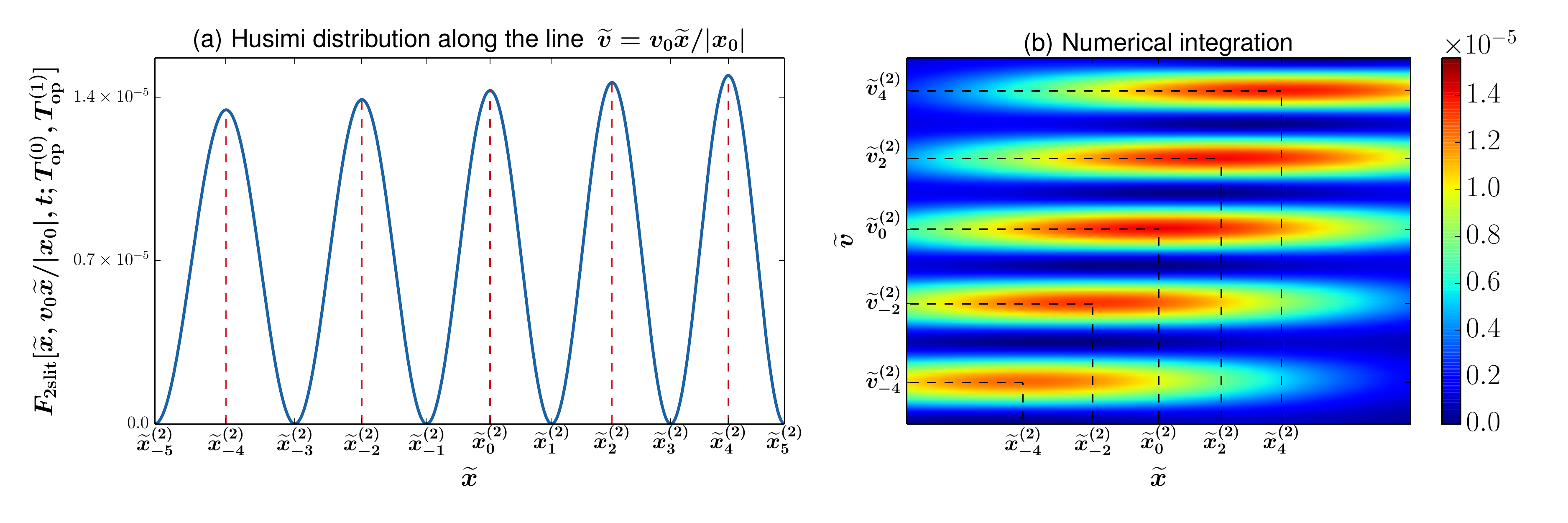}
\caption{Confirmation of the expression~\eref{x_k_v_k_def} of the position of the interference fringes in phase space (same numerical parameters as on figure~\ref{an_vs_num_H_diff_fig}): (a) Double-slit Husimi distribution $F_{\mathrm{2slit}}$, as given by~\eref{Husimi_dist_double_slit_expr}, along the line $\widetilde{v} = v_0 \widetilde{x}/|x_0|$; (b)~Corresponding Husimi distribution obtained by numerically evaluating the integral~\eref{Husimi_ampl_final_expr} for the aperture function~\eref{chi_double_slit_def} and for the same parameters as in (a).}
\label{peaks_fig}
\end{figure}


The validity of our analytic prediction~\eref{x_k_v_k_def} is illustrated on figure~\ref{peaks_fig} (where we use the exact same set of parameters as on figure~\ref{an_vs_num_H_diff_fig} above). We first plot on figure~\ref{peaks_fig}(a) the double-slit Husimi distribution $F_{\mathrm{2slit}}$, as given by~\eref{Husimi_dist_double_slit_expr}, along the line $\widetilde{v} = v_0 \widetilde{x}/|x_0|$. As anticipated, the latter is indeed seen to contain both the dark and the bright interference fringes, whose positions are indeed perfectly described by our analytic result~\eref{x_k_v_k_def}. More precisely, the phase-space points~\eref{x_k_v_k_def} for even values of the index $k$ correspond to peaks of the Husimi distribution $F_{\mathrm{2slit}}$, i.e. to bright interference fringes. On the other hand, the phase-space points~\eref{x_k_v_k_def} for odd values of the index $k$ correspond to dark interference fringes, i.e. points where $F_{\mathrm{2slit}}$ vanishes. To further ensure that~\eref{x_k_v_k_def} indeed accurately describes the phase-space structure of the actual Husimi distribution, we then superimpose on figure~\ref{peaks_fig}(b) the phase-space points~\eref{x_k_v_k_def} for $k=-4,-2,0,2,4$ onto the corresponding Husimi distribution obtained by numerically evaluating the integral~\eref{Husimi_ampl_final_expr} for the aperture function~\eref{chi_double_slit_def} and for the exact same numerical parameters as on figure~\ref{peaks_fig}(a). This clearly confirms the accuracy of our analytic result~\eref{x_k_v_k_def} to describe the position of the diffraction peaks in phase space. 

The above analysis can of course be repeated for generalizations of the double-slit aperture function~\eref{chi_double_slit_def} to aperture functions $\chi_{\mathrm{3slit}}$, $\chi_{\mathrm{4slit}}$, $\ldots$ that are a sum of 3,~4,~$\ldots$ Lorentzian slits all opening at different times. Though the algebra becomes more intricate even in the slit regime $\nu \gg 1$, this effectively allows us to analytically study diffraction in time for general time gratings.


\section{Summary and conclusion}\label{concl_sec}

In this paper we investigated how a particular model of matter-wave absorption, the so-called aperture function model, can be adequately used to obtain an analytic phase-space representation of diffraction in time.

We considered a nonrelativistic, structureless quantum particle that follows a one-dimensional motion along the $x$-axis. The particle is assumed to be free everywhere except at $x=0$ where it is taken to be subjected to a spatially infinitely thin, pointlike time-dependent absorbing barrier. The aperture function model characterizes the transparency of the barrier by a (time-dependent) function $\chi(t)$, termed the aperture function, whose values range between 0 (completely closed barrier) and 1 (fully open barrier). The effect of the barrier on the particle is then taken into account by imposing discontinuous time-dependent matching conditions, which involve $\chi$, on both the wave function $\Psi$ and its spatial derivative $\partial \Psi / \partial x$ at $x=0$.

The advantage of the aperture function model is that it allows to analytically express the wave function $\Psi(x,t)$, evolved from the initial minimum-uncertainty Gaussian state~\eref{Psi_0_def} according to~\eref{transm_state_def}, in the transmission ($x>0$) region in the form of the integral~\eref{transm_state_expr}. This remains true for the Husimi amplitude $f$ (yielding the Husimi distribution $F$ through $F \equiv |f|^2$) associated with $\Psi$, given by~\eref{Husimi_ampl_final_expr}. This integral expression of $f$ proves to be valid for an arbitrary time-dependent aperture function $\chi$. The main aim of our work has then been to explicitly compute the latter integral by means of Cauchy's residue theory.

We found that the main difficulty that arose from this complex-analysis-based approach was the need to compute a residue at an essential singularity, for which no general method exists. Therefore, we had to resort to the very definition of a residue, and thus to construct the relevant part of the Laurent series of the corresponding function (namely, the $1/z$ term if the singularity is $z=0$). We did this for the particular class of aperture functions $\chi_n$ given by~\eref{chi_n_tau_def}, with $n$ an arbitrary even integer. Such functions describe Lorentzian-like barriers that smoothly open around the opening time $T_{\mathrm{op}}$ with a width $1/\nu$. We hence obtained the expression~\eref{h_sc_n_explicit_expr_general}-\eref{Husimi_ampl_n_h_sc_n_def} of the Husimi amplitude $f^{(n)}$ that corresponds to the aperture function $\chi_n$.

The latter expression, valid for any $n$, appears to be rather convoluted, in particular due to the presence of a series of Bessel functions. Therefore, we gave particular attention to the Lorentzian case $n=2$. Furthermore, considering the slit regime $\nu \gg 1$ of very narrow barriers allowed us to reduce the Husimi amplitude $f^{(2)}$ to the considerably simpler expression~\eref{h_sc_2_leading_order_nu_final} in terms of the quantity $f_{\mathrm{1slit}}$ given by~\eref{h_slit_gamma_expr}. We saw in particular that the resulting Husimi distribution $F_{\mathrm{1slit}}$, given by~\eref{Husimi_dist_slit_fin_expr}, exhibits no interference pattern. We then exploited the linearity, obvious on~\eref{Husimi_ampl_final_expr}, of the Husimi amplitude in the aperture function $\chi$ and considered the double Lorentzian $\chi_{\mathrm{2slit}}$ given by~\eref{chi_double_slit_def}. The latter hence describes the time-domain version of the double-slit scenario, with two time slits that open at different times $T_{\mathrm{op}}^{(0)}$ and $T_{\mathrm{op}}^{(1)}$. The general structure of the resulting Husimi distribution $F_{\mathrm{2slit}}$, given by~\eref{Husimi_dist_double_slit_expr}, allows for the appearance of diffraction in time: a clean diffraction pattern indeed arises in the regime of parameters described by~\eref{sym_times_dble_slit}-\eref{t_2_t_c}. This eventually allowed us to derive the analytic expression~\eref{x_k_v_k_def} of the position of the interference fringes in phase space.

In conclusion, the main outcome of our work is to provide a simple and intuitive analytic description of the phase-space structure of diffraction in time that arises from a class of smooth, Lorentzian-like time gratings. An interesting followup question is to investigate the structure of the Husimi distribution $F^{(n)}=|f^{(n)}|^2$ outside of the Lorentzian case, i.e. for $n=4,6,\ldots$ Another followup direction is to explore the phase-space structure of the Husimi distribution out of the narrow-slit regime $\nu \gg 1$ in order for instance to determine the impact of the term that involves Bessel functions, a direct consequence of the essential singularity.


\section*{Acknowledgments}\label{AA}

M. B. acknowledges Pierre Gaspard for useful discussions.


\appendix


\section{Derivation of\texorpdfstring{~\eref{Husimi_ampl_h_sc_def}-\eref{h_sc_def}}{the frozen Gaussian Husimi amplitude}}\label{f_froz_app}

Combining~\eref{Husimi_ampl_final_expr} with~\eref{free_Gaussian_expr}-\eref{S_x_tau_def} and~\eref{alpha_T_def} allows to write $f$ in the form
\begin{eqnarray}
f(\widetilde{x},\widetilde{v},t) = \int_{0}^{t} d\tau \, \widetilde{f}(\tau) \, ,
\label{Husimi_ampl_short_expr}
\end{eqnarray}
with the function $\widetilde{f}(\tau)$ defined by
\begin{eqnarray}
\widetilde{f}(\tau) \equiv \frac{\chi (\tau)}{2} \sqrt{\frac{2}{\pi \alpha_0}} \left( \frac{\widetilde{v}}{\alpha_{\widetilde{t}}} \alpha_{t-\tau}^{\frac{3}{2}} \alpha_{\tau}^{\frac{1}{2}} + \frac{v_0}{\alpha_{t_{\mathrm{c}}}} \alpha_{\tau}^{\frac{3}{2}} \alpha_{t-\tau}^{\frac{1}{2}} \right) e^{\varphi (\tau)} \, .
\label{f_function_def}
\end{eqnarray}
We now use the frozen Gaussian approximation~\eref{frozen_Gaussian_cond} that results from~\eref{semiclassical_reg_def} to Taylor-expand the square roots in~\eref{f_function_def}.

Because~\eref{frozen_Gaussian_cond} can also be alternatively written as
\begin{eqnarray}
0 \leqslant \frac{\hbar \tau}{m \sigma^2} \ll 1 \qquad \mbox{and} \qquad 0 \leqslant \frac{\hbar (t-\tau)}{m \sigma^2} \ll 1 \quad , \quad \forall \tau \in \left[ 0 , t \right] \, ,
\label{frozen_cond_tau}
\end{eqnarray}
combining~\eref{frozen_cond_tau} with the definition~\eref{alpha_T_def} of $\alpha_{\tau}$ hence yields the Taylor expansion
\begin{eqnarray}
\alpha_{\tau}^{\mu} = \alpha_0^{\mu} \left[ 1 + \bigO \left( i \frac{\hbar \tau}{m \sigma^2} \right) \right] \quad , \quad \forall \tau \in \left[ 0 , t \right] \quad , \quad \mu = \frac{1}{2} \, , \frac{3}{2} \, ,
\label{Taylor_alpha_al0_tau_powerMu}
\end{eqnarray}
which, because $\tau \leqslant t$ and by definition of the $\bigO$ notation, is equivalent to
\begin{eqnarray}
\alpha_{\tau}^{\mu} = \alpha_0^{\mu} \left[ 1 + \bigO \left( i \frac{\hbar t}{m \sigma^2} \right) \right] \quad , \quad \forall \tau \in \left[ 0 , t \right] \quad , \quad \mu = \frac{1}{2} \, , \frac{3}{2} \, ,
\label{Taylor_alpha_al0_tau_powerMu_final}
\end{eqnarray}
where the right-hand side has now the advantage of being independent of $\tau$ as compared to the right-hand side in~\eref{Taylor_alpha_al0_tau_powerMu}. Similarly, we can thus write
\begin{eqnarray}
\alpha_{t-\tau}^{\mu} = \alpha_0^{\mu} \left[ 1 + \bigO \left( i \frac{\hbar t}{m \sigma^2} \right) \right] \quad , \quad \forall \tau \in \left[ 0 , t \right] \quad , \quad \mu = \frac{1}{2} \, , \frac{3}{2} \, .
\label{Taylor_alpha_al0_t_minus_tau_powerMu_final}
\end{eqnarray}

Therefore, we combine~\eref{Taylor_alpha_al0_tau_powerMu_final} with~\eref{Taylor_alpha_al0_t_minus_tau_powerMu_final} to get
\begin{eqnarray}
\fl \alpha_{t-\tau}^{\frac{3}{2}} \alpha_{\tau}^{\frac{1}{2}} = \alpha_0^{2} \left[ 1 + \bigO \left( i \frac{\hbar t}{m \sigma^2} \right) \right] \qquad \mbox{and} \qquad \alpha_{\tau}^{\frac{3}{2}} \alpha_{t-\tau}^{\frac{1}{2}} = \alpha_0^{2} \left[ 1 + \bigO \left( i \frac{\hbar t}{m \sigma^2} \right) \right] \, .
\label{products_alpha_al0}
\end{eqnarray}
Substituting now~\eref{products_alpha_al0} into~\eref{f_function_def} hence shows that $\widetilde{f}(\tau)$ can be written in the form
\begin{eqnarray}
\widetilde{f}(\tau) = \left[ 1 + \bigO \left( i \frac{\hbar t}{m \sigma^2} \right) \right] \widetilde{f}_{\mathrm{froz}} (\tau) \, ,
\label{h_function_expansion_app}
\end{eqnarray}
with $\widetilde{f}_{\mathrm{froz}} (\tau)$ given by~\eref{h_sc_def}. As compared to its original definition~\eref{f_function_def}, the resulting expression~\eref{h_function_expansion_app} of $\widetilde{f}(\tau)$ is single valued. This allows for a straightforward extension of the integral in~\eref{Husimi_ampl_short_expr} to a branch-cut-free contour integral in the complex plane.


\section{Upper bound for \texorpdfstring{$\left| I^{(-)} + I^{(+)} \right|$}{the complementary real axis integrals}}\label{upper_bound_app}

This appendix is devoted to deriving a relevant upper bound for $|I^{(-)} + I^{(+)}|$. For clarity, we recall the following standard result about positive powers of positive real numbers:
\begin{eqnarray}
\forall a,b \in \mathbb{R}_+ \quad , \quad \forall \mu \in \mathbb{R}_+^* \quad , \quad a \geqslant b \iff a^{\mu} \geqslant b^{\mu} \, ,
\label{powers_posRealNumb_prop}
\end{eqnarray}
where $\mathbb{R}_+$ ($\mathbb{R}_+^*$) denotes the set of all positive real numbers with 0 included (excluded). Furthermore, combining the definition~\eref{I_minus_I_plus_def} with the triangle inequality yields
\begin{eqnarray}
\fl \left| I^{(-)} \right| \leqslant \int_{- \infty}^{0} d \tau \, \left| \widetilde{f}_{\mathrm{froz}} (\tau) \right| \qquad \mbox{and} \qquad \left| I^{(+)} \right| \leqslant \int_{t}^{\infty} d \tau \, \left| \widetilde{f}_{\mathrm{froz}} (\tau) \right| \, .
\label{modulus_I_minus_plus_triangleIneq}
\end{eqnarray}

We first write an upper bound for $| \widetilde{f}_{\mathrm{froz}} (\tau) |$ in~\ref{bound_modulus_h_fr_sec}. After discussing some technical details in~\ref{f_sec_app} and~\ref{g_sec_app}, we derive upper bounds for $|I^{(-)}|$ and $|I^{(+)}|$ in~\ref{bound_I_minus_sec} and~\ref{bound_I_plus_sec}, respectively.


\subsection{Upper bound for \texorpdfstring{$\left| \widetilde{f}_{\mathrm{froz}} (\tau) \right|$}{the modulus of h in the FGR}}\label{bound_modulus_h_fr_sec}

In view of its definition~\eref{h_sc_def}, the function $\widetilde{f}_{\mathrm{froz}} (\tau)$ can be written as
\begin{eqnarray}
\widetilde{f}_{\mathrm{froz}} (\tau) = \Gamma \chi (\tau) \, e^{\varphi (\tau)} \, ,
\label{h_frozen_short_expr}
\end{eqnarray}
\noindent where $\varphi$ is given by~\eref{varPhi_def}, and $\Gamma$ is, in view of~\eref{alpha_T_def}-\eref{lambdaBar_t_tilde_def}, defined by
\begin{eqnarray}
\Gamma \equiv \frac{1}{2 \sqrt{\pi}} \left[ \frac{\widetilde{v}}{\sigma} \left( 1 +i \frac{\widetilde{\lambdabar} \widetilde{x}}{\sigma^2} \right) + \frac{v_0}{\sigma} \left( 1 +i \frac{\lambdabar \left| x_0 \right|}{\sigma^2} \right) \right] \, .
\label{Gamma_factor_expr}
\end{eqnarray}

We now take the modulus of~\eref{h_frozen_short_expr}. We first note that
\begin{eqnarray}
\left| e^{\varphi (\tau)} \right| = e^{\real \left[ \varphi (\tau) \right]} \, ,
\label{modulus_exp_curlyPhi}
\end{eqnarray}
with $\real \left( z \right)$ denoting the real part of $z$, that is in view of~\eref{alpha_T_def} and~\eref{varPhi_def}
\begin{eqnarray}
\left| e^{\varphi (\tau)} \right| = \widetilde{u}_{\widetilde{x}, \, \widetilde{v}} (\tau) \, u_{x_0, \, v_0} (\tau) \, ,
\label{modulus_exp_curlyPhi_f_and_g}
\end{eqnarray}
where the functions $\widetilde{u}_{\widetilde{x}, \, \widetilde{v}} (\tau)$ and $u_{x_0, \, v_0} (\tau)$ are defined by
\begin{eqnarray}
\fl \widetilde{u}_{\widetilde{x}, \, \widetilde{v}} (\tau) \equiv \mathrm{exp} \left\{ - \frac{\alpha_0 \left[ \widetilde{x} - \widetilde{v} (t-\tau) \right] ^2}{1 + \frac{4 \hbar ^2 \alpha_0^2}{m^2} (t-\tau)^2} \right\} \enspace \mbox{and} \enspace u_{x_0, \, v_0} (\tau) \equiv \mathrm{exp} \left[ - \frac{\alpha_0 \left( x_0 + v_0 \tau \right) ^2}{1 + \frac{4 \hbar ^2 \alpha_0^2}{m^2} \tau^2} \right] \, .
\label{u_xTilde_vTilde_u_x0_v0_def}
\end{eqnarray}
Finally, taking the modulus of~\eref{Gamma_factor_expr} and using the triangle inequality, we can write
\begin{eqnarray}
\left| \Gamma \right| \leqslant \Gamma_{\mathrm{up}} \, ,
\label{modulus_Gamma_bound}
\end{eqnarray}
where $\Gamma_{\mathrm{up}}$ is defined by
\begin{eqnarray}
\Gamma_{\mathrm{up}} \equiv \frac{1}{2 \sqrt{\pi}} \left[ \frac{\widetilde{v}}{\sigma} \sqrt{1 + \left( \frac{\widetilde{\lambdabar} \widetilde{x}}{\sigma^2} \right) ^2} + \frac{v_0}{\sigma} \sqrt{1 + \left( \frac{\lambdabar \left| x_0 \right|}{\sigma^2} \right) ^2} \right] \, .
\label{Gamma_up_def_app5}
\end{eqnarray}

Therefore, combining~\eref{h_frozen_short_expr} with~\eref{modulus_exp_curlyPhi_f_and_g} and~\eref{modulus_Gamma_bound}, we get:
\begin{eqnarray}
\left| \widetilde{f}_{\mathrm{froz}} (\tau) \right| \leqslant \Gamma_{\mathrm{up}} \left| \chi (\tau) \right| \widetilde{u}_{\widetilde{x}, \, \widetilde{v}} (\tau) \, u_{x_0, \, v_0} (\tau) \, .
\label{up_bound_modulus_h_frozen}
\end{eqnarray}
This is our starting point for deriving upper bounds of $|I^{(-)}|$ and $|I^{(+)}|$. We do this in~\ref{bound_I_minus_sec} and~\ref{bound_I_plus_sec}, respectively, after we study in~\ref{f_sec_app} and~\ref{g_sec_app} the general behavior of the two functions $\widetilde{u}_{\widetilde{x}, \, \widetilde{v}} (\tau)$ and $u_{x_0, \, v_0} (\tau)$, respectively. More explicitly, we want to determine their senses of variation for $\tau \in \mathbb{R}$. This is indeed a simple approach of finding upper bounds for these two functions.


\subsection{Behavior of \texorpdfstring{$\widetilde{u}_{\widetilde{x}, \, \widetilde{v}}$}{f}}\label{f_sec_app}

We first find the stationary points of $\widetilde{u}_{\widetilde{x}, \, \widetilde{v}}$, i.e. the points where the derivative $\widetilde{u}_{\widetilde{x}, \, \widetilde{v}}^{\prime} \equiv d \widetilde{u}_{\widetilde{x}, \, \widetilde{v}} / d\tau$ vanishes. Differentiating~\eref{u_xTilde_vTilde_u_x0_v0_def} with respect to $\tau$ yields
\begin{eqnarray}
\fl \widetilde{u}_{\widetilde{x}, \, \widetilde{v}}^{\prime} (\tau) = \left[ \widetilde{v} \tau - \left( \widetilde{v} t - \widetilde{x} \right) \right] \left[ \frac{4 \hbar ^2 \alpha_0^2}{m^2} \, \widetilde{x} \tau - \left( \frac{4 \hbar ^2 \alpha_0^2}{m^2} \, \widetilde{x} t + \widetilde{v} \right) \right] \frac{2 \alpha_0 \widetilde{u}_{\widetilde{x}, \, \widetilde{v}} (\tau)}{\left[ 1 + \frac{4 \hbar ^2 \alpha_0^2}{m^2} (t-\tau)^2 \right] ^2} \, .
\label{fPrime_xTilde_vTilde_expr_app3}
\end{eqnarray}
Since $\alpha_0 \neq 0$ by assumption, and $\widetilde{u}_{\widetilde{x}, \, \widetilde{v}} (\tau) \neq 0$, $\forall \tau \in \mathbb{R}$, we can readily see on~\eref{fPrime_xTilde_vTilde_expr_app3} that the derivative $\widetilde{u}_{\widetilde{x}, \, \widetilde{v}}^{\prime} (\tau)$ admits the two real roots $\tau_{\widetilde{x}, \, \widetilde{v}}^{(1)}$ and $\tau_{\widetilde{x}, \, \widetilde{v}}^{(2)}$ given by
\begin{eqnarray}
\tau_{\widetilde{x}, \, \widetilde{v}}^{(1)} = t - \widetilde{t} \qquad \mbox{and} \qquad \tau_{\widetilde{x}, \, \widetilde{v}}^{(2)} = t + \frac{m^2}{4 \hbar ^2 \alpha_0^2 \widetilde{t}} \, ,
\label{stat_points_f_expr_app3}
\end{eqnarray}
where we used the definition~\eref{lambdaBar_t_tilde_def} of $\widetilde{t}$. Remember that in our case $\widetilde{x}, \, \widetilde{v} > 0$, and hence $\widetilde{t} > 0$. Therefore, a direct consequence of~\eref{stat_points_f_expr_app3} is that
\begin{eqnarray}
\tau_{\widetilde{x}, \, \widetilde{v}}^{(1)} < t < \tau_{\widetilde{x}, \, \widetilde{v}}^{(2)} \, .
\label{stat_points_f_ordering_cond_app3}
\end{eqnarray}

We now study the sign of $\widetilde{u}_{\widetilde{x}, \, \widetilde{v}}^{\prime} (\tau)$ for $\tau \in  (\tau_{\widetilde{x}, \, \widetilde{v}}^{(1)} \, , \tau_{\widetilde{x}, \, \widetilde{v}}^{(2)})$. In view of~\eref{stat_points_f_ordering_cond_app3}, and because $\tau_{\widetilde{x}, \, \widetilde{v}}^{(1)}$ and $\tau_{\widetilde{x}, \, \widetilde{v}}^{(2)}$ are by construction the two only roots of $\widetilde{u}_{\widetilde{x}, \, \widetilde{v}}^{\prime}$, this is equivalent to merely studying the sign of $\widetilde{u}_{\widetilde{x}, \, \widetilde{v}}^{\prime} (t)$. From~\eref{u_xTilde_vTilde_u_x0_v0_def} and~\eref{fPrime_xTilde_vTilde_expr_app3} we get
\begin{eqnarray}
\widetilde{u}_{\widetilde{x}, \, \widetilde{v}}^{\prime} (t) = - 2 \alpha_0 \widetilde{x} \widetilde{v} \, e^{- \alpha_0 \widetilde{x}^2} \, ,
\label{fPrime_xTilde_vTilde_t_expr_app3}
\end{eqnarray}
and thus, since by assumption $\alpha_0, \, \widetilde{x}, \, \widetilde{v} > 0$,
\begin{eqnarray}
\widetilde{u}_{\widetilde{x}, \, \widetilde{v}}^{\prime} (t) < 0 \, .
\label{fPrime_xTilde_vTilde_t_sign_app3}
\end{eqnarray}
Therefore, we deduce from~\eref{fPrime_xTilde_vTilde_t_sign_app3} that 
\begin{eqnarray}
\widetilde{u}_{\widetilde{x}, \, \widetilde{v}}^{\prime} (\tau) < 0 \quad , \quad \forall \tau \in \left( \tau_{\widetilde{x}, \, \widetilde{v}}^{(1)} \, , \tau_{\widetilde{x}, \, \widetilde{v}}^{(2)} \right) \, .
\label{fPrime_xTilde_vTilde_neg_sign_app3}
\end{eqnarray}

We now determine the sign of $\widetilde{u}_{\widetilde{x}, \, \widetilde{v}}^{\prime} (\tau)$ for $\tau \notin [ \tau_{\widetilde{x}, \, \widetilde{v}}^{(1)} \, , \tau_{\widetilde{x}, \, \widetilde{v}}^{(2)} ]$. To do this, we study the behavior of $\widetilde{u}_{\widetilde{x}, \, \widetilde{v}}^{\prime} (\tau)$ as $\tau \to \pm \infty$. From~\eref{u_xTilde_vTilde_u_x0_v0_def} and~\eref{fPrime_xTilde_vTilde_expr_app3} we can write
\begin{eqnarray}
\lim\limits_{\tau \to \pm \infty} \widetilde{u}_{\widetilde{x}, \, \widetilde{v}}^{\prime} (\tau) = \frac{m^2 \widetilde{x} \widetilde{v}}{2 \hbar ^2 \alpha_0} \, e^{- \frac{m^2 \widetilde{v}^2}{4 \hbar ^2 \alpha_0}} \, \lim\limits_{\tau \to \pm \infty} \frac{1}{\tau^2} = 0^+ \, ,
\label{lim_fPrime_xTilde_vTilde_infty_expr_app3}
\end{eqnarray}
since we have $\alpha_0, \, \widetilde{x}, \, \widetilde{v} > 0$. Therefore, because $\tau_{\widetilde{x}, \, \widetilde{v}}^{(1)}$ and $\tau_{\widetilde{x}, \, \widetilde{v}}^{(2)}$ are by construction the two only roots of $\widetilde{u}_{\widetilde{x}, \, \widetilde{v}}^{\prime}$ we deduce from~\eref{lim_fPrime_xTilde_vTilde_infty_expr_app3} that
\begin{eqnarray}
\widetilde{u}_{\widetilde{x}, \, \widetilde{v}}^{\prime} (\tau) > 0 \quad , \quad \forall \tau \notin \left[ \tau_{\widetilde{x}, \, \widetilde{v}}^{(1)} \, , \tau_{\widetilde{x}, \, \widetilde{v}}^{(2)} \right] \, .
\label{fPrime_xTilde_vTilde_pos_sign_app3}
\end{eqnarray}

Combining the results~\eref{fPrime_xTilde_vTilde_neg_sign_app3} and~\eref{fPrime_xTilde_vTilde_pos_sign_app3}, we hence see that the function $\widetilde{u}_{\widetilde{x}, \, \widetilde{v}} (\tau)$ has the following behavior:
\begin{equation}
\left\{\begin{array}{ll}
\widetilde{u}_{\widetilde{x}, \, \widetilde{v}} (\tau) \quad \mathrm{increasing} \quad  \mathrm{,} \quad \forall \tau \in \left( - \infty \, , \, \tau_{\widetilde{x}, \, \widetilde{v}}^{(1)} \right] \\[2mm]
\widetilde{u}_{\widetilde{x}, \, \widetilde{v}} (\tau) \quad \mathrm{decreasing} \quad  \mathrm{,} \quad \forall \tau \in \left[ \tau_{\widetilde{x}, \, \widetilde{v}}^{(1)} \, , \, \tau_{\widetilde{x}, \, \widetilde{v}}^{(2)} \right] \\[2mm]
\widetilde{u}_{\widetilde{x}, \, \widetilde{v}} (\tau) \quad \mathrm{increasing} \quad  \mathrm{,} \quad \forall \tau \in \left[ \tau_{\widetilde{x}, \, \widetilde{v}}^{(2)} \, , \, \infty \right)
\end{array}\right. .
\label{f_behaviour_app3}
\end{equation}


\subsection{Behavior of \texorpdfstring{$u_{x_0, \, v_0}$}{f and g}}\label{g_sec_app}

Now, we first find the stationary points of $u_{x_0, \, v_0}$, i.e. the points where the derivative $u_{x_0, \, v_0}^{\prime} \equiv d u_{x_0, \, v_0} / d\tau$ vanishes. Differentiating~\eref{u_xTilde_vTilde_u_x0_v0_def} with respect to $\tau$ yields
\begin{eqnarray}
u_{x_0, \, v_0}^{\prime} (\tau) = \left( x_0 + v_0 \tau \right) \left( \frac{4 \hbar ^2 \alpha_0^2}{m^2} \, x_0 \tau - v_0 \right) \frac{2 \alpha_0 u_{x_0, \, v_0} (\tau)}{\left( 1 + \frac{4 \hbar ^2 \alpha_0^2}{m^2} \tau^2 \right) ^2} \, .
\label{gPrime_x0_v0_expr_app3}
\end{eqnarray}
Since $\alpha_0 \neq 0$ by assumption, and $u_{x_0, \, v_0} (\tau) \neq 0$, $\forall \tau \in \mathbb{R}$, we can readily see on~\eref{gPrime_x0_v0_expr_app3} that the derivative $u_{x_0, \, v_0}^{\prime} (\tau)$ admits the two real roots $\tau_{x_0, \, v_0}^{(1)}$ and $\tau_{x_0, \, v_0}^{(2)}$ given by
\begin{eqnarray}
\tau_{x_0, \, v_0}^{(1)} = - \frac{m^2}{4 \hbar ^2 \alpha_0^2 t_{\mathrm{c}}} \qquad \mbox{and} \qquad \tau_{x_0, \, v_0}^{(2)} = t_{\mathrm{c}} \, ,
\label{stat_points_g_expr_app3}
\end{eqnarray}
where we used the definition~\eref{semiClassicalTime_def} of $t_{\mathrm{c}}$. Because $t_{\mathrm{c}} > 0$ by assumption, we deduce from~\eref{stat_points_g_expr_app3} that the two stationary points $\tau_{x_0, \, v_0}^{(1)}$ and $\tau_{x_0, \, v_0}^{(2)}$ satisfy
\begin{eqnarray}
\tau_{x_0, \, v_0}^{(1)} < 0 < \tau_{x_0, \, v_0}^{(2)} \, .
\label{stat_points_g_ordering_cond_app3}
\end{eqnarray}

We now study the sign of $u_{x_0, \, v_0}^{\prime} (\tau)$ for $\tau \in ( \tau_{x_0, \, v_0}^{(1)} \, , \tau_{x_0, \, v_0}^{(2)} )$. In view of~\eref{stat_points_g_ordering_cond_app3}, and because $\tau_{x_0, \, v_0}^{(1)}$ and $\tau_{x_0, \, v_0}^{(2)}$ are by construction the two only roots of $u_{x_0, \, v_0}^{\prime}$, this is equivalent to merely studying the sign of $u_{x_0, \, v_0}^{\prime} (0)$. From~\eref{u_xTilde_vTilde_u_x0_v0_def} and~\eref{gPrime_x0_v0_expr_app3} we get
\begin{eqnarray}
u_{x_0, \, v_0}^{\prime} (0) = - 2 \alpha_0 x_0 v_0 \, e^{- \alpha_0 x_0^2} \, ,
\label{gPrime_x0_v0_0_expr_app3}
\end{eqnarray}
and thus, since by assumption $\alpha_0, v_0 > 0$ and $x_0 < 0$,
\begin{eqnarray}
u_{x_0, \, v_0}^{\prime} (0) > 0 \, .
\label{gPrime_x0_v0_0_sign_app3}
\end{eqnarray}
Therefore, we deduce from~\eref{gPrime_x0_v0_0_sign_app3} that
\begin{eqnarray}
u_{x_0, \, v_0}^{\prime} (\tau) > 0 \quad , \quad \forall \tau \in \left( \tau_{x_0, \, v_0}^{(1)} \, , \tau_{x_0, \, v_0}^{(2)} \right) \, .
\label{gPrime_x0_v0_pos_sign_app3}
\end{eqnarray}

We now determine the sign of $u_{x_0, \, v_0}^{\prime} (\tau)$ for $\tau \notin [ \tau_{x_0, \, v_0}^{(1)} \, , \tau_{x_0, \, v_0}^{(2)} ]$. To do this, we study the behavior of $u_{x_0, \, v_0}^{\prime} (\tau)$ as $\tau \to \pm \infty$. From~\eref{u_xTilde_vTilde_u_x0_v0_def} and~\eref{gPrime_x0_v0_expr_app3} we can write
\begin{eqnarray}
\lim\limits_{\tau \to \pm \infty} u_{x_0, \, v_0}^{\prime} (\tau) = \frac{m^2 x_0 v_0}{2 \hbar ^2 \alpha_0} \, e^{- \frac{m^2 v_0^2}{4 \hbar ^2 \alpha_0}} \, \lim\limits_{\tau \to \pm \infty} \frac{1}{\tau^2} = 0^- \, ,
\label{lim_gPrime_x0_v0_infty_expr_app3}
\end{eqnarray}
since we have $\alpha_0, \, v_0 > 0$, $x_0 < 0$. Therefore, because $\tau_{x_0, \, v_0}^{(1)}$ and $\tau_{x_0, \, v_0}^{(2)}$ are by construction the two only roots of $u_{x_0, \, v_0}^{\prime}$ we deduce from~\eref{lim_gPrime_x0_v0_infty_expr_app3} that
\begin{eqnarray}
u_{x_0, \, v_0}^{\prime} (\tau) < 0 \quad , \quad \forall \tau \notin \left[ \tau_{x_0, \, v_0}^{(1)} \, , \tau_{x_0, \, v_0}^{(2)} \right] \, .
\label{gPrime_x0_v0_neg_sign_app3}
\end{eqnarray}

Combining the results~\eref{gPrime_x0_v0_pos_sign_app3} and~\eref{gPrime_x0_v0_neg_sign_app3}, we hence see that the function $u_{x_0, \, v_0} (\tau)$ has the following behavior:
\begin{equation}
\left\{\begin{array}{ll}
u_{x_0, \, v_0} (\tau) \quad \mathrm{decreasing} \quad \mathrm{,} \quad \forall \tau \in \left( - \infty \, , \, \tau_{x_0, \, v_0}^{(1)} \right] \\[2mm]
u_{x_0, \, v_0} (\tau) \quad \mathrm{increasing} \quad \mathrm{,} \quad \forall \tau \in \left[ \tau_{x_0, \, v_0}^{(1)} \, , \, \tau_{x_0, \, v_0}^{(2)} \right] \\[2mm]
u_{x_0, \, v_0} (\tau) \quad \mathrm{decreasing} \quad \mathrm{,} \quad \forall \tau \in \left[ \tau_{x_0, \, v_0}^{(2)} \, , \, \infty \right)
\end{array}\right. \, .
\label{g_behaviour_app3}
\end{equation}


\subsection{Upper bound for \texorpdfstring{$\left| I^{(-)} \right|$}{the negative integral}}\label{bound_I_minus_sec}

In view of~\eref{modulus_I_minus_plus_triangleIneq} we have throughout this section $\tau \leqslant 0$. We first use the results of~\ref{f_sec_app} and~\ref{g_sec_app} to obtain upper bounds of $\widetilde{u}_{\widetilde{x}, \, \widetilde{v}}$ and $u_{x_0, \, v_0}$ for $\tau \in \mathbb{R}_-$.

As is clear from~\eref{f_behaviour_app3}, the point $\tau_{\widetilde{x}, \, \widetilde{v}}^{(1)}$ $( \tau_{\widetilde{x}, \, \widetilde{v}}^{(2)} )$ corresponds to a local maximum (minimum) of $\widetilde{u}_{\widetilde{x}, \, \widetilde{v}} (\tau)$. Since we have both $t > 0$ and $\widetilde{t} > 0$, then we see on~\eref{stat_points_f_expr_app3} that $\tau_{\widetilde{x}, \, \widetilde{v}}^{(2)} > 0$. However, note that $\tau_{\widetilde{x}, \, \widetilde{v}}^{(1)}$ can \textit{a priori} be either positive or negative, and thus, because it is a local maximum of $\widetilde{u}_{\widetilde{x}, \, \widetilde{v}} (\tau)$, we can write
\begin{eqnarray}
\widetilde{u}_{\widetilde{x}, \, \widetilde{v}} (\tau) \leqslant \widetilde{u}_{\widetilde{x}, \, \widetilde{v}} \left[ \mathrm{min} \left( 0 \, , \tau_{\widetilde{x}, \, \widetilde{v}}^{(1)} \right) \right] \quad , \quad \forall \tau \in \mathbb{R}_- \, ,
\label{f_xTilde_vTilde_neg_tau_gen_bound}
\end{eqnarray}
where $\mathrm{min} (\xi_1 \, , \xi_2)$ denotes the minimum between $\xi_1$ and $\xi_2$. Therefore, the upper bound~\eref{f_xTilde_vTilde_neg_tau_gen_bound} of the function $\widetilde{u}_{\widetilde{x}, \, \widetilde{v}}$ depends on the sign of the stationary point $\tau_{\widetilde{x}, \, \widetilde{v}}^{(1)}$, that is, in view of~\eref{stat_points_f_expr_app3}, whether the final time $t$ is larger or smaller than $\widetilde{t} \equiv \widetilde{x}/\widetilde{v}$. For completeness, we explicitly write~\eref{f_xTilde_vTilde_neg_tau_gen_bound} in these two cases.

If we first assume that $\tau_{\widetilde{x}, \, \widetilde{v}}^{(1)}<0$, then~\eref{f_xTilde_vTilde_neg_tau_gen_bound} reads, in view of the definition~\eref{u_xTilde_vTilde_u_x0_v0_def} of $\widetilde{u}_{\widetilde{x}, \, \widetilde{v}}$, the expression~\eref{stat_points_f_expr_app3} of $\tau_{\widetilde{x}, \, \widetilde{v}}^{(1)}$ and the fact that $\widetilde{t} \equiv \widetilde{x}/\widetilde{v}$,
\begin{eqnarray}
\mbox{If } \quad t<\widetilde{t}: \qquad \widetilde{u}_{\widetilde{x}, \, \widetilde{v}} (\tau) \leqslant 1 \quad , \quad \forall \tau \in \mathbb{R}_- \, ,
\label{f_xTilde_vTilde_neg_tau_gen_bound_expl_1}
\end{eqnarray}
which is obviously the most naive upper bound we could write, the function $\widetilde{u}_{\widetilde{x}, \, \widetilde{v}}$ being by construction the exponential of a negative quantity. On the other hand, if we now assume that $\tau_{\widetilde{x}, \, \widetilde{v}}^{(1)}>0$, then~\eref{f_xTilde_vTilde_neg_tau_gen_bound} reads, in view of the definition~\eref{u_xTilde_vTilde_u_x0_v0_def} of $\widetilde{u}_{\widetilde{x}, \, \widetilde{v}}$,
\begin{eqnarray}
\fl \mbox{If } \quad t>\widetilde{t}: \qquad \widetilde{u}_{\widetilde{x}, \, \widetilde{v}} (\tau) \leqslant \mathrm{exp} \left[ - \frac{1}{2} \frac{1}{1 + \left( \frac{\hbar t}{m \sigma^2} \right) ^2} \left( \frac{\widetilde{x} - \widetilde{v} t}{\sigma} \right) ^2 \right] \quad , \quad \forall \tau \in \mathbb{R}_- \, .
\label{f_xTilde_vTilde_neg_tau_gen_bound_expl_2}
\end{eqnarray}

Now, as is clear from~\eref{g_behaviour_app3}, the point $\tau_{x_0, \, v_0}^{(1)}$ $( \tau_{x_0, \, v_0}^{(2)} )$ corresponds to a local minimum (maximum) of $u_{x_0, \, v_0} (\tau)$. Since we have $t_{\mathrm{c}}>0$, then we see on~\eref{stat_points_g_expr_app3} that $\tau_{x_0, \, v_0}^{(1)} < 0$ and $\tau_{x_0, \, v_0}^{(2)} > 0$. Therefore, since $\tau_{x_0, \, v_0}^{(1)}$ is a local minimum of $u_{x_0, \, v_0} (\tau)$ we can write
\begin{eqnarray}
u_{x_0, \, v_0} (\tau) \leqslant \mathrm{max} \left[ \lim\limits_{\tau \to - \infty} u_{x_0, \, v_0} (\tau) \, , u_{x_0, \, v_0} (0) \right] \quad , \quad \forall \tau \in \mathbb{R}_- \, ,
\label{g_x0_v0_neg_tau_gen_bound}
\end{eqnarray}
where $\mathrm{max} (\xi_1 \, , \xi_2)$ denotes the maximum between $\xi_1$ and $\xi_2$. In view of the definition~\eref{u_xTilde_vTilde_u_x0_v0_def} of $u_{x_0, \, v_0} (\tau)$, we have, also using the definition~\eref{lambdaBar_t_tilde_def} of $\lambdabar$,
\begin{eqnarray}
\lim\limits_{\tau \to - \infty} u_{x_0, \, v_0} (\tau) = e^{ - \frac{1}{2} \left( \frac{\sigma}{\lambdabar} \right) ^2} \qquad \mbox{and} \qquad u_{x_0, \, v_0} (0) = e^{ - \frac{1}{2} \left( \frac{x_0}{\sigma} \right) ^2} \, ,
\label{g_x0_v0_value_in_0_and_minus_infty}
\end{eqnarray}
and thus, combining~\eref{g_x0_v0_neg_tau_gen_bound} with~\eref{g_x0_v0_value_in_0_and_minus_infty} we get
\begin{eqnarray}
u_{x_0, \, v_0} (\tau) \leqslant \mathrm{max} \left[ e^{ - \frac{1}{2} \left( \frac{\sigma}{\lambdabar} \right) ^2} \, , e^{ - \frac{1}{2} \left( \frac{x_0}{\sigma} \right) ^2} \right] \quad , \quad \forall \tau \in \mathbb{R}_- \, .
\label{g_x0_v0_neg_tau_explicit_bound}
\end{eqnarray}

Therefore, substituting the results~\eref{f_xTilde_vTilde_neg_tau_gen_bound} and~\eref{g_x0_v0_neg_tau_explicit_bound} into~\eref{up_bound_modulus_h_frozen} yields
\begin{eqnarray}
\fl \left| \widetilde{f}_{\mathrm{froz}} (\tau) \right| \leqslant \Gamma_{\mathrm{up}} \, \widetilde{u}_{\widetilde{x}, \, \widetilde{v}} \left[ \mathrm{min} \left( 0 \, , \tau_{\widetilde{x}, \, \widetilde{v}}^{(1)} \right) \right] \mathrm{max} \left[ e^{ - \frac{1}{2} \left( \frac{\sigma}{\lambdabar} \right) ^2} \, , e^{ - \frac{1}{2} \left( \frac{x_0}{\sigma} \right) ^2} \right] \left| \chi (\tau) \right| \enspace , \enspace \forall \tau \in \mathbb{R}_- \, .
\label{up_bound_modulus_h_frozen_neg_tau}
\end{eqnarray}
Finally, combining~\eref{up_bound_modulus_h_frozen_neg_tau} with~\eref{modulus_I_minus_plus_triangleIneq} yields the following upper bound for $|I^{(-)}|$:
\begin{eqnarray}
\fl \left| I^{(-)} \right| \leqslant \Gamma_{\mathrm{up}} \, \widetilde{u}_{\widetilde{x}, \, \widetilde{v}} \left[ \mathrm{min} \left( 0 \, , \tau_{\widetilde{x}, \, \widetilde{v}}^{(1)} \right) \right] \mathrm{max} \left[ e^{ - \frac{1}{2} \left( \frac{\sigma}{\lambdabar} \right) ^2} \, , e^{ - \frac{1}{2} \left( \frac{x_0}{\sigma} \right) ^2} \right] \int_{- \infty}^{0} d \tau \, \left| \chi (\tau) \right| \, .
\label{modulus_I_minus_UPPER_BOUND_app5}
\end{eqnarray}


\subsection{Upper bound for \texorpdfstring{$\left| I^{(+)} \right|$}{the positive integral}}\label{bound_I_plus_sec}

In view of~\eref{modulus_I_minus_plus_triangleIneq} we have here $\tau \geqslant t$. We first use the results of~\ref{f_sec_app} and~\ref{g_sec_app} to obtain upper bounds of $\widetilde{u}_{\widetilde{x}, \, \widetilde{v}}$ and $u_{x_0, \, v_0}$ for $\tau \in [ t , \infty )$.

In view of~\eref{stat_points_f_ordering_cond_app3}, and since we know from~\eref{f_behaviour_app3} that $\tau_{\widetilde{x}, \, \widetilde{v}}^{(2)}$ is a local minimum of $\widetilde{u}_{\widetilde{x}, \, \widetilde{v}} (\tau)$ we can write
\begin{eqnarray}
\widetilde{u}_{\widetilde{x}, \, \widetilde{v}} (\tau) \leqslant \mathrm{max} \left[ \widetilde{u}_{\widetilde{x}, \, \widetilde{v}} (t) \, , \lim\limits_{\tau \to \infty} \widetilde{u}_{\widetilde{x}, \, \widetilde{v}} (\tau) \right] \quad , \quad \forall \tau \in \left[ t , \infty \right) \, .
\label{f_xTilde_vTilde_pos_tau_gen_bound}
\end{eqnarray}
In view of the definition~\eref{u_xTilde_vTilde_u_x0_v0_def} of $\widetilde{u}_{\widetilde{x}, \, \widetilde{v}} (\tau)$ we have, also using the definition~\eref{lambdaBar_t_tilde_def} of $\widetilde{\lambdabar}$,
\begin{eqnarray}
\widetilde{u}_{\widetilde{x}, \, \widetilde{v}} (t) = e^{ - \frac{1}{2} \left( \frac{\widetilde{x}}{\sigma} \right) ^2} \qquad \mbox{and} \qquad \lim\limits_{\tau \to \infty} \widetilde{u}_{\widetilde{x}, \, \widetilde{v}} (\tau) = e^{ - \frac{1}{2} \left( \frac{\sigma}{\widetilde{\lambdabar}} \right) ^2} \, ,
\label{f_xTilde_vTilde_value_in_t_and_plus_infty}
\end{eqnarray}
and thus, combining~\eref{f_xTilde_vTilde_pos_tau_gen_bound} with~\eref{f_xTilde_vTilde_value_in_t_and_plus_infty} we get
\begin{eqnarray}
\widetilde{u}_{\widetilde{x}, \, \widetilde{v}} (\tau) \leqslant \mathrm{max} \left[ e^{ - \frac{1}{2} \left( \frac{\widetilde{x}}{\sigma} \right) ^2} \, , e^{ - \frac{1}{2} \left( \frac{\sigma}{\widetilde{\lambdabar}} \right) ^2} \right] \quad , \quad \forall \tau \in \left[ t , \infty \right) \, .
\label{f_xTilde_vTilde_pos_tau_explicit_bound}
\end{eqnarray}

Now, combining~\eref{stat_points_g_ordering_cond_app3} with the facts that $\tau_{x_0, \, v_0}^{(2)} = t_{\mathrm{c}}$ [see~\eref{stat_points_g_expr_app3}] and $0<t_{\mathrm{c}}<t$ readily shows in particular that
\begin{eqnarray}
\tau_{x_0, \, v_0}^{(2)} < t \, ,
\label{tau0_2_sign}
\end{eqnarray}
and thus, since it is clear from~\eref{g_behaviour_app3} that $\tau_{x_0, \, v_0}^{(2)}$ is a local maximum of $u_{x_0, \, v_0} (\tau)$,
\begin{eqnarray}
u_{x_0, \, v_0} (\tau) \leqslant u_{x_0, \, v_0} (t) \quad , \quad \forall \tau \in \left[ t , \infty \right) \, .
\label{g_x0_v0_pos_tau_gen_bound}
\end{eqnarray}
Therefore, in view of the definition~\eref{u_xTilde_vTilde_u_x0_v0_def} of $u_{x_0, \, v_0} (\tau)$, we have
\begin{eqnarray}
u_{x_0, \, v_0} (\tau) \leqslant \mathrm{exp} \left[ - \frac{1}{2} \frac{1}{1 + \left( \frac{\hbar t}{m \sigma^2} \right) ^2} \left( \frac{x_0 + v_0 t}{\sigma} \right) ^2 \right] \quad , \quad \forall \tau \in \left[ t , \infty \right) \, .
\label{g_x0_v0_pos_tau_explicit_bound}
\end{eqnarray}

Therefore, substituting the results~\eref{f_xTilde_vTilde_pos_tau_explicit_bound} and~\eref{g_x0_v0_pos_tau_explicit_bound} into~\eref{up_bound_modulus_h_frozen} yields
\begin{eqnarray}
\fl \left| \widetilde{f}_{\mathrm{froz}} (\tau) \right| \leqslant \Gamma_{\mathrm{up}} \, \mathrm{exp} \left[ - \frac{1}{2\sigma^2} \frac{\left( x_0 + v_0 t \right)^2}{1 + \left( \frac{\hbar t}{m \sigma^2} \right) ^2} \right] \mathrm{max} \left[ e^{ - \frac{1}{2} \left( \frac{\widetilde{x}}{\sigma} \right) ^2} \, , e^{ - \frac{1}{2} \left( \frac{\sigma}{\widetilde{\lambdabar}} \right) ^2} \right] \left| \chi (\tau) \right| \, ,
\label{up_bound_modulus_h_frozen_pos_tau}
\end{eqnarray}
for any $\tau \in [ t , \infty)$. Finally, combining~\eref{up_bound_modulus_h_frozen_pos_tau} with~\eref{modulus_I_minus_plus_triangleIneq} yields the following upper bound for $|I^{(+)}|$:
\begin{eqnarray}
\fl \left| I^{(+)} \right| \leqslant \Gamma_{\mathrm{up}} \, \mathrm{exp} \left[ - \frac{1}{2\sigma^2} \frac{\left( x_0 + v_0 t \right)^2}{1 + \left( \frac{\hbar t}{m \sigma^2} \right) ^2} \right] \mathrm{max} \left[ e^{ - \frac{1}{2} \left( \frac{\widetilde{x}}{\sigma} \right) ^2} \, , e^{ - \frac{1}{2} \left( \frac{\sigma}{\widetilde{\lambdabar}} \right) ^2} \right] \int_{t}^{\infty} d \tau \, \left| \chi (\tau) \right| \, .
\label{modulus_I_plus_UPPER_BOUND_app5}
\end{eqnarray}

Finally, substituting~\eref{modulus_I_minus_UPPER_BOUND_app5} and~\eref{modulus_I_plus_UPPER_BOUND_app5} into~\eref{mod_I_minus_I_plus_triangIneq} indeed yields~\eref{sum_I_minus_I_plus_UPPER_BOUND}.


\section{M\"obius transformation for \texorpdfstring{$\mathrm{Res} \, [\widetilde{f}_{\mathrm{froz}} (z) \, , \tau_0]$}{the residue}}\label{Mobius_app}

In this appendix we study how the residue $\mathrm{Res} \, [\widetilde{f}_{\mathrm{froz}} (z) \, , \tau_0]$, given by~\eref{residue_h_fr_tau_0_explicit_def}, can be adequately rewritten by means of a particular M\"obius transformation.

We begin with the change of variable $z' \rightarrow w$ in~\eref{residue_h_fr_tau_0_explicit_def}, where $w$ and $z'$ are related through the M\"obius transformation~\cite{Ablowitz}
\begin{eqnarray}
w = \frac{- z' + \tau_1}{z' - \tau_0} \, ,
\label{Mobius_transform}
\end{eqnarray}
and hence the inverse transformation reads
\begin{eqnarray}
z' = \frac{\tau_0 w + \tau_1}{w + 1} \, .
\label{inverse_Mobius_transform}
\end{eqnarray}
First, the orientation of the integration contour is reversed under the change of variable~\eref{Mobius_transform} (indeed, any M\"obius transformation can be decomposed into two translations and one inverse~\cite{Ablowitz}). We hence formally write
\begin{eqnarray}
\ointctrclockwise \rightarrow \ointclockwise \, .
\label{integral_orientation_Mobius}
\end{eqnarray}
Then, we determine how the integration contour $\mathcal{C} ( \tau_0 \, , r )$ in~\eref{residue_h_fr_tau_0_explicit_def} is mapped under~\eref{Mobius_transform}. By definition, $\mathcal{C} ( \tau_0 \, , r )$ is described by the set of all $z' \in \mathbb{C}$ satisfying the equation
\begin{eqnarray}
z' z'^* - \tau_0^* z' - \tau_0 z'^* + \tau_0 \tau_0^* - r^2 = 0 \, .
\label{equation_circle_before_Mobius}
\end{eqnarray}
Therefore, substituting~\eref{inverse_Mobius_transform} into~\eref{equation_circle_before_Mobius} yields the following equation for $w$:
\begin{eqnarray}
w w^* - (-1)^* w - (-1) w^* + (-1) (-1)^* - \left( \frac{\left| \tau_0-\tau_1 \right|}{r} \right) ^2 = 0 \, ,
\label{equation_circle_after_Mobius}
\end{eqnarray}
which characterizes the circle $\mathcal{C} \left( -1 \, , \left| \tau_0-\tau_1 \right| / r \right)$, with in view of~\eref{r_less_r_tau_0}
\begin{eqnarray}
\frac{\left| \tau_0-\tau_1 \right|}{r} > \frac{\left| \tau_0-\tau_1 \right|}{r_{\tau_0}} \, .
\label{radius_nu_cond_after_Mobius}
\end{eqnarray}
Therefore,~\eref{Mobius_transform} maps the integration contour $\mathcal{C} ( \tau_0 \, , r )$ onto $\mathcal{C} \left( -1 \, , \left| \tau_0-\tau_1 \right| / r \right)$:
\begin{eqnarray}
\mathcal{C} ( \tau_0 \, , r ) \rightarrow \mathcal{C} \left( -1 \, , \frac{\left| \tau_0-\tau_1 \right|}{r} \right) \, .
\label{contours_mapping_Mobius}
\end{eqnarray}
Now, we have in view of~\eref{Mobius_transform} and~\eref{inverse_Mobius_transform}
\begin{eqnarray*}
\frac{dw}{dz'} = \frac{\tau_0 - \tau_1}{\left( z' - \tau_0 \right) ^2} = \frac{\left( w + 1 \right) ^2}{\tau_0 - \tau_1} \, ,
\end{eqnarray*}
so that the Jacobian of the transformation~\eref{Mobius_transform} reads
\begin{eqnarray}
dz' = dw \frac{\tau_0 - \tau_1}{\left( w + 1 \right) ^2} \, .
\label{Jacobian_Mobius}
\end{eqnarray}
Finally, from~\eref{inverse_Mobius_transform} we get
\begin{eqnarray}
\frac{T_0}{z' - \tau_0} + \frac{T_1}{z' - \tau_1} = \frac{1}{\tau_1 - \tau_0} \left( T_0 w - \frac{T_1}{w} \right) + \frac{T_0 - T_1}{\tau_1 - \tau_0} \, .
\label{exponent_argument_Mobius}
\end{eqnarray}
Therefore, combining~\eref{inverse_Mobius_transform},~\eref{integral_orientation_Mobius} and~\eref{contours_mapping_Mobius}-\eref{exponent_argument_Mobius}, we see that under the M\"obius transformation~\eref{Mobius_transform} the residue~\eref{residue_h_fr_tau_0_explicit_def} can be written in the form
\begin{eqnarray}
\fl \mathrm{Res} \left[ \widetilde{f}_{\mathrm{froz}} (z) \, , \tau_0 \right] = \frac{\Omega \left( \tau_1 - \tau_0 \right) \, e^{\frac{T_0 - T_1}{\tau_1 - \tau_0}}}{2 \pi i} \ointctrclockwise_{\mathcal{C} \left( -1 \, , \frac{\left| \tau_0-\tau_1 \right|}{r} \right)} \frac{d w \chi \left( \frac{\tau_0 w + \tau_1}{w + 1} \right)}{\left( w + 1 \right) ^2} \, e^{\frac{1}{\tau_1 - \tau_0} \left( T_0 w - \frac{T_1}{w} \right)} \, .
\label{residue_h_fr_tau_0_after_Mobius}
\end{eqnarray}

Furthermore, noting that
\begin{eqnarray}
T_0 w - \frac{T_1}{w} = \sqrt{T_0 T_1} \left( \sqrt{\frac{T_0}{T_1}} w - \sqrt{\frac{T_1}{T_0}} \frac{1}{w} \right) \, ,
\label{exponent_after_Mobius_alternative}
\end{eqnarray}
we then make the change of variable $w \rightarrow z$ in~\eref{residue_h_fr_tau_0_after_Mobius}, with
\begin{eqnarray}
z = \sqrt{\frac{T_0}{T_1}} w \, ,
\label{change_of_var_after_Mobius}
\end{eqnarray}
which then readily yields~\eref{residue_h_fr_tau_0_after_change_of_var}.


\section{Upper bounds for \texorpdfstring{$\int d\tau \, \chi_n(\tau)$}{}}\label{apod_up_bound_app}

In this appendix we derive upper bounds for the two integrals $\int_{- \infty}^{0} d \tau \left| \chi_n (\tau) \right|$ and $\int_{t}^{\infty} d \tau \left| \chi_n (\tau) \right|$ obtained upon substituting $\chi=\chi_n$ into~\eref{sum_I_minus_I_plus_UPPER_BOUND}.

First, we analytically compute these integrals for $n=2$: using~\eref{chi_n_tau_def} and noting that
\begin{eqnarray}
\chi_2 (\tau) = \frac{1}{\nu} \frac{d}{d\tau} \mathrm{Arctan} \left[ \nu (\tau-T_{\mathrm{op}})  \right] \, ,
\label{ch_2_deriv_atan}
\end{eqnarray}
we have
\begin{eqnarray}
\int_{- \infty}^{0} d \tau \left| \chi_2 (\tau) \right| = \frac{1}{\nu} \left[ \frac{\pi}{2} - \mathrm{Arctan} \left( \nu T_{\mathrm{op}} \right) \right]
\label{neg_integral_chi_2_expr}
\end{eqnarray}
and
\begin{eqnarray}
\int_{t}^{\infty} d \tau \left| \chi_2 (\tau) \right| = \frac{1}{\nu} \left\{ \frac{\pi}{2} - \mathrm{Arctan} \left[ \nu (t-T_{\mathrm{op}}) \right] \right\} \, .
\label{pos_integral_chi_2_expr}
\end{eqnarray}

We then use the equivalence
\begin{eqnarray}
\chi_{2k} (\tau) \leqslant \chi_2 (\tau) \iff \left| \nu (\tau-T_{\mathrm{op}}) \right|^{k-1} \geqslant 1 \, ,
\label{equiv_chi_2k_chi_2}
\end{eqnarray}
that is alternatively
\begin{eqnarray}
\chi_{2k} (\tau) \leqslant \chi_2 (\tau) \qquad , \qquad \forall \tau \notin \left( T_{\mathrm{op}}-\frac{1}{\nu} \, , \, T_{\mathrm{op}}+\frac{1}{\nu} \right) \, ,
\label{chi_2k_chi_2_ineq}
\end{eqnarray}
valid for any positive integer $k$. If in particular $\nu \geqslant 1/T_{\mathrm{op}}$, we get from~\eref{chi_2k_chi_2_ineq}
\begin{eqnarray}
\chi_{2k} (\tau) \leqslant \chi_2 (\tau) \qquad , \qquad \forall \tau \in \left( - \infty \, , \, 0 \right] \, .
\label{chi_2k_chi_2_neg_tau}
\end{eqnarray}
Furthermore, if $\nu \geqslant 1/(t-T_{\mathrm{op}})$ we get from~\eref{chi_2k_chi_2_ineq}
\begin{eqnarray}
\chi_{2k} (\tau) \leqslant \chi_2 (\tau) \qquad , \qquad \forall \tau \in \left[ t \, , \, \infty \right) \, .
\label{chi_2k_chi_2_pos_tau}
\end{eqnarray}
Finally, combining~\eref{chi_2k_chi_2_neg_tau} with~\eref{neg_integral_chi_2_expr} and~\eref{chi_2k_chi_2_pos_tau} with~\eref{pos_integral_chi_2_expr} indeed yields~\eref{neg_integral_chi_2k_up_bound} and~\eref{pos_integral_chi_2k_up_bound}.


\section{Analytic structure of \texorpdfstring{$\chi_n (z)$}{chi} and \texorpdfstring{$g_{\chi_n} (z)$}{f}}\label{struct_chi_n_f_n_app}

Here we study the analytic structure of the functions $\chi_n (z)$ (in~\ref{struct_chi_n_app}) and $g_{\chi_n} (z)$ (in~\ref{struct_f_n_app}), as given by~\eref{chi_n_tau_def} and~\eref{f_n_z_expr}, for an arbitrary even integer $n \geqslant 2$.


\subsection{The function \texorpdfstring{$\chi_n (z)$}{chi}}\label{struct_chi_n_app}

We denote by $Z_j^{(n)}$ the poles of $\chi_n$. As is clear from~\eref{chi_n_tau_def}, they must satisfy
\begin{eqnarray}
1+\nu^n \left[ Z_j^{(n)}-T_{\mathrm{op}} \right]^n = 0 \, ,
\label{poles_chi_n_def_app}
\end{eqnarray}
which we rewrite in the form
\begin{eqnarray*}
\left[ \nu \left( Z_j^{(n)}-T_{\mathrm{op}} \right) \right]^n = \left( e^{\frac{i \pi}{n}} \right)^n \qquad , \qquad 0 \leqslant j \leqslant n-1 \, .
\end{eqnarray*}
We then use the $n$ distinct roots of unity
\begin{eqnarray}
\left( e^{2j \frac{i\pi}{n}} \right)^n = 1 \qquad , \qquad 0 \leqslant j \leqslant n-1 \, ,
\label{n_distinct_roots_unity}
\end{eqnarray}
to get the $n$ poles $Z_j^{(n)}$ of $\chi_n$ given by
\begin{eqnarray}
Z_j^{(n)} = T_{\mathrm{op}} + \frac{1}{\nu} e^{(2j+1) \frac{i \pi}{n}} \qquad , \qquad 0 \leqslant j \leqslant n-1 \, .
\label{poles_chi_n_expr_app}
\end{eqnarray}
These poles can be readily seen to satisfy $Z_j^{(n)} \neq Z_{j'}^{(n)}$ for any $j \neq j'$. Therefore, the function $\chi_n(z)$ admits precisely $n$ simple poles.

Now, the poles $Z_j^{(n)}$ that must be taken into account in~\eref{I_n_expr_with_poles_specified} are, by definition of the contour $\gamma$, those that have a positive imaginary part. It is clear from~\eref{poles_chi_n_expr_app} that
\begin{eqnarray*}
\mathrm{Im} \left( Z_j^{(n)} \right) = \frac{1}{\nu} \sin \left[ \left( 2j+1  \right) \frac{\pi}{n} \right] \qquad , \qquad 0 \leqslant j \leqslant n-1 \, ,
\end{eqnarray*}
where the parameter $\nu$ is by assumption strictly positive. We hence have the equivalence
\begin{eqnarray}
\mathrm{Im} \left( Z_j^{(n)} \right) > 0 \iff \left( 2j+1  \right) \frac{\pi}{n} < \pi \iff j < \frac{n}{2} - \frac{1}{2} \, ,
\label{sign_imag_part_cond}
\end{eqnarray}
where $j$ must be an integer that satisfies $0 \leqslant j \leqslant n-1$. Now, $n$ must be even, hence $n/2$ is an integer, so that the rightmost inequality in~\eref{sign_imag_part_cond} is equivalent to $j \leqslant n/2 - 1$. Therefore, we readily get from~\eref{sign_imag_part_cond} that
\begin{eqnarray}
\fl \mathrm{Im} \left( Z_j^{(n)} \right) > 0 \iff 0 \leqslant j \leqslant \frac{n}{2}-1 \qquad , \qquad \mbox{for any even } \quad n \geqslant 2 \, .
\label{sign_imag_part_even_n}
\end{eqnarray}
Furthermore, we readily check that
\begin{eqnarray}
\fl \mathrm{Im} \left( Z_j^{(n)} \right) < 0 \iff \frac{n}{2} \leqslant j \leqslant n-1 \qquad , \qquad \mbox{for any even } \quad n \geqslant 2 \, .
\label{neg_imag_part_cond}
\end{eqnarray}

The two results~\eref{sign_imag_part_even_n} and~\eref{neg_imag_part_cond} hence ensure that only the poles $Z_j^{(n)}$ for which $0 \leqslant j \leqslant n/2 - 1$ are enclosed by the integration contour in~\eref{I_n_expr_with_poles_specified}. In particular, no pole lies precisely \textit{on} the integration contour.

We now investigate the analytic structure of the function $g_{\chi_n} (z)$.


\subsection{The function \texorpdfstring{$g_{\chi_n} (z)$}{f}}\label{struct_f_n_app}

We now denote by $z_j^{(n)}$ the poles of $g_{\chi_n}$. As is clear from~\eref{f_n_z_expr}, they must satisfy
\begin{eqnarray}
\left[ \frac{z_j^{(n)}}{Z} + 1 \right]^{n} + \nu^n \left[ \frac{\left( \tau_0 - T_{\mathrm{op}} \right) z_j^{(n)}}{Z} + \tau_1 - T_{\mathrm{op}} \right]^n = 0 \, ,
\label{poles_f_n_def_app}
\end{eqnarray}
We here again use the roots of unity~\eref{n_distinct_roots_unity}, and rewrite~\eref{poles_f_n_def_app} in the form
\begin{eqnarray}
\fl \left\{ \nu \left[ \frac{\left( \tau_0 - T_{\mathrm{op}} \right) z_j^{(n)}}{Z} + \tau_1 - T_{\mathrm{op}} \right] \right\}^n = \left[ e^{(2j+1) \frac{i \pi}{n}} \left( \frac{z_j^{(n)}}{Z} + 1 \right) \right]^{n} \enspace , \enspace 0 \leqslant j \leqslant n-1 \, .
\label{polynom_eq_g_chi_n}
\end{eqnarray}
This is readily solved and we get
\begin{eqnarray}
z_j^{(n)} = - Z \, \frac{\tau_1 - T_{\mathrm{op}} - \frac{1}{\nu} e^{(2j+1) \frac{i \pi}{n}}}{\tau_0 - T_{\mathrm{op}} - \frac{1}{\nu} e^{(2j+1) \frac{i \pi}{n}}} \qquad , \qquad 0 \leqslant j \leqslant n-1 \, ,
\label{poles_f_n_temp_expr_app}
\end{eqnarray}
which clearly satisfies $z_j^{(n)} \neq z_{j'}^{(n)}$ for any $j \neq j'$. Therefore, the function $g_{\chi_n}(z)$ admits precisely $n$ simple poles. Finally, combining~\eref{poles_f_n_temp_expr_app} with~\eref{poles_chi_n_expr_app} shows that the $n$ distinct simple poles $z_j^{(n)}$ of $g_{\chi_n} (z)$ are related to the $n$ simple poles $Z_j^{(n)}$ of $\chi_n (z)$ through
\begin{eqnarray}
z_j^{(n)} = - Z \, \frac{\tau_1 - Z_j^{(n)}}{\tau_0 - Z_j^{(n)}} \qquad , \qquad 0 \leqslant j \leqslant n-1 \, .
\label{poles_f_n_final_expr_app}
\end{eqnarray}

The simple relation~\eref{poles_f_n_final_expr_app} now allows us to unambiguously identify which of the poles $z_j^{(n)}$ are enclosed by the integration contour $\mathcal{C} \left( - Z \, , \, \left| Z \right| \left| \tau_0-\tau_1 \right| / r \right)$ in~\eref{residue_f_tilde_froz_n_gen_expr} as a direct consequence of the condition~\eref{r_less_r_tau_0} satisfied by $r$. We do this by showing that the distance between the center $z=-Z$ and any $z_j^{(n)}$ is strictly smaller than the radius $\left| Z \right| \left| \tau_0-\tau_1 \right| / r$. In view of~\eref{poles_f_n_final_expr_app}, the distance $|- Z - z_j^{(n)}|$ is given by
\begin{eqnarray}
\left| - Z - z_j^{(n)} \right| = \left| Z \right| \frac{\left| \tau_0 - \tau_1 \right|}{\left| \tau_0 - Z_j^{(n)} \right|} \qquad , \qquad 0 \leqslant j \leqslant n-1 \, .
\label{distance_center_pole_expr}
\end{eqnarray}
Now, in view of~\eref{r_less_r_tau_0}, $r$ satisfies in particular
\begin{eqnarray*}
r < \left| \tau_0 - Z_j^{(n)} \right| \qquad , \qquad 0 \leqslant j \leqslant n-1 \, ,
\end{eqnarray*}
from which we readily get
\begin{eqnarray}
\left| Z \right| \frac{\left| \tau_0 - \tau_1 \right|}{r} > \left| Z \right| \frac{\left| \tau_0 - \tau_1 \right|}{\left| \tau_0 - Z_j^{(n)} \right|} \qquad , \qquad 0 \leqslant j \leqslant n-1 \, .
\label{r_inequality}
\end{eqnarray}
Finally, combining~\eref{distance_center_pole_expr} with~\eref{r_inequality} yields the strict inequality
\begin{eqnarray}
\left| - Z - z_j^{(n)} \right| < \left| Z \right| \frac{\left| \tau_0 - \tau_1 \right|}{r} \qquad , \qquad 0 \leqslant j \leqslant n-1 \, .
\label{distance_smaller_radius_app}
\end{eqnarray}
This shows that $\mathcal{C} \left( - Z \, , \, \left| Z \right| \left| \tau_0-\tau_1 \right| / r \right)$ indeed encloses \textit{all} the poles $z_j^{(n)}$ of $g_{\chi_n}$.


\section{The residue \texorpdfstring{$\mathrm{Res} \, [g_{\chi_n} (z) g (z) \, , 0 ]$}{at 0}}\label{residue_zero_app}

Here we explicitly compute the residue $\mathrm{Res} \, [ g_{\chi_n} (z) g (z) \, , 0 ]$, which corresponds to the coefficient of the $1/z$ term in the Laurent series of $g_{\chi_n} (z) g (z)$ about $z=0$.

We combine~\eref{f_n_z_Taylor_series} with~\eref{g_Mobius_def} to get
\begin{eqnarray}
g_{\chi_n} (z) g(z) = - \frac{Z^2}{1+\nu^n (\tau_0-T_{\mathrm{op}})^n} \sum_{j=0}^{n-1} \frac{A_j^{(n)}}{z_j^{(n)}} \mathcal{L} \left[ z_j^{(n)} \right] \, ,
\label{f_n_g_z_Laurent_def_app}
\end{eqnarray}
where we introduced the function $\mathcal{L}$ defined by
\begin{eqnarray}
\mathcal{L} (\zeta) \equiv \sum_{k=0}^{\infty} \left( \frac{z}{\zeta} \right)^k \sum_{k'=-\infty}^{\infty} J_{k'} \left( \frac{2\sqrt{T_0 T_1}}{\tau_1 - \tau_0} \right) z^{k'} \, ,
\label{L_zeta_def}
\end{eqnarray}
which we decompose in the form
\begin{eqnarray}
\fl \mathcal{L} (\zeta) = \sum_{k=0}^{\infty} \left( \frac{z}{\zeta} \right)^k \sum_{k'=0}^{\infty} J_{k'} \left( \frac{2\sqrt{T_0 T_1}}{\tau_1 - \tau_0} \right) z^{k'} + \sum_{k=0}^{\infty} \left( \frac{z}{\zeta} \right)^k \sum_{k'=1}^{\infty} J_{-k'} \left( \frac{2\sqrt{T_0 T_1}}{\tau_1 - \tau_0} \right) \frac{1}{z^{k'}} \, .
\label{L_zeta_sum}
\end{eqnarray}

We then rewrite the second term in the right-hand side of~\eref{L_zeta_sum} by means of the Cauchy product of two infinite series, namely
\begin{eqnarray}
\sum_{k=0}^{\infty} a_k \sum_{k'=0}^{\infty} b_{k'} = \sum_{k=0}^{\infty} \sum_{l=0}^{k} a_l b_{k-l} \, .
\label{Cauchy_product}
\end{eqnarray}
Noting that we have, with a change of index $k' \to k'-1$,
\begin{eqnarray}
\sum_{k'=1}^{\infty} J_{-k'} \left( \frac{2\sqrt{T_0 T_1}}{\tau_1 - \tau_0} \right) \frac{1}{z^{k'}} = \sum_{k'=0}^{\infty} J_{-k'-1} \left( \frac{2\sqrt{T_0 T_1}}{\tau_1 - \tau_0} \right) \frac{1}{z^{k'+1}} \, ,
\label{sum_for_Cauchy_prod}
\end{eqnarray}
we hence get, in view of~\eref{Cauchy_product}-\eref{sum_for_Cauchy_prod},
\begin{eqnarray}
\fl \sum_{k=0}^{\infty} \left( \frac{z}{\zeta} \right)^k \sum_{k'=1}^{\infty} J_{-k'} \left( \frac{2\sqrt{T_0 T_1}}{\tau_1 - \tau_0} \right) \frac{1}{z^{k'}} = \sum_{k=0}^{\infty} \sum_{l=0}^{k} \frac{1}{\zeta^l} J_{l-k-1} \left( \frac{2\sqrt{T_0 T_1}}{\tau_1 - \tau_0} \right) \frac{1}{z^{k-2l+1}} \, ,
\label{identity_Cauchy_prod}
\end{eqnarray}
that is, splitting the sum over $k$ in the right-hand side of~\eref{identity_Cauchy_prod} into one sum over even integers only and one over odd integers only,
\begin{eqnarray}
\fl \sum_{k=0}^{\infty} \left( \frac{z}{\zeta} \right)^k \sum_{k'=1}^{\infty} J_{-k'} \left( \frac{2\sqrt{T_0 T_1}}{\tau_1 - \tau_0} \right) \frac{1}{z^{k'}} = \sum_{k=1}^{\infty} \sum_{l=0}^{2k-1} \frac{1}{\zeta^l} J_{l-2k} \left( \frac{2\sqrt{T_0 T_1}}{\tau_1 - \tau_0} \right) \frac{1}{z^{2(k-l)}} \nonumber\\
+ \sum_{k=0}^{\infty} \sum_{l=0}^{2k} \frac{1}{\zeta^l} J_{l-2k-1} \left( \frac{2\sqrt{T_0 T_1}}{\tau_1 - \tau_0} \right) \frac{1}{z^{2(k-l)+1}} \, .
\label{sum_after_Cauchy_prod}
\end{eqnarray}
Substituting now~\eref{sum_after_Cauchy_prod} into~\eref{L_zeta_sum} hence yields
\begin{eqnarray}
\fl \mathcal{L} (\zeta) = \sum_{k=0}^{\infty} \left( \frac{z}{\zeta} \right)^k \sum_{k'=0}^{\infty} J_{k'} \left( \frac{2\sqrt{T_0 T_1}}{\tau_1 - \tau_0} \right) z^{k'} + \sum_{k=1}^{\infty} \sum_{l=0}^{2k-1} \frac{1}{\zeta^l} J_{l-2k} \left( \frac{2\sqrt{T_0 T_1}}{\tau_1 - \tau_0} \right) \frac{1}{z^{2(k-l)}} \nonumber\\[0.3cm]
+ \sum_{k=0}^{\infty} \sum_{l=0}^{2k} \frac{1}{\zeta^l} J_{l-2k-1} \left( \frac{2\sqrt{T_0 T_1}}{\tau_1 - \tau_0} \right) \frac{1}{z^{2(k-l)+1}} \, .
\label{L_zeta_decomposition}
\end{eqnarray}
We now use~\eref{L_zeta_decomposition} to identify the coefficient, which we denote by $\mathcal{L}_{-1}(\zeta)$, of the term $1/z$ in $\mathcal{L} (\zeta)$. It is clear that the latter only arises from the third term in the right-hand side of~\eref{L_zeta_decomposition}, and is obtained from the latter by keeping only the terms of the double sum for which $l=k$. Using in addition the identity $J_{-k} (z) = (-1)^k J_k (z)$~\cite{GradRyz}, we hence obtain for $\mathcal{L}_{-1}(\zeta)$
\begin{eqnarray}
\mathcal{L}_{-1}(\zeta) = \zeta \sum_{k=1}^{\infty} \frac{(-1)^k}{\zeta^k} J_{k} \left( \frac{2\sqrt{T_0 T_1}}{\tau_1 - \tau_0} \right) \, .
\label{L_zeta_residue_expr}
\end{eqnarray}

Since $\mathcal{L}_{-1}$ is by construction the coefficient of the term $1/z$ in the power series $\mathcal{L}$, the desired residue $\mathrm{Res} [ g_{\chi_n} (z) g (z) \, , 0 ]$ is obtained from~\eref{f_n_g_z_Laurent_def_app} and we have
\begin{eqnarray}
\mathrm{Res} \left[ g_{\chi_n} (z) g (z) \, , 0 \right] = - \frac{Z^2}{1+\nu^n (\tau_0-T_{\mathrm{op}})^n} \sum_{j=0}^{n-1} \frac{A_j^{(n)}}{z_j^{(n)}} \mathcal{L}_{-1} \left[ z_j^{(n)} \right] \, .
\label{residue_f_n_g_z_from_Laurent}
\end{eqnarray}
Finally, substituting~\eref{L_zeta_residue_expr} into~\eref{residue_f_n_g_z_from_Laurent} yields the desired residue~\eref{residue_f_n_g_z_final_expr}.


\section{Leading order in \texorpdfstring{$1/\nu$}{nu}}\label{expand_nu_app}

In this appendix we derive the expansion of~\eref{h_sc_2_explicit_expr_general} to the leading order in $1/\nu$. Throughout this appendix the integers $j,j'$ are such that
\begin{eqnarray}
j,j' = 0,1 \, .
\label{j_j_pr_0_1_cond}
\end{eqnarray}
We also explicitly write the poles~\eref{poles_chi_n_expr} and~\eref{poles_f_n_final_expr} for $n=2$, that is
\begin{eqnarray}
Z_0^{(2)} = T + \frac{i}{\nu} \qquad \mbox{and} \qquad Z_1^{(2)} = T - \frac{i}{\nu}
\label{poles_chi_2_expr}
\end{eqnarray}
and
\begin{eqnarray}
z_0^{(2)} = - Z \, \frac{\tau_1 - Z_0^{(2)}}{\tau_0 - Z_0^{(2)}} \qquad \mbox{and} \qquad z_1^{(2)} = - Z \, \frac{\tau_1 - Z_1^{(2)}}{\tau_0 - Z_1^{(2)}} \, .
\label{poles_f_2_expr}
\end{eqnarray}
Furthermore, from~\eref{Taylor_coef_f_n_expr_guess} we have
\begin{eqnarray}
A_0^{(2)} = \frac{1}{z_0^{(2)} - z_{1}^{(2)}} \qquad \mbox{and} \qquad A_1^{(2)} = \frac{1}{z_1^{(2)} - z_{0}^{(2)}} = - A_0^{(2)} \, .
\label{Taylor_coef_f_2_expr}
\end{eqnarray}

First, note that in view of~\eref{poles_chi_2_expr} we have
\begin{eqnarray}
\frac{1}{Z_0^{(2)} - Z_1^{(2)}} = \frac{\nu}{2i}
\label{one_ove_Z_min_Z}
\end{eqnarray}
and
\begin{eqnarray}
\frac{1}{\tau_{j'} - Z_j^{(2)}} = \frac{1}{\tau_{j'} - T_{\mathrm{op}}} \left[ 1 + \frac{e^{\theta_j}}{\tau_{j'} - T_{\mathrm{op}}} \, \frac{1}{\nu} + \bigO \left( \frac{1}{\nu^2} \right) \right] \, ,
\label{one_ov_tau_min_Z_expand}
\end{eqnarray}
where we defined
\begin{eqnarray}
\theta_j \equiv (2j+1) i \frac{\pi}{2} \, ,
\label{theta_j_def}
\end{eqnarray}
as well as
\begin{eqnarray}
\frac{T_0}{ Z_0^{(2)} - \tau_{0}} + \frac{T_1}{ Z_0^{(2)} - \tau_{1}} = \frac{T_0}{ T_{\mathrm{op}} - \tau_{0}} + \frac{T_1}{ T_{\mathrm{op}} - \tau_{1}} +  \frac{\beta}{\nu} + \bigO \left( \frac{1}{\nu^2} \right) \, ,
\label{exponent_expand}
\end{eqnarray}
where we defined
\begin{eqnarray}
\beta \equiv - \frac{T_0 e^{\theta_0}}{\left( T_{\mathrm{op}} - \tau_{0} \right)^2} - \frac{T_1 e^{\theta_1}}{\left( T_{\mathrm{op}} - \tau_{1} \right)^2} \, .
\label{beta_def}
\end{eqnarray}
Therefore, we have
\begin{eqnarray}
e^{\frac{T_0}{Z_0^{(2)} - \tau_0} + \frac{T_1}{Z_0^{(2)} - \tau_1}} = e^{\frac{T_0}{T_{\mathrm{op}} - \tau_0} + \frac{T_1}{T_{\mathrm{op}} - \tau_1}} \left[ 1 + \frac{\beta}{\nu} + \bigO \left( \frac{1}{\nu^2} \right) \right] \, ,
\end{eqnarray}
and thus finally, also using~\eref{one_ove_Z_min_Z},
\begin{eqnarray}
\fl \frac{1}{\nu^2} \, \frac{1}{Z_0^{(2)} - Z_{1}^{(2)}} e^{\frac{T_0}{Z_0^{(2)} - \tau_0} + \frac{T_1}{Z_0^{(2)} - \tau_1}} = \frac{1}{2i\nu} e^{\frac{T_0}{T_{\mathrm{op}} - \tau_0} + \frac{T_1}{T_{\mathrm{op}} - \tau_1}} \left[ 1 + \frac{\beta}{\nu} + \bigO \left( \frac{1}{\nu^2} \right) \right] \nonumber\\
= \frac{1}{2i} e^{\frac{T_0}{T_{\mathrm{op}} - \tau_0} + \frac{T_1}{T_{\mathrm{op}} - \tau_1}} \frac{1}{\nu} + \bigO \left( \frac{1}{\nu^2} \right) \, .
\label{first_term_expand}
\end{eqnarray}

Then, we have in view of~\eref{poles_f_2_expr} and~\eref{one_ov_tau_min_Z_expand}
\begin{eqnarray}
z_j^{(2)} = - Z \frac{\tau_1 - T_{\mathrm{op}}}{\tau_0 - T_{\mathrm{op}}} + \frac{\mu_j}{\nu} + \bigO \left( \frac{1}{\nu^2} \right) \, ,
\label{poles_f_2_expand}
\end{eqnarray}
where
\begin{eqnarray}
\mu_j \equiv - Z \frac{\tau_1 - T_{\mathrm{op}}}{\tau_0 - T_{\mathrm{op}}} \frac{\tau_1 - \tau_0}{\left( \tau_0 - T_{\mathrm{op}} \right) \left( \tau_1 - T_{\mathrm{op}} \right)} e^{\theta_j} \, ,
\label{mu_j_def}
\end{eqnarray}
and similarly
\begin{eqnarray}
\frac{1}{z_j^{(2)}} = - \frac{1}{Z} \frac{\tau_0 - T_{\mathrm{op}}}{\tau_1 - T_{\mathrm{op}}} + \frac{\eta_j}{\nu} + \bigO \left( \frac{1}{\nu^2} \right) \, 
\label{inv_poles_f_2_expand}
\end{eqnarray}
where
\begin{eqnarray}
\eta_j \equiv - \frac{1}{Z} \frac{\tau_0 - T_{\mathrm{op}}}{\tau_1 - T_{\mathrm{op}}} \frac{\tau_0 - \tau_1}{\left( \tau_0 - T_{\mathrm{op}} \right) \left( \tau_1 - T_{\mathrm{op}} \right)} e^{\theta_j} \, .
\label{eta_j_def}
\end{eqnarray}
From~\eref{poles_f_2_expand} we hence get
\begin{eqnarray}
z_0^{(2)} - z_{1}^{(2)} = \frac{\mu_0 - \mu_{1}}{\nu} + \bigO \left( \frac{1}{\nu^2} \right) \, ,
\label{diff_poles_f_2_expand}
\end{eqnarray}
and thus, also using~\eref{Taylor_coef_f_2_expr},
\begin{eqnarray}
A_0^{(2)} = - A_1^{(2)} =\frac{1}{z_0^{(2)} - z_{1}^{(2)}} = \frac{\nu}{\mu_0 - \mu_{1}} \left[ 1 + \bigO \left( \frac{1}{\nu} \right) \right] \, .
\label{inv_diff_poles_f_2_expand}
\end{eqnarray}

Therefore, combining~\eref{inv_poles_f_2_expand} with~\eref{inv_diff_poles_f_2_expand} yields
\begin{eqnarray}
\frac{A_0^{(2)}}{z_0^{(2)}} + \frac{A_1^{(2)}}{z_1^{(2)}} = A_0^{(2)} \left[ \frac{1}{z_0^{(2)}} - \frac{1}{z_1^{(2)}} \right] = \frac{\eta_0 - \eta_1}{\mu_0 - \mu_1} + \bigO \left( \frac{1}{\nu} \right) \, ,
\label{A_ov_z_id}
\end{eqnarray}
then
\begin{eqnarray}
\fl \frac{A_0^{(2)}}{\left[ z_0^{(2)} \right]^2} + \frac{A_1^{(2)}}{\left[ z_1^{(2)} \right]^2} = A_0^{(2)} \left\{ \frac{1}{\left[ z_0^{(2)} \right]^2} - \frac{1}{\left[ z_1^{(2)} \right]^2} \right\} = \frac{2}{Z} \frac{\tau_0 - T_{\mathrm{op}}}{\tau_1 - T_{\mathrm{op}}} \frac{\eta_1 - \eta_0}{\mu_0 - \mu_1} + \bigO \left( \frac{1}{\nu} \right) \, ,
\label{A_ov_z_sq_id}
\end{eqnarray}
and more generally
\begin{eqnarray}
\frac{A_0^{(2)}}{\left[ z_0^{(2)} \right]^k} + \frac{A_1^{(2)}}{\left[ z_1^{(2)} \right]^k} = A_0^{(2)} \left\{ \frac{1}{\left[ z_0^{(2)} \right]^k} - \frac{1}{\left[ z_1^{(2)} \right]^k} \right\} = \bigO \left( 1 \right) \, ,
\label{A_ov_z_k_id}
\end{eqnarray}
for any integer $k \geqslant 1$. Furthermore, it is clear that
\begin{eqnarray}
\frac{Z^2}{1+\nu^2 (\tau_0-T_{\mathrm{op}})^2} = \frac{Z^2}{\nu^2 (\tau_0-T_{\mathrm{op}})^2} \left[ 1 + \bigO \left( \frac{1}{\nu^2} \right) \right] = \bigO \left( \frac{1}{\nu^2} \right) \, .
\label{fraction_expand}
\end{eqnarray}
Therefore, combining~\eref{A_ov_z_k_id} and~\eref{fraction_expand} readily shows that
\begin{eqnarray}
\frac{Z^2}{1+\nu^2 (\tau_0-T_{\mathrm{op}})^2} \sum_{j=0}^{1} A_j^{(2)} \sum_{k=1}^{\infty} \frac{(-1)^k}{\left[ z_j^{(2)} \right]^k} J_{k} \left( \frac{2\sqrt{T_0 T_1}}{\tau_1 - \tau_0} \right) = \bigO \left( \frac{1}{\nu^2} \right) \, ,
\label{second_term_expand}
\end{eqnarray}
where we used the fact that the Bessel functions $J_k$ are independent of $\nu$.

Furthermore, combining~\eref{poles_f_2_expand} and~\eref{inv_poles_f_2_expand} yields
\begin{eqnarray}
z_j^{(2)} - \frac{1}{z_j^{(2)}} = - Z \frac{\tau_1 - T_{\mathrm{op}}}{\tau_0 - T_{\mathrm{op}}} + \frac{1}{Z} \frac{\tau_0 - T_{\mathrm{op}}}{\tau_1 - T_{\mathrm{op}}} + \frac{\mu_j - \eta_j}{\nu} + \bigO \left( \frac{1}{\nu^2} \right) \, ,
\label{z_min_z_poles_f_2_expand}
\end{eqnarray}
so that we have
\begin{eqnarray}
e^{\frac{\sqrt{T_0 T_1}}{\tau_1 - \tau_0} \left[ z_j^{(2)} - \frac{1}{z_j^{(2)}} \right]} = e^{\Xi} \left[ 1 + \bigO \left( \frac{1}{\nu} \right) \right] \, ,
\label{third_exp_expand}
\end{eqnarray}
where the quantity
\begin{eqnarray}
\Xi \equiv \frac{\sqrt{T_0 T_1}}{\tau_1 - \tau_0} \left( - Z \frac{\tau_1 - T_{\mathrm{op}}}{\tau_0 - T_{\mathrm{op}}} + \frac{1}{Z} \frac{\tau_0 - T_{\mathrm{op}}}{\tau_1 - T_{\mathrm{op}}} \right)
\label{Xi_def}
\end{eqnarray}
is independent of $j$. Therefore, combining~\eref{inv_diff_poles_f_2_expand},~\eref{fraction_expand} and~\eref{third_exp_expand} shows that
\begin{eqnarray}
\fl \sum_{j=0}^{1} \frac{Z^2 A_j^{(2)} \, e^{\frac{\sqrt{T_0 T_1}}{\tau_1 - \tau_0} \left[ z_j^{(2)} - \frac{1}{z_j^{(2)}} \right]}}{1+\nu^2 (\tau_0-T_{\mathrm{op}})^2} = \frac{Z^2 e^{\Xi} \left[ A_0^{(2)} + A_1^{(2)} \right]}{1+\nu^2 (\tau_0-T_{\mathrm{op}})^2} \left[ 1 + \bigO \left( \frac{1}{\nu} \right) \right] = \bigO \left( \frac{1}{\nu^2} \right) \, .
\label{third_term_expand}
\end{eqnarray}

Therefore,~\eref{first_term_expand} shows that the first term in~\eref{h_sc_2_explicit_expr_general} is of order $1/\nu$, while~\eref{second_term_expand} and~\eref{third_term_expand} show that the other terms in~\eref{h_sc_2_explicit_expr_general} are of order $1/\nu^2$. Hence~\eref{h_sc_2_explicit_expr_general} reads
\begin{eqnarray}
f_{\mathrm{froz}}^{(2)} (\widetilde{x},\widetilde{v},t;T_{\mathrm{op}}) = \frac{\pi \Omega}{\nu} e^{\frac{T_0}{T_{\mathrm{op}} - \tau_0} + \frac{T_1}{T_{\mathrm{op}} - \tau_1}} - I^{(-)} - I^{(+)} + \bigO \left( \frac{1}{\nu^2} \right) \, .
\label{h_sc_2_leading_order_nu_app}
\end{eqnarray}

Now, since~\eref{neg_integral_chi_2k_up_bound}-\eref{pos_integral_chi_2k_up_bound} are exact for $n=2$,  we get
\begin{eqnarray}
\fl \int_{- \infty}^{0} d \tau \, \chi_{2} (\tau) = \frac{1}{\nu^2 T_{\mathrm{op}}} + \bigO \left( \frac{1}{\nu^4} \right) \enspace \mbox{and} \enspace\int_{t}^{\infty} d \tau \, \chi_{2} (\tau) = \frac{1}{\nu^2 \left( t - T_{\mathrm{op}} \right)} + \bigO \left( \frac{1}{\nu^4} \right) \, ,
\label{neg_integral_chi_2_up_bound_app}
\end{eqnarray}
so that
\begin{eqnarray}
\int_{- \infty}^{0} d \tau \, \chi_{2} (\tau) = \bigO \left( \frac{1}{\nu^2} \right) \qquad \mbox{and} \qquad \int_{t}^{\infty} d \tau \, \chi_{2} (\tau) = \bigO \left( \frac{1}{\nu^2} \right) \, .
\label{chi_2_integrals_expand}
\end{eqnarray}
Substituting~\eref{chi_2_integrals_expand} into~\eref{sum_I_minus_I_plus_UPPER_BOUND} then readily shows that
\begin{eqnarray}
\left| I^{(-)} + I^{(+)} \right| = \bigO \left( \frac{1}{\nu^2} \right) \, .
\label{sum_I_minus_I_plus_UPPER_BOUND_expand}
\end{eqnarray}
Therefore, combining~\eref{h_sc_2_leading_order_nu_app} with~\eref{sum_I_minus_I_plus_UPPER_BOUND_expand} readily yields~\eref{h_sc_2_leading_order_nu_final}-\eref{h_slit_gamma_expr}.


\section{Explicit expressions of \texorpdfstring{$g_2$}{g}, \texorpdfstring{$f_1$}{g} and \texorpdfstring{$f_2$}{f}}\label{expr_funct_app}

In this appendix we give the explicit expressions of the functions~\eref{g_2_def}-\eref{f_2_def} that result from~\eref{sym_times_dble_slit} and~\eref{t_2_t_c}. We get, also using~\eref{real_gamma_0_def}-\eref{imag_gamma_1_def},
\begin{eqnarray}
\fl g_2 \left[ \widetilde{x},\widetilde{v},2t_{\mathrm{c}};T_{\mathrm{op}}^{(0)},T_{\mathrm{op}}^{(1)} \right] = \frac{1}{2 \sigma^2} \, \frac{1}{1 + \hbar^2 ( T_{\mathrm{op}}^{(0)} )^2 / m^2 \sigma^4} \, \frac{1}{1 + \hbar^2 ( T_{\mathrm{op}}^{(1)} )^2 / m^2 \sigma^4} \nonumber\\
\fl \times \Bigg\{ 2 \left[ T_{\mathrm{op}}^{(0)} + T_{\mathrm{op}}^{(1)} \right] \left[ 1 + \frac{\hbar^2 T_{\mathrm{op}}^{(0)} T_{\mathrm{op}}^{(1)}}{m^2 \sigma^4} \right] \left( |x_0| v_0 + \widetilde{x} \widetilde{v} \right) \nonumber\\
\fl + \left( 2 + \frac{\hbar^2}{m^2 \sigma^4} \left[ (T_{\mathrm{op}}^{(0)})^2 +  (T_{\mathrm{op}}^{(1)})^2 \right] \right) \left[ \frac{m^2 \sigma^4}{\hbar^2} \left( \widetilde{v}^2 + v_0^2 \right) - \left( \widetilde{x}^2 + x_0^2 \right) \right] \Bigg\} \nonumber\\
- \frac{m^2 \sigma^2}{\hbar^2} \left( \widetilde{v}^2 + v_0^2 \right) \, ,
\label{g_2_expr}
\end{eqnarray}
then
\begin{eqnarray}
\fl f_1 \left[ \widetilde{x},\widetilde{v},2t_{\mathrm{c}};T_{\mathrm{op}}^{(0)},T_{\mathrm{op}}^{(1)} \right] = \frac{1}{2 \sigma^2} \, \frac{1}{1 + \hbar^2 ( T_{\mathrm{op}}^{(0)} )^2 / m^2 \sigma^4} \, \frac{1}{1 + \hbar^2 ( T_{\mathrm{op}}^{(1)} )^2 / m^2 \sigma^4} \left[ T_{\mathrm{op}}^{(1)} - T_{\mathrm{op}}^{(0)} \right] \nonumber\\
\fl \times \Bigg\{ 2 \left[ - 1 + \frac{\hbar^2 T_{\mathrm{op}}^{(0)} T_{\mathrm{op}}^{(1)}}{m^2 \sigma^4} \right] \left( |x_0| v_0 - \widetilde{x} \widetilde{v} \right) \nonumber\\
\fl + \left[ T_{\mathrm{op}}^{(0)} + T_{\mathrm{op}}^{(1)} \right] \left[ v_0^2 - \widetilde{v}^2 + \frac{\hbar^2}{m^2 \sigma^4} \left( \widetilde{x}^2 - x_0^2 \right) \right] \Bigg\}
\label{f_1_expr}
\end{eqnarray}
and
\begin{eqnarray}
\fl f_2 \left[ \widetilde{x},\widetilde{v},2t_{\mathrm{c}};T_{\mathrm{op}}^{(0)},T_{\mathrm{op}}^{(1)} \right] = \frac{m}{2 \hbar} \, \frac{1}{1 + \hbar^2 ( T_{\mathrm{op}}^{(0)} )^2 / m^2 \sigma^4} \, \frac{1}{1 + \hbar^2 ( T_{\mathrm{op}}^{(1)} )^2 / m^2 \sigma^4} \left[ T_{\mathrm{op}}^{(1)} - T_{\mathrm{op}}^{(0)} \right] \nonumber\\
\fl \times \Bigg\{ \frac{2 \hbar^2}{m^2 \sigma^4} \left[ T_{\mathrm{op}}^{(0)} + T_{\mathrm{op}}^{(1)} \right] \left( |x_0| v_0 - \widetilde{x} \widetilde{v} \right) \nonumber\\
\fl + \left[ - 1 + \frac{\hbar^2 T_{\mathrm{op}}^{(0)} T_{\mathrm{op}}^{(1)}}{m^2 \sigma^4} \right] \left[ \widetilde{v}^2 - v_0^2 + \frac{\hbar^2}{m^2 \sigma^4} \left( x_0^2 - \widetilde{x}^2 \right) \right] \Bigg\} \, .
\label{f_2_expr}
\end{eqnarray}


\section{Phase-space structure of the diffraction peaks}\label{peaks_app}

Here we derive the result~\eref{x_k_v_k_def} that describes the phase-space structure of the interference fringes exhibited by the Husimi distribution $F_{\mathrm{2slit}}$ in the case of the double-slit scenario.

Our strategy to infer, from the mathematical structure~\eref{Husimi_dist_double_slit_expr} of $F_{\mathrm{2slit}}$, an analytic expression of the position of the interference fringes in phase space is then the following. We first note on figure~\ref{an_vs_num_H_diff_fig} that the peaks of $F_{\mathrm{2slit}}$ seem to be arranged on a line $\widetilde{v}(\widetilde{x}) = \alpha \widetilde{x} + \beta$, for some $\alpha > 0$ and $\beta \in \mathbb{R}$. We determine these parameters $\alpha$ and $\beta$ in~\ref{line_subsec}, by i) substituting the ansatz $\widetilde{v} = \alpha \widetilde{x} + \beta$ into the expression~\eref{f_1_expr} of $f_1$, and ii) requiring the resulting expression of $f_1$ to vanish, i.e. $ f_1 \left[ \widetilde{x},\alpha \widetilde{x} + \beta \right] = 0$. Then, as we discuss in~\ref{fringes_subsec}, we substitute the resulting ansatz $\widetilde{v} = \alpha \widetilde{x} + \beta$ into the expression~\eref{f_2_expr} of $f_2$, and require $\sin f_2$ to vanish, i.e. $ \sin \left[ f_2 \left( \widetilde{x},\alpha \widetilde{x} + \beta \right) \right] = 0$. Since the latter condition hence sets $f_2 \left( \widetilde{x},\alpha \widetilde{x} + \beta \right) = k \pi$ with $k \in \mathbb{Z}$, it thus yields a countable family of solutions $\{\widetilde{x}_k^{(2)},\widetilde{v}_k^{(2)}\}$: the latter precisely describe the position of the interference fringes in the phase space.


\subsection{The line \texorpdfstring{$\widetilde{v}(\widetilde{x})$}{v}}\label{line_subsec}

A numerical analysis of the double-slit Husimi distribution (see figure~\ref{an_vs_num_H_diff_fig}) strongly suggests that i) the peaks of $F_{\mathrm{2slit}}$ are arranged on a line $\widetilde{v}(\widetilde{x}) = \alpha \widetilde{x} + \beta$, for some $\alpha > 0$ and $\beta \in \mathbb{R}$, and ii) that $f_1$ vanishes along this line $\widetilde{v}(\widetilde{x})$: we hence take the latter as our starting point in order to determine the parameters $\alpha$ and $\beta$.

First, our numerical results suggest that $(x_t,v_0)$, i.e. $(|x_0|,v_0)$ in view of~\eref{t_2_t_c}, belongs to the line $\widetilde{v}(\widetilde{x})$, i.e. $\widetilde{v}(|x_0|) = \alpha |x_0| + \beta = v_0$, which hence readily yields for $\beta$
\begin{eqnarray}
\beta = v_0 - \alpha |x_0| \, .
\label{beta_gen_expr}
\end{eqnarray}
We hence get for $\widetilde{v}(\widetilde{x})$
\begin{eqnarray}
\widetilde{v}(\widetilde{x}) = \alpha \left( \widetilde{x} - |x_0| \right) + v_0 \, .
\label{v_tilde_x_tilde_def}
\end{eqnarray}
We then introduce the variables $X$ and $V$ defined by
\begin{eqnarray}
X \equiv \widetilde{x} - |x_0| \qquad \mbox{and} \qquad V \equiv \widetilde{v} (\widetilde{x}) - v_0 \, ,
\label{X_V_def}
\end{eqnarray}
which in view of~\eref{v_tilde_x_tilde_def} are thus related through
\begin{eqnarray}
V = \alpha X \, .
\label{X_V_rel}
\end{eqnarray}
Substituting~\eref{X_V_def}-\eref{X_V_rel} into~\eref{f_1_expr} hence yields
\begin{eqnarray}
\fl f_1 \left[ \widetilde{x},\widetilde{v}(\widetilde{x}),2t_{\mathrm{c}};T_{\mathrm{op}}^{(0)},T_{\mathrm{op}}^{(1)} \right] = \frac{1}{2 \sigma^2} \frac{1}{1 + \hbar^2 ( T_{\mathrm{op}}^{(0)} )^2 / m^2 \sigma^4} \frac{1}{1 + \hbar^2 ( T_{\mathrm{op}}^{(1)} )^2 / m^2 \sigma^4} \left[ T_{\mathrm{op}}^{(1)} - T_{\mathrm{op}}^{(0)} \right] \nonumber\\
\times \left( \bar{\alpha} X + \bar{\beta} \right) X \, ,
\label{f_1_X_V_def}
\end{eqnarray}
where the quantities $\bar{\alpha}$ and $\bar{\beta}$ are defined by
\begin{eqnarray}
\bar{\alpha} \equiv 2 \alpha \left[ 1 - \frac{\hbar^2 T_{\mathrm{op}}^{(0)} T_{\mathrm{op}}^{(1)}}{m^2 \sigma^4} \right] - \alpha^2 \left[ T_{\mathrm{op}}^{(0)} + T_{\mathrm{op}}^{(1)} \right] + \frac{\hbar^2 \left[ T_{\mathrm{op}}^{(0)} + T_{\mathrm{op}}^{(1)} \right]}{m^2 \sigma^4}
\label{alpha_bar_def}
\end{eqnarray}
and
\begin{eqnarray}
\fl \bar{\beta} \equiv 2 \left( |x_0| \alpha + v_0 \right) \left[ 1 - \frac{\hbar^2 T_{\mathrm{op}}^{(0)} T_{\mathrm{op}}^{(1)}}{m^2 \sigma^4} \right] - 2 v_0 \alpha \left[ T_{\mathrm{op}}^{(0)} + T_{\mathrm{op}}^{(1)} \right] + 2 |x_0| \frac{\hbar^2 \left[ T_{\mathrm{op}}^{(0)} + T_{\mathrm{op}}^{(1)} \right]}{m^2 \sigma^4} \, .
\label{beta_bar_def}
\end{eqnarray}

We now require that $\bar{\alpha} = 0$, that is in view of~\eref{alpha_bar_def}
\begin{eqnarray}
\alpha^2 - \frac{2}{T_{\mathrm{op}}^{(0)} + T_{\mathrm{op}}^{(1)}} \left[ 1 - \frac{\hbar^2 T_{\mathrm{op}}^{(0)} T_{\mathrm{op}}^{(1)}}{m^2 \sigma^4} \right] \alpha - \frac{\hbar^2}{m^2 \sigma^4} = 0 \, .
\label{alpha_bar_zero_cond}
\end{eqnarray}
The quadratic (in $\alpha$) equation~\eref{alpha_bar_zero_cond} hence admits the two solutions $\alpha_{\pm}$ given by
\begin{eqnarray}
\alpha_{\pm} \equiv \frac{1}{2} \left\{ \frac{2}{T_{\mathrm{op}}^{(0)} + T_{\mathrm{op}}^{(1)}} \left[ 1 - \frac{\hbar^2 T_{\mathrm{op}}^{(0)} T_{\mathrm{op}}^{(1)}}{m^2 \sigma^4} \right] \pm \sqrt{\Delta} \right\}
\label{alpha_pm_def}
\end{eqnarray}
in terms of the discriminant
\begin{eqnarray}
\fl \Delta \equiv \frac{4}{\left[ T_{\mathrm{op}}^{(0)} + T_{\mathrm{op}}^{(1)} \right]^2} \left\{ 1 + \frac{\hbar^2 \left[ (T_{\mathrm{op}}^{(0)})^2 + (T_{\mathrm{op}}^{(1)})^2 \right]}{m^2 \sigma^4} + \left[ \frac{\hbar^2 T_{\mathrm{op}}^{(0)} T_{\mathrm{op}}^{(1)}}{m^2 \sigma^4} \right]^2 \right\} \, ,
\label{Delta_def}
\end{eqnarray}
which is thus positive by construction. Furthermore, substituting~\eref{sym_times_dble_slit} into~\eref{Delta_def} [and remembering~\eref{epsilon_small}] readily shows that
\begin{eqnarray}
\Delta = \frac{1}{t_{\mathrm{c}}^2} \left[ \left( 1 + \frac{\hbar^2 t_{\mathrm{c}}^2}{m^2 \sigma^4} \right)^2 + \bigO \left( \frac{\hbar^2 t_{\mathrm{c}}^2}{m^2 \sigma^4} \epsilon^2 \right) \right] \, .
\label{Delta_expr}
\end{eqnarray}
Therefore, taking the square root of~\eref{Delta_expr} and using $\sqrt{1+y} = 1 + \bigO (y)$ we get
\begin{eqnarray}
\sqrt{\Delta} = \frac{1}{t_{\mathrm{c}}} \left[ 1 + \frac{\hbar^2 t_{\mathrm{c}}^2}{m^2 \sigma^4} + \bigO \left( \frac{\hbar^2 t_{\mathrm{c}}^2}{m^2 \sigma^4} \epsilon^2 \right) \right] \, .
\label{Delta_expand}
\end{eqnarray}
Substituting~\eref{Delta_expand} into~\eref{alpha_pm_def} hence yields the two roots
\begin{eqnarray}
\alpha_+ = \frac{1}{t_{\mathrm{c}}} \left[ 1 + \bigO \left( \frac{\hbar^2 t_{\mathrm{c}}^2}{m^2 \sigma^4} \epsilon^2 \right) \right]
\label{alpha_plus_expr}
\end{eqnarray}
and
\begin{eqnarray}
\alpha_- = - \frac{\hbar^2 t_{\mathrm{c}}}{m^2 \sigma^4} \left[ 1 + \bigO \left( \epsilon^2 \right) \right] \, .
\label{alpha_minus_expr}
\end{eqnarray}

Note on~\eref{alpha_minus_expr} that $\alpha_-$ is negative: since it is clear from our numerical results (see figure~\ref{an_vs_num_H_diff_fig}) that the peaks of the Husimi distribution are arranged along a line that has a positive slope, the solution $\alpha_-$ must thus be discarded. Therefore, the slope $\alpha$ of the line~\eref{X_V_rel} is given by the solution $\alpha_+$, that is in view of~\eref{alpha_plus_expr}
\begin{eqnarray}
\alpha = \alpha_+ = \frac{1}{t_{\mathrm{c}}} \left[ 1 + \bigO \left( \frac{\hbar^2 t_{\mathrm{c}}^2}{m^2 \sigma^4} \epsilon^2 \right) \right] \, .
\label{alpha_expr_app}
\end{eqnarray}
Furthermore, substituting~\eref{alpha_expr_app} into~\eref{beta_bar_def} yields, using again~\eref{sym_times_dble_slit},
\begin{eqnarray}
\bar{\beta} = 4 v_0 \bigO \left( \frac{\hbar^2 t_{\mathrm{c}}^2}{m^2 \sigma^4} \epsilon^2 \right) \, .
\label{beta_bar_zero}
\end{eqnarray}
Furthermore, substituting~\eref{alpha_expr_app} into~\eref{beta_gen_expr} yields
\begin{eqnarray}
\beta = v_0 \bigO \left( \frac{\hbar^2 t_{\mathrm{c}}^2}{m^2 \sigma^4} \epsilon^2 \right) \, ,
\label{beta_final_expr}
\end{eqnarray}
so that $\beta$ vanishes to first order in $\epsilon$. Therefore, substituting~\eref{alpha_expr_app} into~\eref{f_1_X_V_def} and using~\eref{sym_times_dble_slit} and~\eref{epsilon_small} indeed ensures that we have
\begin{eqnarray}
f_1 \left[ \widetilde{x}, \widetilde{v} (\widetilde{x}),2t_{\mathrm{c}};T_{\mathrm{op}}^{(0)},T_{\mathrm{op}}^{(1)} \right] = \bigO \left( \frac{\hbar^2 t_{\mathrm{c}}^2}{m^2 \sigma^4} \epsilon^3 \right) \, ,
\label{f_1_zero}
\end{eqnarray}
showing that $f_1$ indeed vanishes, to first order in $\epsilon$, on the line
\begin{eqnarray}
\widetilde{v}(\widetilde{x}) = \frac{\widetilde{x}}{t_{\mathrm{c}}} + \bigO \left( \frac{\hbar^2 t_{\mathrm{c}}^2}{m^2 \sigma^4} \epsilon^2 \right) \, .
\label{v_tilde_x_tilde_expr}
\end{eqnarray}

We now derive the phase-space positions of the interference fringes.


\subsection{Derivation of \texorpdfstring{$\widetilde{x}_k^{(2)}$}{} and \texorpdfstring{$\widetilde{v}_k^{(2)}$}{the fringes}}\label{fringes_subsec}

We determined in~\ref{line_subsec} the line~\eref{v_tilde_x_tilde_expr} that we expect contains the phase-space points $\{ \widetilde{x}_k^{(2)},\widetilde{v}_k^{(2)} \}$ that characterize the positions of the interference fringes. Our strategy to obtain the latter is to impose that the sine of the function $f_2$ along this line $\widetilde{v}(\widetilde{x})$ must vanish (to first order in $\epsilon$), that is
\begin{eqnarray}
\sin \left\{ f_2 \left[ \widetilde{x}, \widetilde{v} (\widetilde{x}),2t_{\mathrm{c}};T_{\mathrm{op}}^{(0)},T_{\mathrm{op}}^{(1)} \right] \right\} = \bigO \left( \epsilon^2 \right) \, .
\label{sine_f_2_zero}
\end{eqnarray}

Therefore, we first compute $f_2$ along the line $\widetilde{v} (\widetilde{x})$. Noting, in view of~\eref{sym_times_dble_slit}, that
\begin{eqnarray}
\frac{1}{1 + \hbar^2 ( T_{\mathrm{op}}^{(0)} )^2 / m^2 \sigma^4} = 1 - \frac{\hbar^2 t_{\mathrm{c}}^2}{m^2 \sigma^4} (1-\epsilon)^2 + \bigO \left[ \frac{\hbar^4 t_{\mathrm{c}}^4}{m^4 \sigma^8} (1-\epsilon)^4 \right]
\label{fraction_T_op_0_expand}
\end{eqnarray}
and
\begin{eqnarray}
\frac{1}{1 + \hbar^2 ( T_{\mathrm{op}}^{(1)} )^2 / m^2 \sigma^4} = 1 - \frac{\hbar^2 t_{\mathrm{c}}^2}{m^2 \sigma^4} (1+\epsilon)^2 + \bigO \left[ \frac{\hbar^4 t_{\mathrm{c}}^4}{m^4 \sigma^8} (1+\epsilon)^4 \right] \, ,
\label{fraction_T_op_1_expand}
\end{eqnarray}
we hence get upon substituting~\eref{v_tilde_x_tilde_expr} and~\eref{fraction_T_op_0_expand}-\eref{fraction_T_op_1_expand} into~\eref{f_2_expr}
\begin{eqnarray}
f_2 \left[ \widetilde{x},\widetilde{v} (\widetilde{x}),2t_{\mathrm{c}};T_{\mathrm{op}}^{(0)},T_{\mathrm{op}}^{(1)} \right] = - \frac{m \epsilon}{\hbar t_{\mathrm{c}}} \left( \widetilde{x}^2 - x_0^2 \right) \left[ 1 + \bigO \left( \frac{\hbar^2 t_{\mathrm{c}}^2}{m^2 \sigma^4} \epsilon^2 \right) \right] \, .
\label{f_2_expr_expand}
\end{eqnarray}
Combining now~\eref{sine_f_2_zero} with~\eref{f_2_expr_expand} hence requires that
\begin{eqnarray}
\frac{m \epsilon}{\hbar t_{\mathrm{c}}} \left( \widetilde{x}^2 - x_0^2 \right) \left[ 1 + \bigO \left( \frac{\hbar^2 t_{\mathrm{c}}^2}{m^2 \sigma^4} \epsilon^2 \right) \right] = k \pi \, ,
\label{sine_zero_expl_cond}
\end{eqnarray}
for $k \in \mathbb{Z}$. Since we have
\begin{eqnarray*}
\frac{1}{1 + \bigO \left( \frac{\hbar^2 t_{\mathrm{c}}^2}{m^2 \sigma^4} \epsilon^2 \right)} = 1 + \bigO \left( \frac{\hbar^2 t_{\mathrm{c}}^2}{m^2 \sigma^4} \epsilon^2 \right) \, ,
\end{eqnarray*}
we hence get from~\eref{sine_zero_expl_cond}
\begin{eqnarray}
\widetilde{x}^2 = x_0^2 \left\{ 1 + \frac{\hbar}{m v_0^2 t_{\mathrm{c}} \epsilon} k \pi \left[ 1 + \bigO \left( \frac{\hbar^2 t_{\mathrm{c}}^2}{m^2 \sigma^4} \epsilon^2 \right) \right] \right\} \, .
\label{x_tilde_fringes_cond}
\end{eqnarray}
Taking the square root of~\eref{x_tilde_fringes_cond} hence readily yields the positions $\widetilde{x}_k^{(2)}$ of the interference fringes, namely
\begin{eqnarray}
\widetilde{x}_k^{(2)} = |x_0| \sqrt{1 + \frac{\hbar}{m v_0^2 t_{\mathrm{c}} \epsilon} k \pi} \, ,
\label{position_interf_fringes_expand}
\end{eqnarray}
that is alternatively, recognizing that in view of~\eref{sym_times_dble_slit} we have $t_{\mathrm{c}} \epsilon =  [T_{\mathrm{op}}^{(1)} - T_{\mathrm{op}}^{(0)}]/2$,
\begin{eqnarray}
\widetilde{x}_k^{(2)} = \left| x_0 \right| \sqrt{1 + \frac{2 \hbar}{m v_0^2 \left[ T_{\mathrm{op}}^{(1)} - T_{\mathrm{op}}^{(0)} \right]} k \pi} \, .
\label{x_k_def_app}
\end{eqnarray}
The latter readily yields the corresponding velocities $\widetilde{v}_k^{(2)}$, since in view of~\eref{v_tilde_x_tilde_expr} we can write (neglecting the $\epsilon^2$ terms)
\begin{eqnarray}
\widetilde{v}_k^{(2)} = \frac{\widetilde{x}_k^{(2)}}{t_{\mathrm{c}}} \, .
\label{v_tilde_k_x_tilde_k_expr}
\end{eqnarray}
Substituting~\eref{x_k_def_app} into~\eref{v_tilde_k_x_tilde_k_expr} hence yields (also recalling that $t_{\mathrm{c}} \equiv |x_0|/v_0$) 
\begin{eqnarray}
\widetilde{v}_k^{(2)} = v_0 \sqrt{1 + \frac{2 \hbar}{m v_0^2 \left[ T_{\mathrm{op}}^{(1)} - T_{\mathrm{op}}^{(0)} \right]} k \pi} \, .
\label{v_k_def_app}
\end{eqnarray}

The expressions~\eref{x_k_def_app} and~\eref{v_k_def_app} of the positions $\widetilde{x}_k^{(2)}$ and velocities $\widetilde{v}_k^{(2)}$ of the interference fringes exhibited by the double-slit Husimi distribution~\eref{Husimi_dist_double_slit_expr} have been obtained by focusing only on the terms $\cosh f_1$ and $\cos f_2$ in the expression~\eref{Husimi_dist_double_slit_expr} of $F_{\mathrm{2slit}}$. Therefore, we emphasize that our analysis up to this point does not yet ensure that the phase-space points $\{ \widetilde{x}_k^{(2)}, \widetilde{v}_k^{(2)} \}$ indeed correspond to actual critical points of $F_{\mathrm{2slit}}$, which would require to show that
\begin{eqnarray}
\left. \frac{\partial F_{\mathrm{2slit}}}{\partial \widetilde{x}} \right|_{\{ \widetilde{x}, \widetilde{v} \} = \{ \widetilde{x}_k^{(2)}, \widetilde{v}_k^{(2)} \}} = \left. \frac{\partial F_{\mathrm{2slit}}}{\partial \widetilde{v}} \right|_{\{ \widetilde{x}, \widetilde{v} \} = \{ \widetilde{x}_k^{(2)}, \widetilde{v}_k^{(2)} \}} = 0 \, ,
\label{crit_points_def}
\end{eqnarray}
to the leading order in $\epsilon$. While this can technically be done, it would require some tedious algebra. Therefore, here we choose a pragmatic approach: we test the validity of our analytic expressions~\eref{x_k_def_app} and~\eref{v_k_def_app} by numerically evaluating the Husimi distribution $F_{\mathrm{2slit}}$. As is clear from figure~\ref{peaks_fig}, the agreement between our expressions~\eref{x_k_def_app} and~\eref{v_k_def_app} and the numerical results is excellent, as we are indeed able to predict, with an excellent precision, the positions of the dark and bright interference fringes, i.e. the minima and maxima, respectively, of the Husimi distribution.


\section*{References}

\bibliographystyle{unsrt}
\bibliography{DATABASE_Residue}


\end{document}